# The peculiar spectral evolution of the new X-ray transient MAXI J0637−430

R. C. Ma,[1,2] R. Soria,[3,4] L. Tao,[1]★ W. Zhang,[1,2] J. L. Qu,[1] S. N. Zhang,[1,2] L. Zhang,[1] E. L. Qiao,[2,5] S. J. Zhao,[1,2] M. Y. Ge,[1] X. B. Li,[1] Y. Huang,[1] L. M. Song,[1] S. Zhang,[1] Q. C. Bu,[6] Y. N. Wang,[7] X. Ma,[1] S. M. Jia[1]

[1]*Key Laboratory of Particle Astrophysics, Institute of High Energy Physics, Chinese Academy of Sciences, Beijing 100049, People's Republic of China*
[2]*University of Chinese Academy of Sciences, Chinese Academy of Sciences, Beijing 100049, People's Republic of China*
[3]*College of Astronomy and Space Sciences, University of the Chinese Academy of Sciences, Beijing 100049, People's Republic of China*
[4]*Sydney Institute for Astronomy, School of Physics A28, The University of Sydney, Sydney, NSW 2006, Australia*
[5]*Key Laboratory of Space Astronomy and Technology, National Astronomical Observatories, Chinese Academy of Sciences, Beijing 100012, People's Republic of China*
[6]*Institut für Astronomie und Astrophysik, Kepler Center for Astro and Particle Physics, Eberhard Karls Universität, 72076 Tübingen, Germany*
[7]*Physics and Astronomy, University of Southampton, Southampton, Hampshire SO17 1BJ, UK*

**ABSTRACT**

We studied the transient Galactic black hole candidate MAXI J0637−430 with data from *Insight-HXMT*, *Swift* and *XMM-Newton*. The broad-band X-ray observations from *Insight-HXMT* help us constrain the power-law component. MAXI J0637−430 is located at unusually high Galactic latitude; if it belongs to the Galactic thick disk, we suggest a most likely distance ≲7 kpc. Compared with other black hole transients, MAXI J0637−430 is also unusual for other reasons: a fast transition to the thermal dominant state at the start of the outburst; a low peak temperature and luminosity (we estimate them at ≈0.7 keV and ≲0.1 times Eddington, respectively); a short decline timescale; a low soft-to-hard transition luminosity (≲0.01 times Eddington). We argue that such properties are consistent with a small binary separation, short binary period ($P \sim 2$ hr), and low-mass donor star ($M_2 \sim 0.2\ M_\odot$). Moreover, spectral modelling shows that a single disk-blackbody component is not a good fit to the thermal emission. Soft spectral residuals, and deviations from the standard $L_{\rm disk} \propto T_{\rm in}^4$ relation, suggest the need for a second thermal component. We propose and discuss various scenarios for such component, in addition to those presented in previous studies of this source. For example, a gap in the accretion disk between a hotter inner ring near the innermost stable orbit, and a cooler outer disk. Another possibility is that the second thermal component is the thermal plasma emission from an ionized outflow.

## 1 INTRODUCTION

Black hole (BH) candidates in a binary system with a low-mass donor star go through cycles of outburst and quiescence. The general consensus is that outbursts are caused by a thermal-viscous instability in the accretion disk (Lasota 2001). The X-ray spectra in outburst are primarily composed of an accretion disk component and a corona component. The accretion disk is optically thick and geometrically thin (Shakura & Sunyaev 1973) if the accretion rate is not too close to the Eddington limit; its emission spectrum is approximated by a multicolour blackbody. The hot, geometrically thick corona produces a power-law spectrum with a high-energy cutoff.

Outbursts proceeds through a sequence of accretion states. In the canonical sequence (Fender, Belloni, & Gallo 2004; Belloni et al. 2005; Remillard & McClintock 2006; Fender & Muñoz-Darias 2016), an outburst starts in the hard state, in which the power-law component carries most of the X-ray flux, the inner disk is truncated far from the innermost stable circular orbit (ISCO), and a compact flat-spectrum jet is detected. The transition to the soft state corresponds to the condensation, collapse or ejection of the corona, the decrease of the inner disk radius down to ISCO, and the quenching of the jet. The spectrum becomes dominated by the multicolour disk emission. After accretion has drained most of the mass from the disk, a reverse transition occurs back to the power-law dominated hard state, with the inner disk moving away from ISCO and the jet turning on again. In this process, the X-ray source follows a characteristic counter-clockwise q-shaped track in the hardness-intensity diagram (HID) (Fender, Belloni, & Gallo 2004; Fender, Homan, & Belloni 2009; Belloni & Motta 2016).

One way to improve our understanding of BH accretion is to look at outbursts that do *not* pass through all the canonical steps. For example, some systems (at least on some occasions) do not reach the soft state, and remain dominated by the hard power-law component; such events are known as "failed outbursts" (Brocksopp et al. 2004). Examples of failed outbursts are those of H1743−322 in 2008 (Capitanio et al. 2009) and 2018 (Grebenev et al. 2020), that of GX 339-4 in 2013 (Fürst et al. 2015) and 2017 (García et al. 2019), and that of XTE J1550−564 in 2003 (Sturner & Shrader 2005). Failed outburst are not a rare occurrence, and account for ≈40% of all outbursts (Tetarenko et al. 2016). A plausible reason for the incomplete outburst cycle is that the accretion rate (and hence the luminosity) is too low, below the threshold (at $L_X \approx 0.1 L_{Edd}$) where the inner disk reaches ISCO and the corona cools and collapses (Capitanio et al. 2009; Tetarenko et al. 2016).

★ E-mail: taolian@ihep.ac.cn (IHEP)

© 2022 The Authors

An apparently opposite (and rarer) situation is an outburst that does not have an initial hard state, or at least one in which the initial luminosity rise in the hard state happens too quickly to be detected. This is the case in 4U 1630−472, a frequently outbursting source where the initial hard rise is usually either missed or lasts for <1 day (Capitanio et al. 2009; Baby et al. 2020). Unfortunately, 4U 1630−472 is seen through a high absorbing column ($N_H \approx 8 \times 10^{22}$ cm$^{-2}$: Baby et al. 2020), which makes it difficult to study the physical parameters of the soft thermal component.

Here, we report on the spectral properties and evolution of another X-ray transient, MAXI J0637−430, which went into the thermal dominant state almost immediately at the start of its outburst in 2019. This source is seen through a much lower absorbing column, which facilitates a study of its soft thermal emission. In particular, we will discuss whether the accretion disk properties of MAXI J0637−430 are the same as those of "canonical" BH transients that exhibit a longer hard state during outburst rise, or instead the faster switch to a thermal state signals some underlying physical differences.

MAXI J0637−430 was discovered by the *Monitor of All-sky X-ray Image* Gas Slit Camera (*MAXI/GSC*; Matsuoka et al. 2009) on 2019 November 2 = MJD 58789 (Negoro et al. 2019). Its X-ray outburst was then promptly monitored by the *Neil Gehrels Swift Observatory* X-ray Telescope (*Swift/XRT*; Burrows et al. 2005) from November 3 (Kennea et al. 2019), by the *Neutron star Interior Composition ExploreR* (*NICER*; Gendreau, Arzoumanian, & Okajima 2012)) also from November 3 (Remillard et al. 2020), by *Insight-HXMT* (Zhang et al. 2020) from November 4, by the *Nuclear Spectroscopic Telescope Array* (*NuSTAR*; Harrison et al. 2013) from November 5 (Tomsick et al. 2019), and by *Astrosat* (Singh et al. 2014) from November 8 (Thomas et al. 2019, 2022).

The most obvious property of MAXI J0637−430 is that it was not in the hard state at the beginning of the outburst, or, at least, that the residence time in the hard state was too short to be caught (Tetarenko et al. 2021; Lazar et al. 2021; Jana et al. 2021; Baby et al. 2021). From the beginning of X-ray monitoring until its final decline and transition to the low/hard state, about 80 days later (Remillard et al. 2020; Tetarenko et al. 2021), the source was dominated by a thermal component with a temperature peaking at ≈0.6–0.7 keV, plus a power-law or Comptonization component with a photon index $\gtrsim 2$. This is consistent (to a first approximation) with a transient BH candidate in the canonical high/soft state. However, more unusually for a high/soft state spectrum, the *NuSTAR* and *Swift* data show (Lazar et al. 2021) that the soft spectrum cannot be well fitted with a single multicolour disk component. Residuals in the soft band suggest the presence of at least two components. Lazar et al. (2021) speculated that the additional soft component is either emission from the plunging region beyond ISCO, or a combination of thermal Comptonization and relativistic disk reflection of blackbody returning radiation.

In addition to the previously mentioned spectral studies, MAXI J0637−430 has also been the target of time-variability studies. Power density spectra from the *NICER* data show that the power decreases sharply at frequencies >10 Hz (Jana et al. 2021). There is no evidence of kilo-Hz quasi periodic oscillations or thermonuclear bursts at any stage during the outburst. The lack of those typical NS signatures, as well as the relatively low temperature of the thermal component(s), are consistent with the interpretation of this source as a BH candidate (Lazar et al. 2021; Jana et al. 2021).

In the optical/UV bands, the counterpart of MAXI J0637−430 was identified by *Swift/UVOT* (Kennea et al. 2019) at a location not previously associated with any optical stars; this suggests a low-mass donor, with an optical flux dominated by the irradiated accretion disk. The optical brightness at outburst peak was $U \approx 15$ mag, $V \approx B \approx 16.2$ mag (Tetarenko et al. 2021). The optical counterpart was bright enough to enable spectroscopic observations, first on November 3 with the Southern Astrophysical Research (SOAR) telescope (Strader et al. 2019) and then with Gemini in December (Tetarenko et al. 2021). The optical spectra reveal strong, broad, double-peaked H$\alpha$ and He II $\lambda$4686 emission lines, a tell-tale sign of X-ray-irradiated accretion disks in BH low-mass X-ray binaries (*e.g.,* Soria, Wu, & Hunstead 2000; Dubus et al. 2001; Charles & Coe 2006; Casares 2015, 2016).

MAXI J0637−430 was also detected in the radio bands (5.5 GHz and 9 GHz) by the Australia Telescope Compact Array on 2019 November 6 (Russell et al. 2019). The radio flux and spectral index are consistent with optically thin flaring or ejections at the onset of the soft state, rather than with a steady compact jet, typical of the hard state (Tetarenko et al. 2021). From the radio detections, the most precise position of the source was determined as R.A.(J2000) = 06$^h$36$^m$23$^s$.7 ±0$''$.2, Dec.(J2000) = −42°52$'$04$''$.1 ± 0$''$.7 (Russell et al. 2019).

In this paper, first we present the results of a previously unpublished series of high-cadence monitoring observations with *Insight-HXMT*. Then, we re-analyze the *Swift*/XRT data, to re-examine the claims that the spectral properties and evolution of the thermal emission are somewhat unusual compared with typical BH candidates in the high/soft state. We also use supporting data from the European Photon Imaging Camera (EPIC) on board *XMM-Newton*, taken at two different epochs of the outburst. We discuss possible interpretations of the soft X-ray emission, and how it may be related to the unusually short outburst-rise phase and low peak luminosity.

## 2 OBSERVATIONS AND DATA REDUCTION

### 2.1 *Insight-HXMT*

Following the *MAXI* and *Swift* detections, *Insight-HXMT*, China's first X-ray astronomy satellite (Zhang et al. 2020), triggered a Target of Opportunity (ToO) observation on 2019 November 4 (MJD 58791) and monitored the source for 17 times until November 29 (MJD 58816). The observation log is in Table A1. We used the *Insight-HXMT* Data Analysis Software (HXMTDAS) Version 2.02 for data reduction. In order to eliminate the effect of charged particles and limit the background level, we chose the following screening conditions for the good time intervals: (1) Earth elevation angle > 10°; (2) geomagnetic cutoff rigidity > 10 GV; (3) pointing offset angle < 0.04°. Background rates were estimated with a linear correlation coefficient between detectors with a small field of view and blind detectors (Li et al. 2019; Liao et al. 2020).

*Insight-HXMT* carries three detectors: high energy (HE: 35–150 keV; Liu et al. 2020), medium energy (ME: 10–35 keV; Cao et al. 2020) and low energy (LE: 1–10 keV; Chen et al. 2020). Because MAXI J0637−430 is a relatively weak and soft source, the count rates of the HE are too low for useful analysis. Thus, we only used the LE and ME data. The response files of LE and ME were generated with the tasks *lerspgen* and *merspgen* in the data reduction pipeline. We regrouped the extracted spectra to a minimum of 20 counts per bin, using *grppha* task in the FTOOLS package (Blackburn 1995). We then modelled them with the XSPEC software Version 12.10.1f (Arnaud 1996), using the $\chi^2$ statistics.

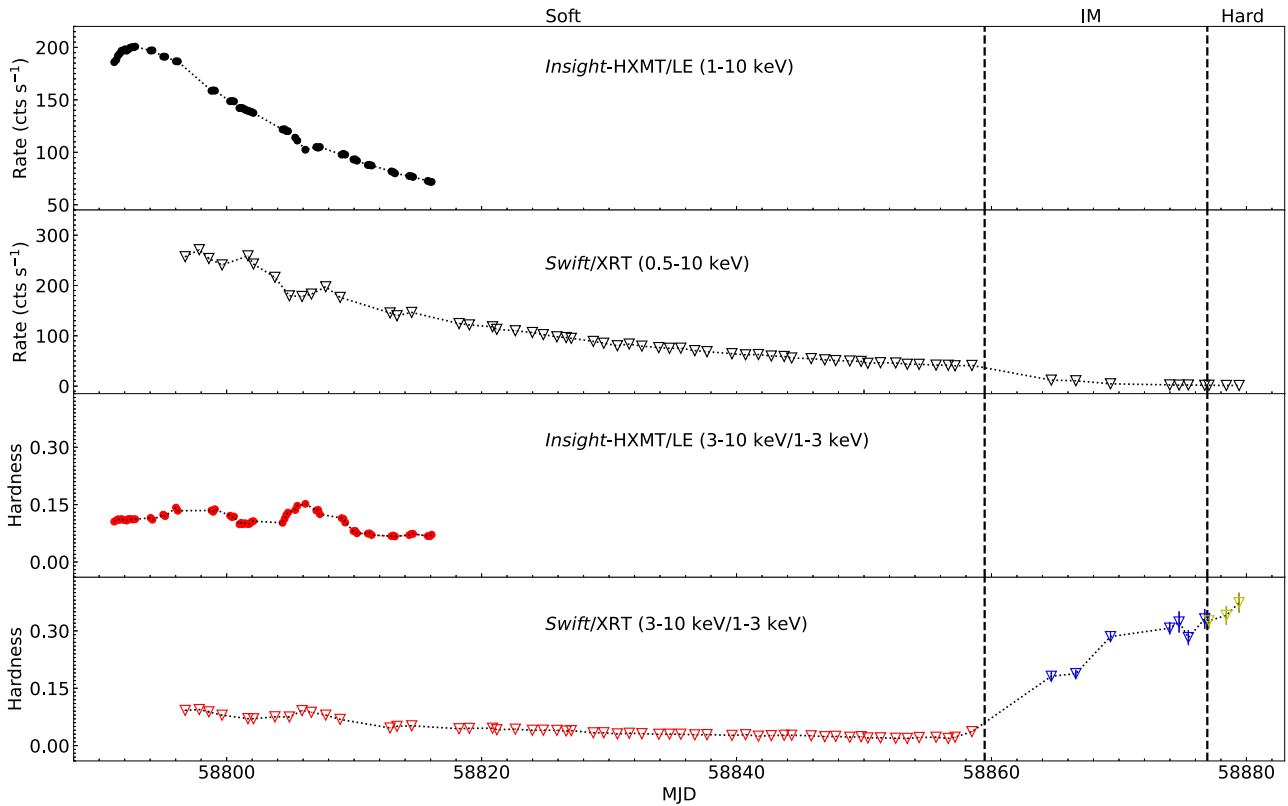

**Figure 1.** From top to bottom: *Insight-HXMT*/LE light curve; *Swift*/XRT light curve; *Insight-HXMT*/LE hardness ratio (3–10 keV over 1–3 keV count rate ratio); *Swift*/XRT hardness ratio (also 3–10 keV over 1–3 keV). Filled circles represent *Insight-HXMT* measurements, and open triangles are for *Swift*/XRT. In the *Swift*/XRT hardness ratio plot, the soft state (Soft), intermediate state (IM) and hard state (Hard) are displayed with red, blue and yellow symbols, respectively. Note that the error bars are so small as to not be clearly indicated in the figure.

### 2.2 *Swift*/XRT

*Swift*/XRT observed MAXI J0637−430 starting from 2019 November 3 (MJD 58790), and monitored the whole outburst. In this work, five observations (ObsIDs 00012162007, 00012163001, 00012167001, 00012168001 and 00012172001) are not used, as the target pointing offset is large. Moreover, in four other observations (ObsIDs 00012172002, 00012172007, 00012172015 and 00012172016), the source falls on bad columns of the CCD, which may cause the loss of counts and is difficult to correct[1]; thus, those observations were also discarded. In total, we retained for our analysis 65 XRT observations, taken in Windowed Timing (WT) mode, from 2019 November 9 (MJD 58796) to 2020 January 31 (MJD 58879) (Table A2).

We ran the *xrtpipeline* tool in FTOOLS to reprocess the data and used XSELECT (Blackburn 1995) to filter the event files, with the *Swift*/XRT calibration database (CALDB) version 20190910. In WT mode, if the count rates are >100 ct s$^{-1}$, the data are affected by pile-up (Romano et al. 2006). This is the case for at least the first half of the outburst. To avoid this problem, we defined an annulus as the source extraction region, excluding the piled-up central region. The inner radius of the annulus was selected to make the source count rate less than 100 ct s$^{-1}$, and the outer radius was fixed at 20 pixels. The background region was an annulus centered at the source position, with inner and outer radii of 90 pixels and 110 pixels, respectively. All spectra were regrouped to at least 20 counts per bin, with *grppha* in FTOOLS. Finally, we modelled the spectra in the 0.5–10.0 keV band and computed fluxes and luminosities with the XPEC software.

### 2.3 *XMM-Newton*/EPIC

*XMM-Newton* observed MAXI J0637−430 on 2019 November 17 (MJD 58804; ObsID 0853980801, 23.2 ks) and December 2 (MJD 58819; ObsID 0853981301, 22.3 ks). Both observations are in Burst Mode; however, for the purpose of this paper, we are interested in the spectral information. We downloaded the EPIC-pn Observation Data Files from NASA's High Energy Astrophysics Science Archive Research Center (HEASARC)[2]. We used the *XMM-Newton* Science Analysis System (SAS) v20.0 to process the data. We filtered out intervals of high particle background, using standard SAS routines.

---

[1] https://www.swift.ac.uk/analysis/xrt/digest_sci.php

[2] https://heasarc.gsfc.nasa.gov/cgi-bin/W3Browse/w3browse.pl

The good-time-interval considered for further analysis was 20.0 ks for ObsID 0853980801 and 19.2 ks for ObsID 0853981301. We ignored bad pixels with FLAG==0 and only used single and double events with PATTERN ≤ 4. Follow the recommendation of Kirsch et al. (2006), we extracted source events in RAWY < 140, and 29 ≤ RAWX ≤ 47, using the sas task *xmmselect*. We extracted the local background from 3 ≤ RAWX ≤ 5, near the edge of the chip, away from the source.

We then generated response and ancillary response files with *rmfgen* and *arfgen*. For EPIC-pn burst mode data, the instrumental calibration is reliable only over the 0.7–12 keV range[3]; therefore, we limited our spectral modelling to that band. Finally, we rebinned the spectra with the sas task *specgroup*, with the following two requirements: to have at least 25 counts for each background-subtracted spectral bin, and not to oversample the intrinsic energy resolution by a factor larger than 3. Finally, we used xspec to determine best-fitting spectral parameters, fluxes and luminosities.

## 3 RESULTS

### 3.1 Time variability and hardness ratios

The 1–10 keV background-subtracted light curve from *Insight-HXMT*/LE is shown in the top panel of Figure 1, and the 0.5–10 keV background-subtracted light curve from *Swift*/XRT (with pile-up correction) is shown in the second panel. The initial hard-state rising phase of the outburst was not caught either by *Insight-HXMT* or *Swift*. The LE count rate increased from 186 ct s$^{-1}$ in the first observation (MJD 58791 = 2019 November 4) to the peak value of 201 ct s$^{-1}$ on MJD 58793. Then, it decreased exponentially to 72 ct s$^{-1}$ over the next 22 days, until it became too faint to be detected by *Insight-HXMT*. The XRT count rate decreased from 257 ct s$^{-1}$ on MJD 58797 to 41 ct s$^{-1}$ on MJD 58858. After that, the count rate dropped: at MJD 58865, it was less than a third (12 ct s$^{-1}$) of the rate at MJD 58858. Finally, the flux declined as low as ≈1 ct s$^{-1}$ on MJD 58879.

The hardness ratio of the *Insight-HXMT*/LE data (Figure 1, third panel from the top), defined as the ratio of count rates in the hard band (3–10 keV) over those in the soft band (1–3 keV), varied between ≈0.05–0.15 but remained in the range expected for a soft state throughout the series of observations. For *Swift*/XRT (Figure 1, bottom panel), the hardness ratio over the same bands is ≲0.1 (typical of the soft state) until MJD 58858; then, in the next observing epoch, six days later, the source was caught in the middle of the transition towards the hard state, where it stayed until the end of the outburst. This outline of the outburst evolution is consistent with the spectral evolution discussed by Tetarenko et al. (2021).

Timing analysis for the *Insight-HXMT*/LE data shows a power-law-like power density spectrum at all epochs, rather than band-limited noise. The fractional root-mean-square noise in the 0.1–16 Hz band is ≈3%. This is again consistent with the timing properties of a typical high/soft state (Belloni et al. 2010). (The *Insight-HXMT*/ME and HE data are not usable for fast timing analysis, because their count rates are too low. The *Swift*/XRT data are also not usable for timing analysis, because of the strong pile-up.)

An initial rising phase is seen in the *MAXI* lightcurve (top panel of Figure 2, and Negoro et al. 2019). The source was detected for the first time above 3$\sigma$ significance, at a 2–20 keV flux of 41 ± 13 mCrab, on MJD 58789.3 (November 2), while it was still undetected in the previous orbit's scan (MJD 58789.2). The hardness ratio at the time of discovery was already only marginally higher than during the bright soft state (bottom panel in Figure 2). This suggests that the hard-to-soft state transition was also missed by *MAXI*. Monitoring observations with *NuSTAR* (Tomsick et al. 2019), *AstroSat* (Thomas et al. 2019; Baby et al. 2021) and *NICER* (Remillard et al. 2020; Jana et al. 2021; Baby et al. 2021) also confirm the soft nature at the beginning of the outburst. The lack or the very short duration of the initial hard state is unusual among BH LMXBs, which tend to have either full outbursts or hard-only outbursts.

---
[3] https://xmmweb.esac.esa.int/docs/documents/CAL-TN-0018.pdf

### 3.2 Basic spectral modelling

In the first step of our spectral analysis, we started from a simple standard model for BH transients: a multicolor disk-blackbody plus a power-law (diskbb+powerlaw in xspec). The neutral absorption was modelled with the Tuebingen-Boulder model (TBabs in xspec), with cross-sections from Verner et al. (1996) and abundances from Wilms, Allen, & McCray (2000).

The *Swift*/XRT spectra (0.5–10 keV band) have been extensively discussed by Tetarenko et al. (2021) and Lazar et al. (2021). It is not necessary to revisit every spectral property in this paper. We shall focus only on two specific issues. One (discussed in this Section) is the value of the inner-disk radius, derived from the normalization of the diskbb component; the other is the presence and origin of soft residuals (Section 3.3). The average of the apparent inner-disk radius over the first 10 *Swift* observations is $r_{\rm in}\sqrt{\cos i} \approx (33.0\pm1.3)\,d_{10}$ km, where the uncertainty is the 1-$\sigma$ dispersion of the best-fitting values, $i$ is the viewing angle to the disk plane, and $d_{10}$ is the (unknown) distance to the source in units of 10 kpc. An estimate of $i = 64° \pm 6°$ was provided by Lazar et al. (2021), based on the best-fitting parameters of a reflection component in their *NuSTAR* spectra (3–79 keV band). Including also the usual correction factor of 1.19 (Kubota et al. 1998), the physical inner-disk radius from the *Swift*/XRT spectra is $R_{\rm in} \approx 39\,(\cos i)^{-1/2}\,d_{10}$ km. This is consistent with the results of Tetarenko et al. (2021), who find, from their analysis of the same subset of *Swift*/XRT spectra, $r_{\rm in}\sqrt{\cos i} \approx (37 \pm 3)\,d_{10}$ km when using a diskbb+powerlaw model, or $r_{\rm in}\sqrt{\cos i} \approx (35 \pm 2)\,d_{10}$ km when using a diskir model. The peak colour temperature $kT_{\rm in} \approx 0.7$ keV found in our spectral modelling, and the general evolutionary trend during outburst decline, are also consistent with the results of Tetarenko et al. (2021).

For *Insight-HXMT*/LE+ME, our basic two-component model (TBabs*(diskbb+powerlaw)) provides a good fit to all datasets over the 1–35 keV range (Figure 3 and Table A3). The disk-blackbody component contributes ≈90% of the unabsorbed flux above 1 keV, as expected, during the observations of November 4–7 (Table A3, Column 7). After that, the disk contribution declines to ≈70–80%, with a variability from epoch to epoch corresponding to the hard excursions seen in the Figure 1. We also tried replacing the power-law with a cutoff power-law (cutoffpl in xspec). However, the cutoff energy is always unconstrained (>100 keV), and there is no statistical improvement compared with the simpler model. The absorbing column density $N_{\rm H}$ is not well constrained because of the lack of coverage below 1 keV: thus, we fixed it at the average value ($N_{\rm H} = 2 \times 10^{21}$ cm$^{-2}$) found from the *Swift*/XRT spectra in the soft state. The apparent inner disk radius is stable from observation to observation (Table A3), especially near outburst peak. Taking the best-fitting values of the first 10 days of *Insight-HXMT* observations (*i.e.*, all the data up to and including ObsID P0214057011, Table A3), we estimate $r_{\rm in}\sqrt{\cos i} \approx (55 \pm 2)\,d_{10}$ km. The colour temper-ature of the inner disk at the peak of the outburst is $kT_{\rm in} \approx 0.6$ keV. Thus, we notice a discrepancy between the *Swift*/XRT and the *Insight-HXMT* results. The inner disk radius inferred from *Swift* ≈ 40 per cent lower, while the inner-disk temperature is about 20 per cent higher, at corresponding epochs.

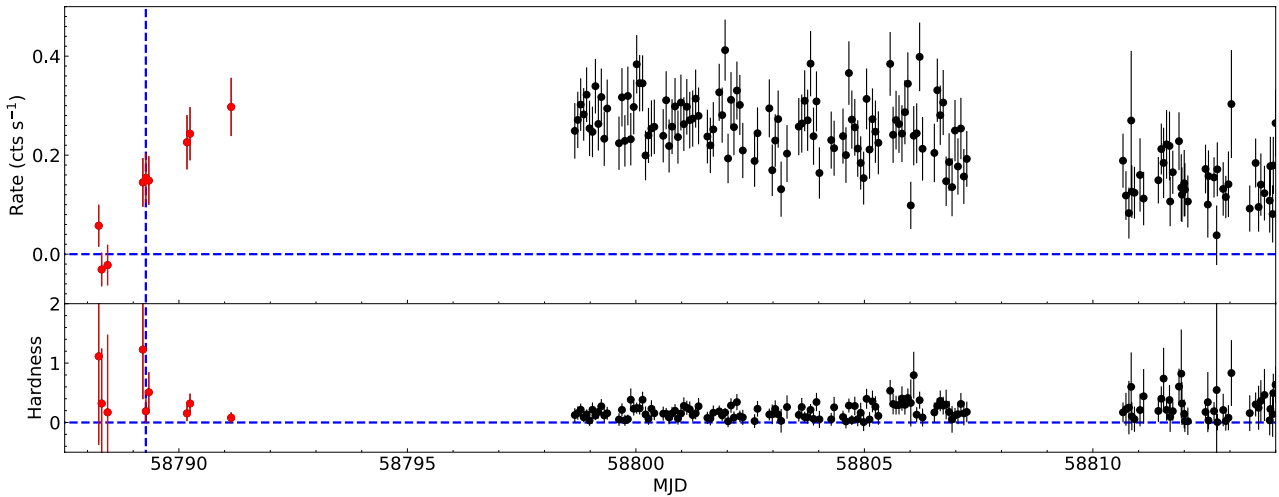

**Figure 2.** Top panel: *MAXI* light curve in the 2–20 keV band, binned to one datapoint per orbit. The dashed vertical line represents the time of discovery, and the datapoints plotted in red correspond to a time interval of ±6 hr before/after the discovery. Bottom panel: corresponding hardness ratios (4–20 keV over 2–4 keV count rate ratio).

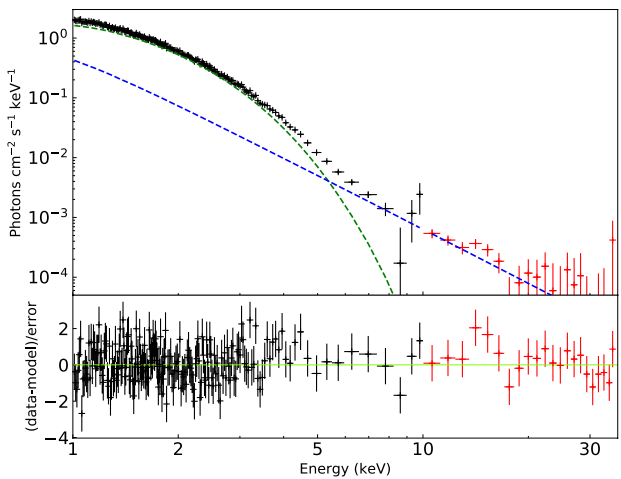

**Figure 3.** *Insight-HXMT* spectra of MAXI J0637−430 in the 1–35 keV band, from ObsID P021405700402 (MJD 58795, near the peak of the soft state), fitted with a `TBabs*(diskbb+powerlaw)` model (see Table A3 for the parameter values). The LE and ME data are represented by black and red symbols, respectively. The spectrum has been rebinned for display purpose only.

We then fitted the spectra from the two *XMM-Newton*/EPIC observations, again with `TBabs*(diskbb+powerlaw)`, over the 0.7–12 keV range. We obtained a radius $r_{\rm in}\sqrt{\cos i} \approx (48 \pm 2)\, d_{10}$ km and $r_{\rm in}\sqrt{\cos i} \approx (50 \pm 2)\, d_{10}$ km, respectively: a value intermediate between those found with *Swift* and *Insight-HXMT*. Finally, we recall that the average apparent radius derived from the *NICER* observations (0.7–10 keV band), computed over the same epochs we used to estimate the radius from *Insight-HXMT* and *Swift*, is $r_{\rm in}\sqrt{\cos i} \approx (55 \pm 1)\, d_{10}$ km (Jana et al. 2021).

Since the measurement of $r_{\rm in}$ is an important constraint on the BH mass and, hence, on the peak Eddington ratio of the outburst, it is worth investigating this curious discrepancy between different instruments in more detail. We have discussed the issue with the *Insight-HXMT* calibration group, who acknowledged that the cross-calibration of the LE response below ≈2 keV with other X-ray missions is still work-in-progress, and is affected by uncertainties in the intrinsic shape of the soft X-ray spectrum of the Crab nebula (calibration source for *Insight-HXMT*) (Kirsch et al. 2005; Li et al. 2020). In addition, if the intrinsic shape of the thermal emission component in MAXI J0637−430 is not a perfect disk-blackbody (as we shall discuss later), a *diskbb* approximation will give different values of $T_{\rm in}$ and $r_{\rm in}$ depending on the lower limit of the energy band used for fitting (0.5 keV for *Swift*/XRT, 0.7 keV for *XMM-Newton*/EPIC in timing mode, 1 keV for *Insight-HXMT*/LE) and on the different energy resolution and sensitivity of the various instruments.

### 3.3 Soft residuals

When fitted with our standard two-component model (`TBabs*(diskbb+powerlaw)`), the *Swift*/XRT spectra before MJD 58858 show significant systematic residuals at the low-energy end (≈0.5–1.0 keV). A typical example of such residuals is shown in Figure 4a. From MJD 58864 to MJD 58877, as the count rate and signal-to-noise ratio decrease, the soft residuals are no longer significant and a simple disk-blackbody component suffices to describe the thermal emission. After MJD 58877, the disk-blackbody component itself is no longer significant and the spectrum is consistent with a simple power-law. In our modelling, the neutral hydrogen column density was left as a free parameter over most of the epochs (typical values are $N_{\rm H} \approx 1.5$–$3 \times 10^{21}$ cm$^{-2}$) except for the last few epochs in the hard state, when it was fixed at $N_{\rm H} = 9 \times 10^{20}$ cm$^{-2}$, which is the average of the best-fitting values after MJD 58864. However, the soft residuals are present also if we fix $N_{\rm H}$ at a constant value throughout the outburst (which was the choice made for example by Tetarenko et al. 2021).

The apparent presence of soft residuals or of an additional soft thermal component is unusual for Galactic BH transients. We care-fully tested the possibility that the residuals are artifacts from the spectral extraction process. We compared our spectra (extracted with XSELECT) with those extracted from the same datasets with the on-line *Swift*/XRT data products generator[4] (Evans et al. 2009). We found that the spectra are statistically identical (red and black data-points in Figure 4a). Moreover, we consulted the Swift Science Data Center, and they confirmed that the residuals are not caused by calibration issues. Our results agree with Lazar et al. (2021), who also found residuals below 1 keV, near 6 keV and above 20 keV from their *Swift*/XRT and *NuSTAR* analysis. Residuals below 1 keV appear also in the *NICER* spectra in the soft-intermediate and soft state, and in the *AstroSat* spectra in the soft state (Fig. 4 of Jana et al. (2021) and Fig. 4 of Baby et al. (2021), respectively), when the spectra are fitted with two-component models.

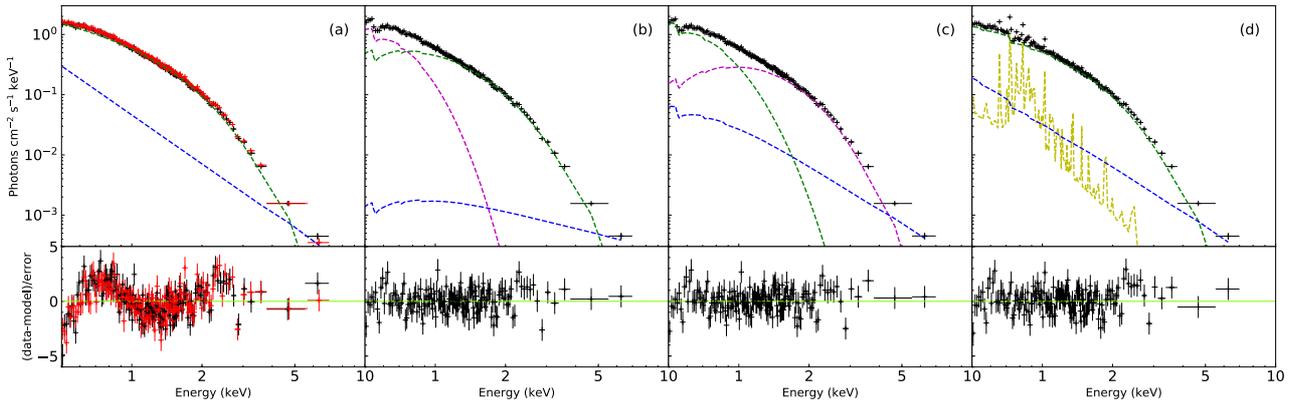

**Figure 4.** Unfolded *Swift*/XRT spectrum from ObsID 00012172039 (MJD 58837), fitted with alternative models in the 0.5–10.0 keV band. (a) The basic two-component model `TBabs*(diskbb+powerlaw)` shows strong systemic residuals in the 0.5–1.0 keV band. In this sub-panel, the black datapoints are the result of our independent data analysis, while the red ones come from the online *Swift*/XRT data products generator. (b) A significant improvement is obtained when we use a double thermal model: in this case, `TBabs*(bbodyrad_colder+diskbb_hotter+powerlaw)`. (c) A similarly good fit statistics is also obtained with `TBabs*(bbodyrad_hotter+diskbb_colder+powerlaw)`. (d) The soft residuals can also be reproduced with a combination of ionized wind absorption and emission, *i.e.*, `absori*TBabs*(apec+diskbb+powerlaw)`; the thermal-plasma emission component is the more important one, for a good fit statistics. In all the upper panels, the green, blue, purple and yellow dashed lines represent the `diskbb`, `powerlaw`, `bbodyrad` and `apec` components, respectively.

Next, we tried fitting the spectra with alternative two-component models: i) with the physical comptonization model `nthcomp` instead of `powerlaw`; ii) with the *p*-free disk model `diskpbb` instead of `diskbb`; iii) with `diskir` instead of `diskbb+powerlaw`. In all cases, we obtained a very similar structure of 0.5–1.0 keV residuals. This suggests that they are a real physical property, and that a single thermal component is not a good approximation of the thermal emission for this source.

To model the soft residuals, we considered three simple scenarios: 1) an additional blackbody component, cooler than the inner disk; 2) an additional blackbody component, hotter than the inner disk; 3) an additional ionized absorber plus optically thin thermal plasma emission. Scenario 1 may correspond to a reprocessing region, such as warps or tidal bulges on the disk surface, or downscattered emission in a wind (which would look blackbody-like at low spectral resolution). In Scenario 2, the main contribution to the soft emission would be from a truncated disk, with an additional source of hotter thermal photons from a smaller radius, at or near the innermost stable orbit. Scenario 3 is expected if there is a dense wind around the X-ray source. An ionized absorber removes photons mostly in the ≈1–2 keV band, creating the impression of a soft excess below 1 keV; at the same time, velocity-broadened emission lines from $\alpha$ elements around 0.5–1 keV can appear as a soft excess. In narrow-line Seyfert I galaxies (a class of active galaxies with a relatively low-mass nuclear BH in the high/soft state), there is a long-standing debate on whether the soft excess below 1 keV is produced by an optically thick emitter, or is the result of relativistically smeared absorption in an ionized outflow (as proposed by Gierliński & Done 2004b, 2006). A similar effect was noted for example in the spectrum of the transient ultraluminous X-ray source NGC 1365 X-1 near outburst peak (Soria et al. 2007). More generally, the spectral residuals in the soft X-ray spectra of ultraluminous X-ray sources are explained as a combination of smeared emission and absorption lines in fast outflows (Middleton et al. 2015; Pinto, Middleton, & Fabian 2016).

For the *Swift*/XRT spectra of MAXI J0637−430, each of the three alternative solutions substantially improves the quality of the fits, removing the residuals, at the expense of two additional free parameters (temperature and normalization of the additional thermal component) for Scenarios 1 and 2, and four free parameters (column density and ionization parameter of the ionized absorber, and temperature and normalization of the optically thin emitter) for Scenario 3. For example, we illustrate the situation of ObsID 00012172039 (Figure 4), in which the reduced $\chi^2_\nu$ decreases from 1.48 for 307 degrees of freedom (single disk component) to 0.96 for 305 degrees of freedom (Scenario 1), 0.98 for 305 degrees of freedom (Scenario 2) and 1.03 for 303 degrees of freedom (Scenario 3). In Section 4, we will discuss which of the three solutions may have a plausible physical interpretation.

### 3.3.1 *Scenario 1: colder blackbody + hotter disk*

We successfully modelled the *Swift*/XRT soft state spectra with `TBabs*(bbodyrad_colder+diskbb_hotter+powerlaw)` (Figure 4b). The evolution of the best-fitting parameters is reported in Figure 5 and Table A4. The colour temperature $kT_{\rm in}$ of the hotter component decreases from 0.7 keV to 0.4 keV, while the apparent inner disk radius $r_{\rm in}\sqrt{\cos i}$ is almost stable at ≈40 $d_{10}$ km. The temperature $kT_{\rm BB}$ of the colder component increases from 0.11 keV to 0.14 keV during the outburst decline. Its characteristic radius $R_{\rm BB}$ decreases from $\approx 1.9 \times 10^3 \, d_{10}$ km to $\approx 0.4 \times 10^3 \, d_{10}$ km (an order of magnitude larger than the innermost stable orbit).

---

[4] https://www.swift.ac.uk/user_objects/index.php

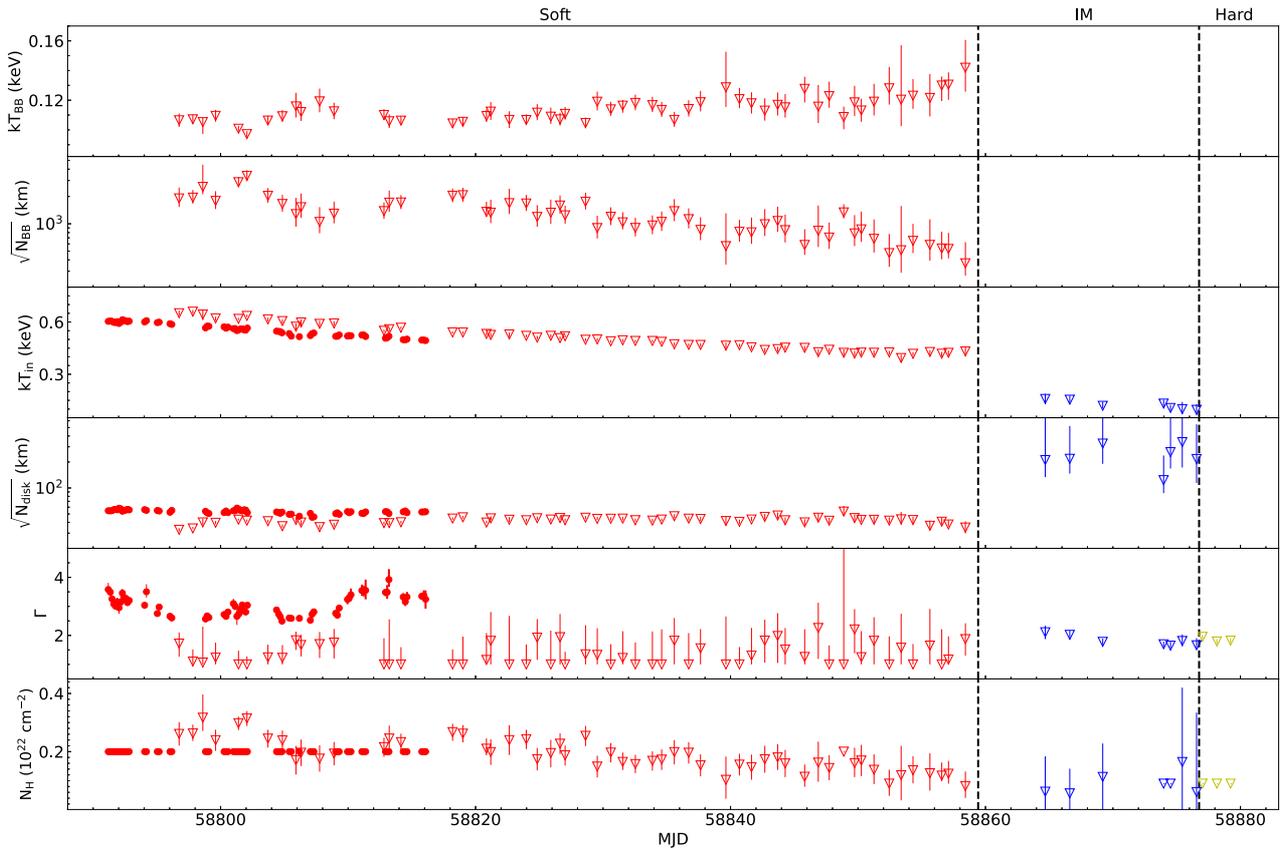

**Figure 5.** Time evolution of the model parameters in Scenario 1 (colder blackbody plus hotter disk-blackbody plus powerlaw). Open triangles represent the best-fitting values for the *Swift*/XRT data. For comparison, we also plotted the best-fitting parameters of a disk-blackbody plus powerlaw model for the *Insight-HXMT* data (filled circles). A second thermal component is not required for the *Insight-HXMT* spectra, because of its band limit at 1 keV; conversely, the powerlaw component is not required or not well constrained by the *Swift*/XRT data, but is well determined by *Insight-HXMT*. Specifically, $kT_{BB}$ is the temperature of the bbodyrad component, and $\sqrt{N_{BB}}$ is the square root of its normalization in XSPEC. $kT_{in}$ is the inner disk temperature, and $\sqrt{N_{disk}}$ is the square root of the diskbb normalization in XSPEC (units of km). $\Gamma$ is the powerlaw photon index. We set a lower limit of $\Gamma > 1$ for the *Swift*/XRT spectra, to avoid unphysically low values when the weak powerlaw was essentially undetected or unconstrained. $N_H$ is the neutral absorption column density, which is fixed at $2 \times 10^{21}$ cm$^{-2}$ for *Insight-HXMT* data. Different colors of the plotted symbols correspond to different states, as in Figure 1.

In the intermediate state, there is no longer a statistical improvement with two thermal components compared with a single one. If we model the intermediate-state thermal component as a disk-blackbody, its temperature is $kT_{in} \approx 0.10$–0.15 keV, and its apparent radius is $r_{in}\sqrt{\cos i} \approx 200$–300 $d_{10}$ km (a factor of 5 times larger than in the soft state). Thus, another way of interpreting this evolution is that the characteristic sizes and temperatures of the two thermal components seen in the soft state converge during outburst decline until they become indistinguishable.

### 3.3.2 Scenario 2: hotter blackbody + colder disk

We then tried a colder disk model and a hotter blackbody model for the two thermal components, namely TBabs*(bbodyrad$_{hotter}$+diskbb$_{colder}$+powerlaw). The results are shown in Figure 4c, Figure 6, and Table A5. The fit statistics are as good as those obtained for Scenario 1. The blackbody temperature $kT_{BB}$ gradually decreases from 0.47 keV to 0.32 keV, while the best-fitting size $R_{BB}$ fluctuates between $\approx$(60–80) $d_{10}$ km without a well-defined trend. The inner-disk colour temperature $kT_{in}$ remains almost unchanged at $\approx 0.18$ keV. The apparent inner disk radius gradually decreases from $r_{in}\sqrt{\cos i} \approx 800 \, d_{10}$ km to $r_{in}\sqrt{\cos i} \approx 250 \, d_{10}$ km. In other words, in this scenario, the truncation radius of the disk moves inwards during the outburst but remains always significantly larger (and colder) than the additional blackbody emitter, and presumably larger than the innermost stable orbit. A decrease of the inner disk radius without a corresponding increase in its temperature suggests a decline in the accretion rate through the disk. In the intermediate state at the end of the outburst, the single thermal component needed to fit the spectrum is consistent with a survival of the colder disk component; instead, the hotter component initially located closer to the innermost stable orbit disappears.

### 3.3.3 Scenario 3: ionized outflows

Thirdly, we considered the possibility of an optically thin wind, contributing both in emission and absorption. We modelled the spectra with TBabs*absori*(apec+diskbb+powerlaw). For the neutral absorber, we set a lower limit of $N_H = 2 \times 10^{20}$ cm$^{-2}$, which is half of the Galactic column density in the direction of the source, and is consistent with the value obtained by Lazar et al. (2021) from their joint *Swift* plus *NuSTAR* modelling. For simplicity, we assumed a single-temperature plasma. Temperatures ~0.7–1.0 keV for the emitting plasma, and column densities ~$10^{22}$ cm$^{-2}$ for the ionized absorber, provide a good fit to the spectral residuals (Figure 4d, Figure 7, Table A6). The disk parameters around outburst peak are $r_{in}\sqrt{\cos i} \approx 35\,d_{10}$ km, $kT_{in} \approx 0.65$ keV. During outburst decline, the inner-disk temperature decreases by a factor of 2, while the apparent inner-disk radius increases to $r_{in}\sqrt{\cos i} \approx 50\,d_{10}$ km before the transition to the intermediate state.

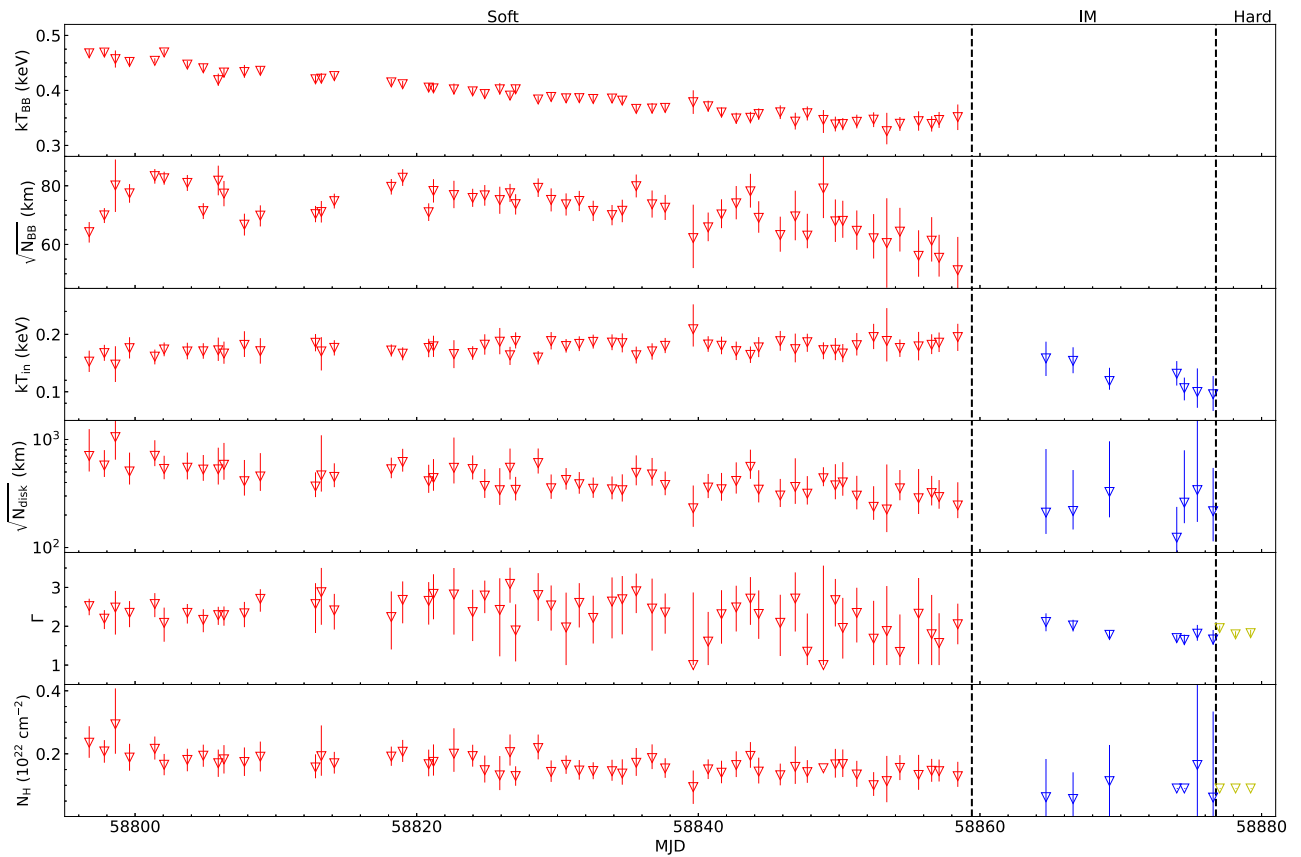

**Figure 6.** Time evolution of the best-fitting model parameters in Scenario 2 (hotter blackbody plus colder disk-blackbody plus powerlaw), for the *Swift*/XRT data. The physical parameters and their plotting colors are the same as in Figure 5.

### 3.4 Parameter evolution assuming constant $N_H$

In our spectral modelling reported so far, the column density of the cold absorber was left as a free parameter whenever possible, at least during the soft state. There are of course degeneracies between the value of $N_H$ and the best-fitting disk parameters. Different choices of $N_H$ do not remove the soft X-ray residuals, but can change the value of $r_{in}$ and hence the BH mass estimates. One common feature of all our spectral models is that the best-fitting value of $N_H$ tends to decrease near the end of the soft state (Figures 5, 6, 7) and is significantly lower in the intermediate and hard state. This can be a real effect, if some of the intrinsic absorption is caused by gas in the accretion inflow/outflows, which is reduced as the outburst declines.

On the other hand, we cannot rule out that the apparent decrease in $N_H$ masks a degeneracies with other parameters. To estimate the severity of this uncertainty, we repeated our modelling in all three scenarios, but this time assuming a constant $N_H$.

Two possible generic choices for the fixed value of $N_H$ are the Galactic line-of-sight value (inferred from optical/UV reddening), or an average or median of the best-fitting values of $N_H$ when left as a free parameter in the soft state. In the former case (adopted for example by Tetarenko et al. 2021), $N_H \sim 4 \times 10^{20}$ cm$^{-2}$. In the latter case, a characteristic value is $N_H \sim 2 \times 10^{21}$ cm$^{-2}$.

Here, we illustrate and briefly summarize the results of a constant-$N_H$ modelling and the best-fitting parameter comparison for Scenario 2 (Figure 8). The results for the other two scenarios are qualitatively similar. We find that the characteristic size of the colder thermal component (disk-blackbody in this case) increases significantly when $N_H$ is fixed at the lower value. The apparent inner-disk radius near outburst peak becomes $r_{in}\sqrt{\cos i} \approx 150$–$200\,d_{10}$ km for $N_H = 4 \times 10^{20}$ cm$^{-2}$, as opposed to $r_{in}\sqrt{\cos i} \approx 800\,d_{10}$ km for $N_H = 2 \times 10^{21}$ cm$^{-2}$. The inner-disk temperature correspondingly increases to $\approx 0.25$ keV. Even in this case, though, it is necessary to add a second, hotter thermal component, with a characteristic size of $\approx 60$–$75\,d_{10}$ km and a temperature of $\approx 0.5$ keV. In other words, regardless of our choice of $N_H$, both thermal components are still required, and there is still an apparent gap between their locations. The effect of a change in $N_H$ on the best-fitting parameters and evolutionary trend for the hotter component is very small, especially near outburst peak (Figure 8). Similar considerations apply to the modelling for Scenario 1 and Scenario 3, in which the disk-blackbody component is assumed to represent the hotter thermal emitter. In particular, in Scenario 1, the apparent inner-disk radius near outburst peak is $r_{in}\sqrt{\cos i} \approx 36 \, d_{10}$ km for $N_H = 4 \times 10^{20}$ cm$^{-2}$, as opposed to $r_{in}\sqrt{\cos i} \approx 37 \, d_{10}$ km for $N_H = 2 \times 10^{21}$ cm$^{-2}$. In Scenario 3, the inner-disk radii are $r_{in}\sqrt{\cos i} \approx 44 \, d_{10}$ km and $r_{in}\sqrt{\cos i} \approx 43 \, d_{10}$ km, for the two choices of $N_H$.

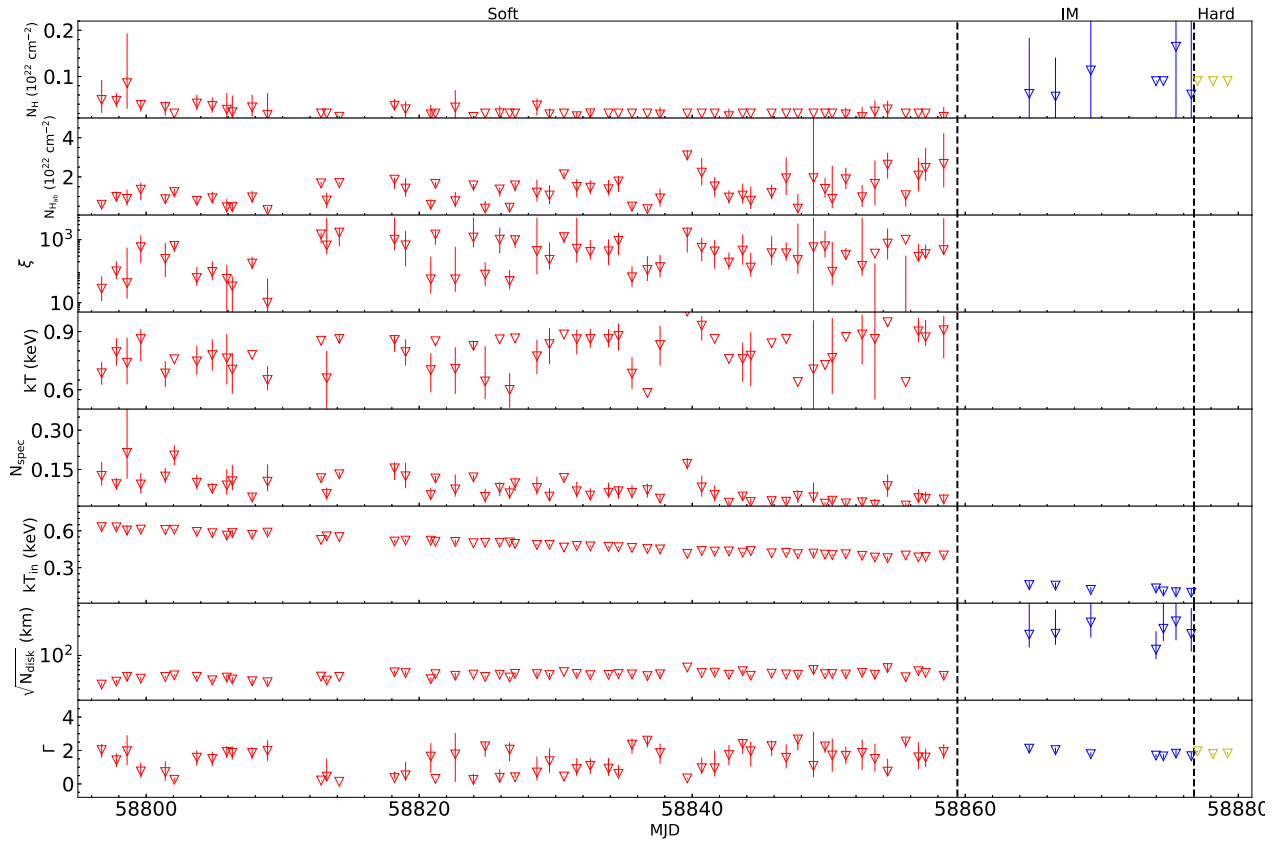

**Figure 7.** Time evolution of the best-fitting model parameters in Scenario 3 (ionized absorber plus thermal plasma emitter plus disk-blackbody plus powerlaw), for the *Swift*/XRT data. $N_{H_{ab}}$ is the hydrogen column density of the ionized absorber (`absori` model in XSPEC), and $\xi$ is its ionization parameter. $kT$ and $N_{apec}$ are the plasma temperature and the normalization of the `apec` model, respectively. The other physical parameters and their plotting colors are the same as in Figure 5. For the neutral absorber, we imposed a lower limit of $N_H = 2 \times 10^{20}$ cm$^{-2}$ in the fitting process, and we fixed $N_H$ at that value for the observations in which it could not be significantly determined.

### 3.5 Thermal and non-thermal flux

Based on the *Swift*/XRT data alone, with its narrower band coverage, the relative contribution of the thermal and non-thermal components and the power-law photon index are not well constrained (see for example the alternative models in Figure 4). Moreover, both *Swift*/XRT (light curve in the second panel from the top of Figure 1) and *MAXI* (Figure 2) may have missed the peak of the outburst. Instead, *Insight-HXMT* provides stronger constraints on those values.

From the *Insight-HXMT* data (Tables A1, A3) we estimate that outburst peak occurred between MJD 58792–58794 (ObsIDs P021405700105– P021405700302). The bolometric, unabsorbed thermal flux at outburst peak (single disk-blackbody model) was $F_{th} \approx 8.9 \times 10^{-9}$ erg cm$^{-2}$ s$^{-1}$. The power-law photon index at outburst peak was $\Gamma \approx 3.20 \pm 0.16$, where the error range is the 1-$\sigma$ scatter of the best-fit values over all the subexposures at MJD 58792–58794. In fact, $\Gamma > 3$ in every dataset prior to MJD 58795, and starts to decline slightly afterwards ($\Gamma \approx 2.9$ on MJD 58794, $\Gamma \approx 2.6$ on MJD 58796). The unabsorbed power-law flux in the 1–100 keV band at outburst peak was $F_{pow,1-100} \approx 0.63 \times 10^{-9}$ erg cm$^{-2}$ s$^{-1}$; we can take this value as a lower limit to the Comptonized emission. If we assume that the power-law component extends without a break down to 0.3 keV, we estimate $F_{pow,0.3-100} \approx 2.7 \times 10^{-9}$ erg cm$^{-2}$ s$^{-1}$; we take this as an upper limit to the total power-law flux.

To reduce the uncertainty on the power-law flux at the low-energy end, we also re-fitted the *Insight-HXMT* spectra at peak outburst with a `diskir` model. We found an unabsorbed 0.3–100 keV flux $F_{0.3-100} \approx 9.2 \times 10^{-9}$ erg cm$^{-2}$ s$^{-1}$, and a bolometric flux (estimated with an extrapolation to the 0.01–500 keV band) of $F_{bol} \approx 1.1 \times 10^{-8}$ erg cm$^{-2}$ s$^{-1}$. In the approximation of isotropic emission, this corresponds to a peak bolometric luminosity $L_{bol} \approx 1.3 \, (d_{10})^2 \, 10^{38}$ erg s$^{-1}$. Alternatively, if the flux is split into a thermal component (for which $L \propto 2\pi d^2 F/\cos\theta$) and a Comptonized component (for which $L \propto 4\pi d^2 F$), the peak bolometric luminosity is $L_{bol} \approx 1.5 \, (d_{10})^2 \, 10^{38}$ erg s$^{-1}$.

The unabsorbed luminosity depends only very weakly on the photon index of the Comptonized component: it changes by $\lesssim 10\%$ for the range $1.7 \leq \Gamma \leq 3$. However, a more accurate estimate of the relative contributions of thermal and non-thermal flux requires a more complex modelling of the thermal component(s): for example, splitting it into a disk-blackbody and a blackbody component.

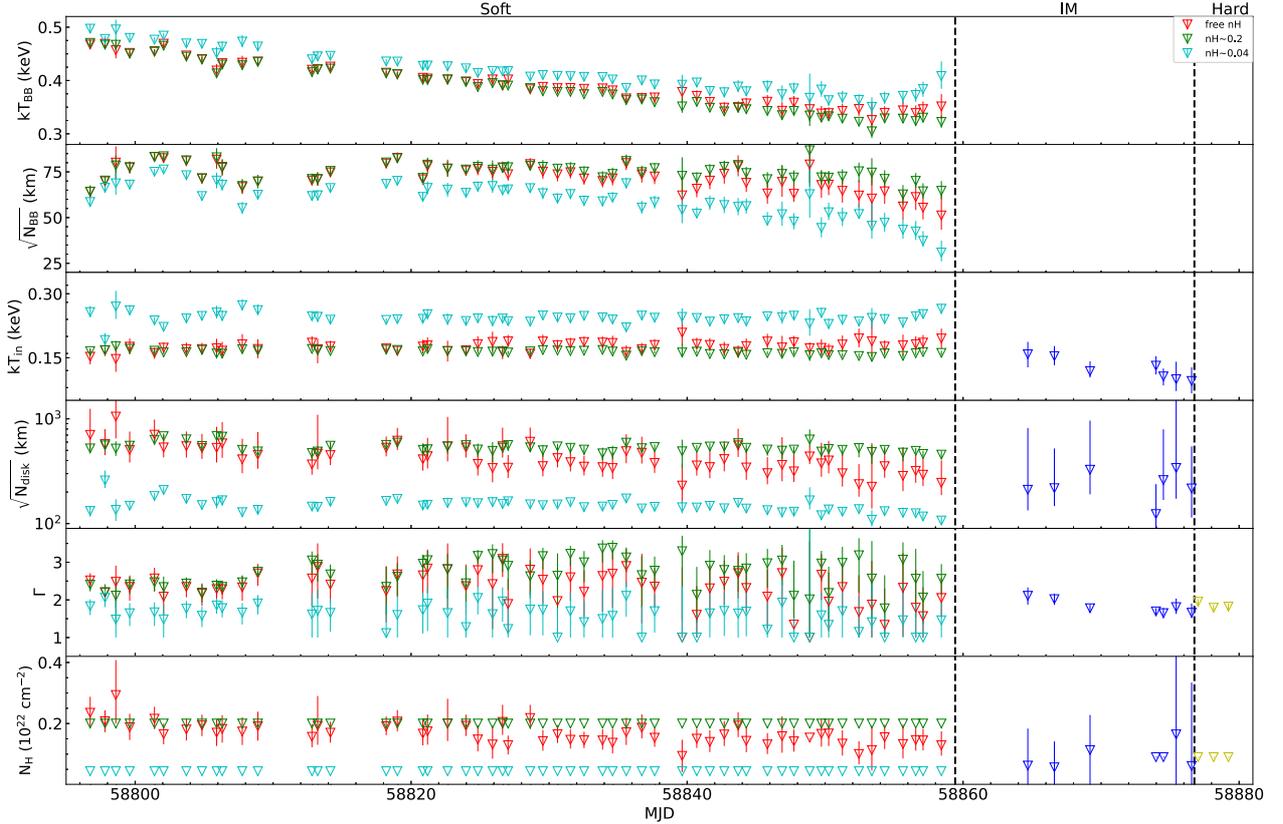

**Figure 8.** Comparison between the modelled spectral evolution of the soft state in Scenario 2 (hotter blackbody plus colder disk plus powerlaw) from the *Swift*/XRT data, when we leave $N_H$ as a free parameter (red datapoints), when we fix it at $N_H = 2 \times 10^{21}$ cm$^{-2}$ (green datapoints), and when we fix it at $N_H = 4 \times 10^{20}$ cm$^{-2}$ (cyan datapoints). Different choices of $N_H$ lead to substantially different properties of the second (colder) thermal component, but have only a minimal effect of the normalization of the hotter thermal component at the start of the outburst, which is a useful proxy for the size of ISCO.

Our *Insight*-HXMT peak flux is consistent with the peak flux estimates of Tetarenko et al. (2021), namely an unabsorbed X-ray flux $F_{0.5-10} \approx 9 \times 10^{-9}$ erg cm$^{-2}$ s$^{-1}$ (based on a disk-blackbody fit to the earliest *Swift*/XRT data) and a bolometric flux $F_{bol} \approx 1.3 \times 10^{-8}$ erg cm$^{-2}$ s$^{-1}$ (based on a `diskir` fit).

## 4 DISCUSSION

### 4.1 A BH candidate in the thick disk population

The distance to MAXI J0637−430, and therefore a reliable estimate of its luminosity, remain elusive. However, we attempt here to place some plausible constraints. The system is located at high Galactic latitude: $b = -20°.67$. Out of the 68 dynamical BHs and BH candidates listed in the latest version of the BlackCAT catalog (Corral-Santana et al. 2016), only two have a higher (in absolute value) Galactic latitude: Swift J1357.2−0933 (Rau, Greiner, & Filgas 2011) and XTE J1118+480 (Remillard et al. 2000). Thus, MAXI J0637−430 likely belongs to the thick-disk population of Milky Way BHs, a group that currently includes at least seven sources (Mata Sanchez et al. 2015). Observations (Corral-Santana et al. 2016) and population synthesis models (Zuo, Li, & Liu 2008) show that very few (if any) Galactic BH LMXBs are located at a vertical distance >2.5 kpc from the Galactic plane. This is consistent with the scale-height of the stellar thick disk, which is variously estimated at ≈0.6–1.0 kpc at different radial locations and for different metal abundances (*e.g.*, de Jong et al. 2010; Carollo et al. 2010; Bovy et al. 2016; Mateu & Vivas 2018; Pieres et al. 2020). Only one of the 68 BH candidates in BlackCAT, Swift J1357.2−0933, may be located at a higher distance from the Galactic plane, ≳4.5 kpc (Charles et al. 2019). For now, we assume that the distance to the Galactic plane of MAXI J0637−430 is ≲ 2.5 kpc. This implies that its distance from us is $d \lesssim 7$ kpc. We will use this constraint to assess whether MAXI J0637−430 behaves according to the canonical BH outburst sequence.

### 4.2 A low-luminosity thermal dominant state

Let us now assess whether the outburst peak and its early decline correspond to the canonical thermal dominant state (high/soft state) of most other transient BH candidates. First we shall compare their inner disk radii, then their peak temperatures and luminosities.

A defining property of the thermal dominant state (Remillard & McClintock 2006) is that the inner-disk radius $R_{in}$ remains approximately constant at ISCO, with $R_{in} \approx GM/c^2$ for a maximally spinning BH and $R_{in} \approx 6GM/c^2$ for a non-spinning BH. In our case, we showed in Sections 3.2 and 3.3 that from the *Swift*/XRT data, when at least the hottest part of the thermal emission is modelled with a disk-blackbody, $r_{in}\sqrt{\cos i} \approx 35–40\,d_{10}$ km in the early part of the outburst. For $i = 64°$ and a standard hardening factor of 1.7, this corresponds to $R_{in} \approx 60$–$70\,d_{10}$ km. If the hottest part of the thermal emission is modelled with a simple blackbody, $R_{bb} \approx 60$–$80\,d_{10}$ km. Such radii are indeed typical of what is observed in the thermal dominant state of stellar-mass BHs ($M \sim 5$–$20\,M_\odot$).

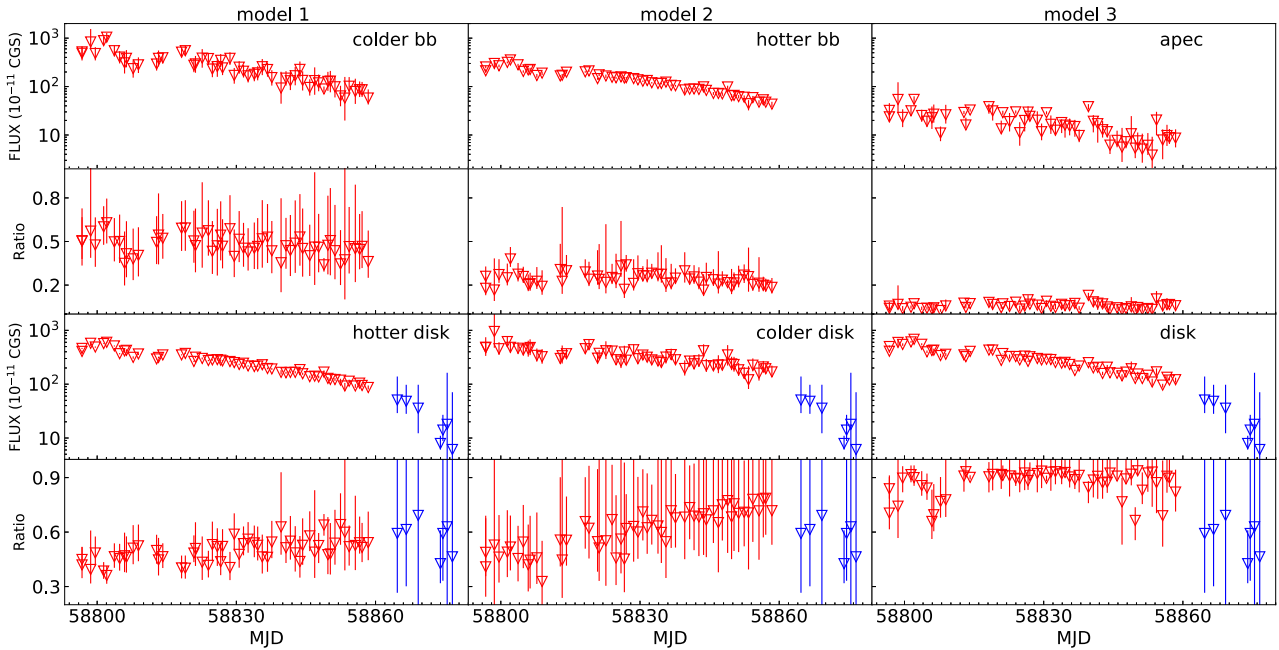

**Figure 9.** Evolution of the unabsorbed flux for three models with double thermal components (Sections 3.3.1, 3.3.2, 3.3.3, respectively). For each of the three models, we plot the 0.1–10 keV flux of each thermal component and its ratio to the total 0.1–10 keV unabsorbed flux (including the Comptonized component). The sum of the two thermal components is ≈90% of total flux (0.1–10 keV band) in models 1 and 3. For model2, the power-law contribution is more significant, accounting for ≈30% of the flux near outburst peak, before gradually declining to ≈10% like in the other two models. Blue datapoints after MJD 58860 represent the system in the HIMS, in which only one thermal component is needed in the fit.

The bolometric luminosity of a standard disk is defined as $L_{disk} \approx 4\pi r_{in}^2 \sigma T_{in}^4 \propto M^2 T_{in}^4$ (Shakura & Sunyaev 1973; Mitsuda et al. 1984; Makishima et al. 1986). For a non-spinning BH, the inner-disk temperature is $kT_{in} \approx 1.4(M/7M_\odot)^{-0.25}(\dot{M}/\dot{M}_{Edd})^{0.25}$ keV (Kubota et al. 1998; Makishima et al. 2000; Soria 2007; Done et al. 2012), where $\dot{M}$ and $\dot{M}_{Edd}$ are the mass accretion rate and the Eddington accretion rate, respectively. We want to check whether the outburst of MAXI J0637−430 occupies a different region of the parameter space, compared with "typical" stellar-mass BHs in the high/soft state. Just for the sake of visual illustration (Figure 10), we compared the $L_{disk}$–$T_{in}^4$ relation observed in MAXI J0637−430 with the analogous relation in the Galactic BH LMC X-3, which is often used as a "gold standard" for a thermal dominant state with constant $R_{in}$ (Gierliński & Done 2004a; Steiner et al. 2010; Orosz et al. 2014). For LMC X-3, we used the disk parameters from (Done & Gierliński 2003). For $L_{disk}$ and $T_{in}$ of MAXI J0637−430, we used three alternative choices of disk parameters: those obtained from a single `diskbb` model fit; those from a hotter `diskbb` plus cooler `diskbb` (Scenario 1, Section 3.3.1 ); and those from `diskbb` plus `apec` (Scenario 3, Section 3.3.3). In those three models, the disk is expected to reach ISCO and contain most of the thermal luminosity, and the $L_{disk}$–$T_{in}^4$ relation should apply.

In order to place the $L_{disk}$ datapoints in the temperature-luminosity plot, we need to assume a distance. If we choose $d = 7$ kpc (Section 4.1), we notice (Figure 10a,b,c) that MAXI J0637−430 and LMC X-3 evolve along a similar track, which suggests a similar combination of BH mass and spin. (The BH mass in LMC X-3 is $M = (6.9 \pm 0.6)\,M_\odot$: Orosz et al. 2014; the BH spin parameter is $a = 0.25^{+0.20}_{-0.29}$: Steiner et al. 2014.) However, the inner-disk temperature range spanned by MAXI J0637−430 is lower: $kT_{in} \approx 0.4$–$0.7$ keV, a factor of 2 lower than the disk temperatures of LMC X-3 in its high/soft state. At that assumed distance, the disk luminosity remains much below Eddington: $L_{disk} \approx (0.005$–$0.05)\,L_{Edd}$ if the BH mass in MAXI J0637−430 is ≈$7\,M_\odot$. The corresponding peak accretion rate is only ≈$5 \times 10^{17}$ g s$^{-1}$ (assuming a disk efficiency $\eta \approx 0.1$). Even after including the Comptonized emission component, the peak bolometric luminosity (Section 3.5) is only $L_{bol} \approx 0.12\,(d_{10})^2\,(M_7)^{-1}\,L_{Edd}$.

It is fairly common for BH transients to reach outburst peak luminosities of ≈$0.1\,L_{Edd}$ without leaving the hard state or hard-intermediate states, that is without reaching the thermal dominant state ("failed outbursts": Brocksopp et al. 2004; Capitanio et al. 2009; Tetarenko et al. 2016). However, here we have the (much rarer) opposite situation of a BH transient that remains in the thermal dominant state throughout the outburst without apparently even reaching (or barely reaching) a luminosity of ≈$0.1\,L_{Edd}$.

The low peak luminosity of MAXI J0637−430 is probably related to its small binary separation and short orbital period. From their `diskir` fit to the combined *Swift*/XRT and Ultraviolet and Optical Telescope data, Tetarenko et al. (2021) estimated an outer disk radius $R_{out} \approx 10^4 R_{in} \approx 6 \times 10^{10}\,d_{10}$ cm. Assuming that the outer disk extends to $R_{out} \approx 0.8\,R_{RL}$ (e.g., Paczynski 1977; Whitehurst & King 1991; Warner 1995; Frank et al. 2002), where $R_{RL}$ is the volume radius of the BH Roche lobe, and using standard approximations for $R_{RL}$ (Eggleton 1983), we estimate a binary separation $a \approx 1.3 \times 10^{11}\,d_{10}$ cm, for a characteristic mass ratio $q = M_2/M_1 = 0.1$, or, more generally, $a \sim 1.2$–$1.4 \times 10^{11}\,d_{10}$ cm for $q \sim 0.05$–$0.2$. From Kepler's law, for $M_1 \approx 7\,M_\odot$ and $q \sim 0.1$, this implies a binaries period $P$ as short as $P \approx 2.5\,d_{10}^{3/2}$ hr. This is consistent with the result of Tetarenko et al. (2021), who also estimate a binary period of ∼2–4 hr. Tetarenko et al. (2021) searched for evidence of periodicity in the optical and UV data, but did not find any; thus, indirect estimates of the binary parameters from the inner and outer disk size is the only available way at this stage to constrain the period. A period of ∼2 hr for MAXI J0637−430 may be the shortest one known to-date among transient Galactic BH candidates (Coriat, Fender, & Dubus 2012; Corral-Santana et al. 2016); at the moment, the record belongs to MAXI J1659−152, with $P_{\rm orb} = (2.414 \pm 0.005)$ hr (Kuulkers et al. 2013; Corral-Santana et al. 2018; Torres et al. 2021). Finally, the period-density relation (Frank et al. 2002) implies a donor star with a mass of ∼0.2 $M_\odot$ (roughly corresponding to an M4 main-sequence star). A more detailed discussion of the disk size and binary period will be presented in further work (Soria et al., submitted).

In the framework of the thermal-viscous instability model (Lasota 2001), a small outer disk radius implies a small outburst peak luminosity, because the peak accretion rate $\dot{M} \propto R_{\rm out}^{2.39}$ (using the approximations of Lasota et al. 2015 for irradiated disks), or $\dot{M} \propto R_{\rm out}^{2.1}$ (using the slightly different scalings from Hameury & Lasota 2020). Peak accretion rates as low as $\dot{M} \sim 10^{17}$ g s$^{-1}$ are indeed predicted by the thermal-viscous instability model for a system as compact as MAXI J0637−430, in agreement with the observed peak luminosity. For the same reason, such a system can remain in the soft state (that is, the disk can remain on the hot branch of the S-curve in the temperature-surface density plane) for accretion rates as low as $\dot{M} \sim 10^{16}$ g s$^{-1}$ (Dubus et al. 1999), in agreement with the observed luminosity (∼a few $10^{-3}\,L_{\rm Edd}$) at the soft-to-hard transition. Thus, the example of MAXI J0637−430 shows that for a generic BH candidate, we cannot always assume that the luminosity of its soft-to-hard transition is between ≈0.01–0.04 $L_{\rm Edd}$ (Maccarone 2003) and use that value for a mass estimate.

### 4.3 Ruling out the intermediate mass BH scenario

Just for the sake of argument, let us consider instead the possibility that the outburst of MAXI J0637−430 does reach a peak luminosity of ≈0.8 $L_{\rm Edd}$ (similar to the peak luminosity of LMC X-3 and other Galactic BH transients) but we have severely underestimated its distance and mass (which would explain its low disk temperature). This assumption would imply a BH mass ≈140 $M_\odot$ at a distance of ≈140 kpc, with a peak luminosity ≈ 1.5 ×$10^{40}$ erg s$^{-1}$ (Figure 10d). In this case, MAXI J0637−430 would be the first intermediate-mass BH seen in the Local Group. Galaxy halos are one environment where intermediate-mass BHs from accreted satellite dwarfs may still exit; therefore, we cannot dismiss this scenario without proper investigations.

The nearest Milky Way satellite galaxy in this part of the sky is the Carina Galaxy (8 degrees away), at a distance of ≈100 kpc (Pietrzyński et al. 2009). Carina has a heliocentric velocity of about 220 km s$^{-1}$ (Muñoz et al. 2006). Instead, MAXI J0637−430 has essentially zero systemic velocity, based on the central position of the He II 4686 emission lines measured by Tetarenko et al. (2021). Thus, an association (*e.g.*, tidal stripping) with the Carina galaxy is unlikely.

Although disk temperature and luminosity are self-consistent, the intermediate-mass BH scenario is strongly disfavoured by another measurement, namely the exponential decay timescale $\tau$ of the outburst, proportional to the viscous timescale $t_{\rm vis}$ at the outer disk radius ($\tau = 3 t_{\rm vis}$: King & Ritter 1998). For a standard thin disk,

$$t_{\rm vis} \approx 7.4\,(\alpha/0.3)^{-4/5}\,\dot{M}_{17}^{-3/10}\,(M/7M_\odot)^{1/4}\,R_{10}^{5/4}\,{\rm d} \quad (1)$$

(Frank et al. 2002), where $\alpha$ is the viscosity parameter (Shakura & Sunyaev 1973), $\dot{M}_{17}$ is the accretion rate in units of $10^{17}$ g s$^{-1}$, and $R_{10}$ is the outer disk radius in units of $10^{10}$ cm. A reference value of $\alpha \approx 0.3$ is based on the typical observed values for BH transients in the high state (King, Pringle, & Livio 2007; Tetarenko et al. 2018). For the outer disk radius $R_{\rm out}$, we use again the diskir estimates of Tetarenko et al. (2021), and take representative values $R_{\rm out} \sim 5 \times 10^{10}$ cm for a 7-$M_\odot$ BH and $R_{\rm out} \sim 10^{12}$ cm for a 140-$M_\odot$ BH. This would imply a viscous timescale $t_{\rm vis} \sim 40$ d for the stellar-mass BH solution, and $t_{\rm vis} \sim 540$ d for the intermediate-mass BH solution. The viscous timescale actually observed in this outburst of MAXI J0637−430 is $t_{\rm vis} \approx 53$ d (Tetarenko et al. 2021), consistent with an ordinary stellar-mass BH.

### 4.4 Physical origin of the soft excess

Although the general properties of the system are now established (*e.g.*, $R_{\rm ISCO} \approx 60\,d_{10}$ km, $L_{\rm bol} \approx 0.1\,L_{\rm Edd}$ at outburst peak, binary separation $a \sim 10^{11}\,d_{10}$ cm), we still need to explain the soft X-ray residuals, particularly significant in the *Swift* and *NuSTAR* data (Section 3.3 of our paper, and Lazar et al. 2021). The temperature-luminosity diagram also suggests that a single diskbb thermal component is not behaving as expected for a canonical high/soft state, that is the fit parameters deviate from an $L_{\rm disk} \propto T_{\rm in}^4$ slope (Figure 10a). Our best fit to the datapoints yields a slope $L_{\rm disk} \propto T_{\rm in}^{2.2}$. Instead, when a second thermal component is added to the model, the main disk component does follow the standard relation (Figure 10b,c). Understanding the reason for this discrepancy is particularly important because continuum-fitting of the disk-blackbody component is one of the two techniques for the simultaneous measurement of BH spin and inclination angle (or, alternatively, of mass and distance) in BH X-ray binaries (Zhang, Cui, & Chen 1997; Steiner et al. 2009, 2010, 2011; McClintock et al. 2014; Parker et al. 2019). Residuals such as those seen in MAXI J0637−430 may throw a spanner in the works, or, at the very least, bias the results for some classes of BHs.

Two explanations were proposed by Lazar et al. (2021): i) reflection on the inner disk surface of blackbody photons emitted by the disk itself ("returning disk radiation"), deflected by relativistic light bending; ii) emission from material inside ISCO (a location known as the "plunging region"). Scenario i works better for BHs near maximal spin (Cunningham 1976), in which the inner accretion disk is closest to the event horizon (maximizing the light bending effects). In the case of MAXI J0637−430, the condition $R_{\rm in} \approx GM/c^2$ implies a BH mass $\gtrsim 50 M_\odot$ and a distance > 30 kpc (Lazar et al. 2021). Both conditions are implausible, based on the BH X-ray binaries distribution in the Milky Way and (as discussed in Section 4.3) for the short viscous timescale inferred during outburst decline. Scenario ii would be a surprising and ground-breaking discovery. Photon emission from the plunging region has long been sought for (*e.g.*, Krolik & Hawley 2002; Zhu et al. 2012; Schnittman, Krolik, & Noble 2013; Wilkins, Reynolds, & Fabian 2020), without empirical proof so far; it would provide the most stringent tests for general relativity and for the existence of BH horizons. The consensus so far, based on magneto-hydrodynamical simulations, is that the contribution of the emission from the plunging region to the time-averaged soft X-ray continuum is negligible; photons emitted from that region may be detectable from the lag-frequency spectrum (Wilkins, Reynolds, & Fabian 2020) or from a steep, weak high-energy tail (Zhu et al. 2012).

Thus, plunging-region photons do not seem a plausible cause of the soft X-ray emission in MAXI J0637−430.

Given the difficulties of those two scenarios, we investigated alternatives phenomenological scenarios in this paper: an additional colder or hotter blackbody emitter, and the effect of ionized outflows. We will now briefly assess whether they are physically plausible.

### 4.4.1 Colder blackbody plus hotter disk

Additional blackbody-like emission may come from accretion streams outside the disk plane, from hot spots, or from the illuminated surface of disk bulges, warps or spiral-arm features. The main difficulty of this scenario is that the fitting results suggest a large surface radius ($R_{bb} \approx 1.9 \times 10^3 \, d_{10}$ km $\approx 30 \, R_{ISCO}$; Figure 5) and high luminosity ($L_{bb} \gtrsim L_{disk}$; Figure 9). The location of the blackbody emitter must be at least as far from the BH as its characteristic radius. Thus, the efficiency of gravitational energy release in the blackbody-emitting structure would be at least 30 times lower than at the innermost radius of the disk. If the blackbody emission is powered by viscous heat, it would require a mass feeding above the critical Eddington limit (obviously absurd in this system) to release such high luminosity at such low efficiency. The sudden disappearance of the blackbody component at the end of the soft state also rules out its origin in a separate structure outside the disk. If we suppose instead that the additional blackbody emission comes from reprocessing of inner-disk photons off some parts of the disk surface, we would have to assume that the emitting structure intercepts about half of the directly emitted thin-disk photons without at the same time blocking our view of the inner disk; the existence of such putative structure is highly unlikely. No such reprocessing structures have been seen or predicted before for any other BH candidates in the high/soft state, even at higher Eddington ratios. In summary, we cannot find any plausible physical interpretation for this phenomenological scenario.

### 4.4.2 Hotter blackbody + colder disk

This scenario (sketched in Figure 11) is an adaptation of the plunging region hypothesis. Instead of a disk truncated at ISCO, plus additional emitting material inside that radius, let us suppose that the disk is truncated moderately far from ISCO and there is an additional ring (or narrow disk) of optically thick, blackbody-emitting material near ISCO, with a possible gap between the two structures. The inner structure is well modelled with a single-temperature blackbody because of its small radial width. Our phenomenological fitting results (Section 3.3.2) suggest that the outer disk could be truncated at $R_{in} \approx 20 \, R_{ISCO}$ near the peak of the outburst, and gradually move inwards, reaching $R_{in} \approx 7 \, R_{ISCO}$ before the transition to the hard intermediate state, thus reducing or closing the gap. It is the presence of this gap that makes the integrated spectrum of outer disk and inner ring different from a standard disk-blackbody spectrum. The relative contribution of the hot blackbody ring decreases during the outburst decline, as the outer disk moves inwards (Figure 9). In the canonical picture of BH outbursts, the hard-to-soft state transitions is caused by an increased accretion flow through the disk. However, condensation of the corona onto a small inner disk has also been invoked (Liu et al. 2007; Taam et al. 2008; Meyer-Hofmeister, Liu, & Meyer 2009; Qiao & Liu 2017), especially if the inner disk was already present in the hard state. The range of relatively low Eddington ratios spanned by MAXI J0637−430 in its outburst is consistent with the range in which an inner disk formed from condensation can coexist with a hot corona (Liu, Done, & Taam 2011).

A difficulty of this gap scenario is that somehow, the optically thick inner ring must continue to be fed during the outburst, perhaps via continuous condensation from a hot corona. This would be consistent with a two-component accretion flow (Chakrabarti and Titarchuk 1995; Smith et al. 2002), with a lower angular momentum component feeding the hot inner region and condensed inner ring, and the higher angular momentum component feeding the outer thin disk. Another consequence of this scenario is that if the transition between hot corona and thin disk is not complete at all radii, we expect a substantial contribution from the power-law component throughout the outburst. Here, the flux contribution of the steep power-law component is $\approx 30\%$ of the total (thermal plus non-thermal) flux near outburst peak (Figure 9), if we assume that the power-law extends down to 0.1 keV; or $\approx 10\%$ (like in the other two scenarios), if truncated at $\approx 0.5$ keV. The gradual relative decline of the power-law component during outburst evolution is consistent with the gap getting slowly filled by the optically thick disk. However, even the initial $\approx 30\%$ contribution is less than what is seen in other BH transients that show incomplete transitions to the thermal dominant state (*e.g.*, MAXI J1836−194: Russell et al. 2014).

In summary, we cannot rule out a condensed inner ring immersed in a hot flow, with a gap and an outer standard disk, but there is no compelling evidence for it. Other physical scenarios, for example a mis-aligned (warped) inner disk, can produce an integrated continuum spectrum different from a simple disk-blackbody. In some cases, the warping can lead to disk tearing and a gap between inner and outer disk (Raj & Nixon 2021), which can be modelled with a double thermal component. Further exploration of those scenarios is beyond the scope of this observational study.

### 4.4.3 Ionized outflows

X-ray binaries in the super-Eddington regime (ultraluminous X-ray sources) are well known to have strong line emission in the soft X-ray band from hot, semi-relativistic disk winds (Pinto, Middleton, & Fabian 2016; Pinto et al. 2017; Kosec et al. 2021), with luminosities of a few percent of the underlying continuum luminosity. This is similar to the relative luminosity of the thermal plasma (`apec`) component required to fit the soft residuals in the *Swift*/XRT spectra of MAXI J0637−430 ($L_{apec} \approx 0.07 \times L_{disk}$; Figure 9). Among Galactic BH X-ray binaries, strong outflows have been observed in systems that were also (temporarily) exceeding their Eddington limit (GRO J1655−40: Neilsen et al. 2016; Shidatsu, Done, & Ueda 2016; GRS 1915+105: Miller et al. 2016), as well as in at least three candidate BH transients in the high/soft state (IGR J17091−3624: King et al. 2012; 4U 163047 and H 1732−322: Miller et al. 2015), at Eddington ratios comparable to the peak Eddington ratio of MAXI J0637−430. The strongly magnetized neutron star X-ray binary Her X-1 also shows ionized emission and absorption lines from a strong disk wind (mass loss rate similar to or exceeding the mass accretion rate onto the compact object) (Kosec et al. 2020). As an alternative to unbound outflows, bound massive flows along the disk surface have also been proposed for those systems (Nixon & Pringle 2020). The kinetic energy stored in such flows may substantially reduce the radiative energy output from the disk annuli where the outflows are launched (*i.e.*, they may locally cool the disk).

In summary, a contribution from massive outflows via photon scattering or absorption/re-emission is a viable way to modify the disk-blackbody continuum in the inner part of the disk, but the reason why such outflows may be stronger in some systems (including perhaps MAXI J0637−430) and weaker in others remains unclear. Two selection effects may also be at work, in the sense that the distortion of the soft X-ray continuum from the standard disk-blackbody shape may be evident only in systems with the following two properties: i) they are seen through low $N_{\rm H}$, for example because located in the thick disk or halo, well outside the Galactic plane; ii) the disk luminosity is relatively low and does not completely wash out other contributions at the ∼$10^{37}$ erg s$^{-1}$ level in the soft band, either from a reflection component, or from outflows.

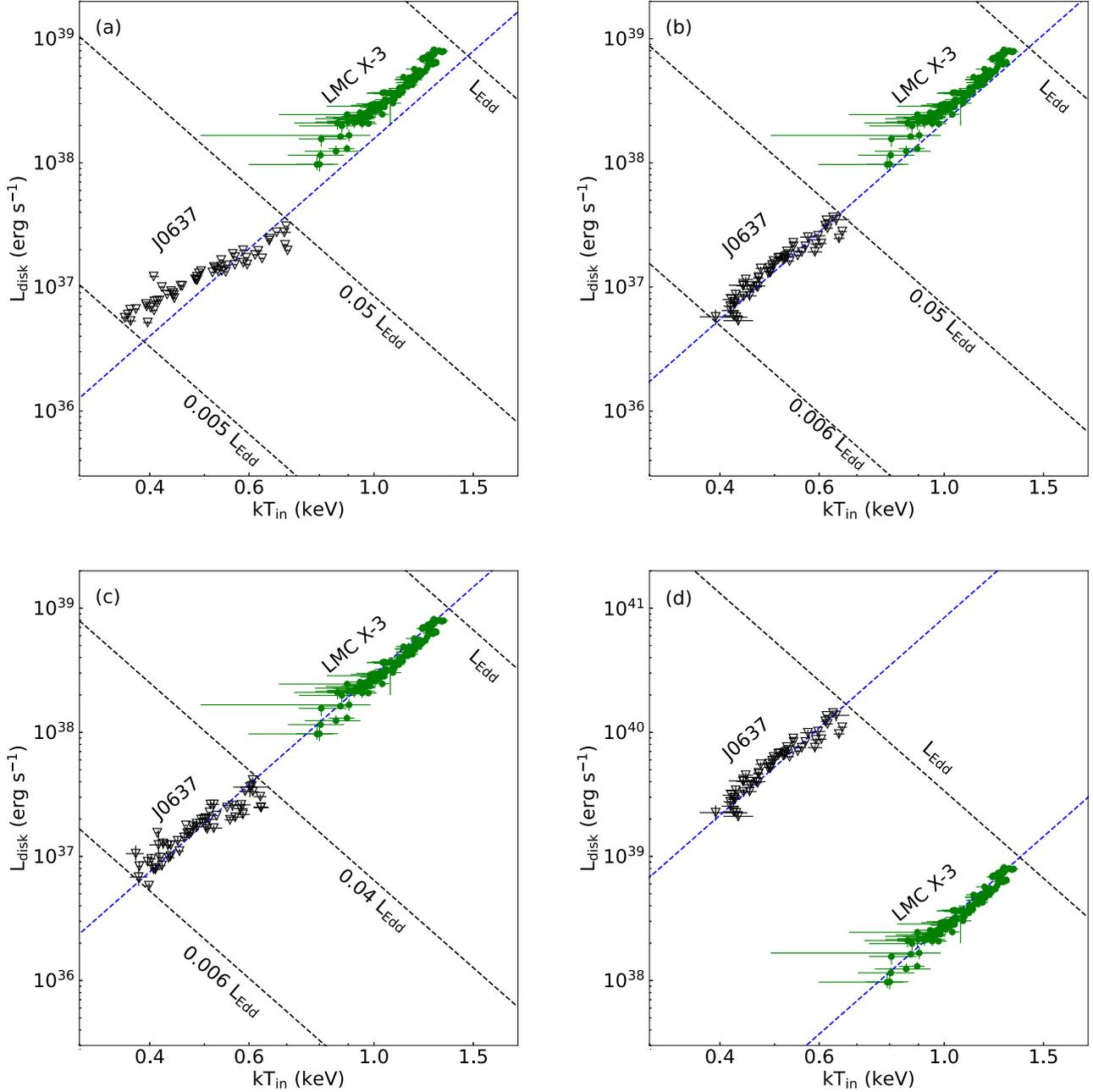

**Figure 10.** Relationship between the bolometric disk luminosity ($L_{\rm disk}$) and the peak colour temperature ($kT_{\rm in}$) for MAXI J0637−430, for alternative models. In all panels, black symbols represent the best-fitting values for the *Swift*/XRT data, as derived in this work. Green symbols represent the temperature-luminosity relation for an outburst of LMC X-3 Done & Gierliński (2003), presented here as a comparison with a "typical" high/soft state behaviour of a stellar-mass BH. The different model assumptions for MAXI J0637−430 in the four panels are: a) single thermal model, and an assumed distance of 7 kpc; b) disk parameters in a double thermal model (Scenario 1, hotter disk-blackbody plus colder blackbody components), and $d$ = 7 kpc; c) disk parameters in a double thermal model (Scenario 3, disk-blackbody plus thermal plasma), and $d$ = 7 kpc; d) same as in panel b), but this time assuming that the peak Eddington ratio in MAXI J0637−430 is the same as in LMC X-3 (≈0.8) and increasing the distance to satisfy this condition. The distance required is $d \approx 140$ kpc. All the scenarios at a distance of 7 kpc suggest a BH mass comparable to the BH mass of LMC X-3, but a significantly sub-luminous outburst, peaking at $L_{\rm disk} \approx 0.05 L_{\rm Edd}$ and remaining in the high/soft state well below 0.01 $L_{\rm Edd}$. The scenario at a distance of 140 kpc requires a BH mass of ≈140 $M_\odot$; we consider this possibility unphysical, based on the arguments outlined in Section 4.3.

## 5 CONCLUSIONS

We re-examined the X-ray spectral evidence available for the 2019 outburst of MAXI J0637−430, including previously unpublished broad-band data from *Insight-HXMT*. It was already well known that the initial rise in the hard state was absent or too quick to be detected, and the outburst was dominated from the very beginning by thermal emission. One unsolved question is whether this soft outburst has the same properties of a canonical high/soft state: for example, whether the thermal spectrum is consistent with the standard thin disk, and whether the fitted inner disk radius and the soft-to-hard transition luminosity at the end of the outburst can be used as indicators of the BH mass.

We argued that the high Galactic latitude of the source suggests a distance ≲ 7 kpc if the BH belongs to the thick disk population. At that distance (and for a viewing angle ≈ 64°), the bolometric luminosity at outburst peak is ≈ 7 ×$10^{37}$ erg s$^{-1}$. For a typical 7-$M_{\odot}$ BH, that corresponds to a peak Eddington ratio of ≈ 0.08. At that peak luminosity, other BH transients typically show failed (hard-only) outbursts, in which the hot corona is not completely removed and the inner edge of the thin disk may not reach ISCO. In MAXI J0637−430, the soft-to-hard transition at the end of the outburst begins at a bolometric luminosity of ≈ 6 ×$10^{36}$ erg s$^{-1}$ (at 7 kpc), corresponding to an Eddington ratio of ≈ 0.007, lower than the range of values (0.01–0.04) usually assumed for stellar-mass BHs. We argued that the small binary separation and short orbital period in MAXI J0637−430 (and, therefore, the small size of the BH Roche lobe and the low total mass stored in the accretion disk) explain this low outburst luminosity, in the thermal-viscous disk instability model.

We then examined another aspect of the outburst that may not look canonical for the soft state: the presence of substantial spectral residuals below ∼1 keV. We showed that a single disk-blackbody component is not a good approximation to the continuum and deviates from the standard $L_{\rm disk} \propto T_{\rm in}^4$ relation. A reflection component is probably the least contrived solution proposed in the literature (Lazar et al. 2021) for this feature of MAXI J0637−430. As an alternative, we examined other possible scenarios. An additional, cooler blackbody component at $kT$ ∼ 0.1 keV can fit the residuals but it is physically implausible because of the large emitting area implied by the fit parameters, and because its temperature appears to increase during the outburst decline while the temperature and luminosity of the primary disk components decline. Thermal-plasma emission and absorption from an ionized disk outflow can also fit the residuals. However, the outflow scenario struggles to explain why the wind should be so strong in this system, at such a modest Eddington ratio. We also suggested a more speculative scenario, which also fits the spectral data, in a purely phenomenological sense: that the disk has a gap between an innermost ring at $R$ ∼ $R_{\rm ISCO}$, and an outer section truncated at $R$ ∼ 10 $R_{\rm ISCO}$. The inner part of the disk may be formed and fed directly *in situ* from condensation of the surrounding hot corona; the gap between inner and outer disk part may be related to the low mass accretion rate from the outer disk, which does not enable the complete cooling of the corona and the formation of a full disk. In this scenario, the 2019 outburst of MAXI J0637−430 could be considered a (different type of) failed outburst.

Regardless of the correct explanation for the soft X-ray residuals, we showed that various alternative models agree on $R_{\rm in}$ ∼ 60 $d_{10}$ km as the innermost source of optically-thick thermal photons around the peak of the outburst, and therefore the most likely location of $R_{\rm ISCO}$ (disregarding the unlikely possibility of significant emission from the plunging region). This measurement is not sufficient to estimate the BH mass yet, because of the degeneracy between the mass and spin dependencies, and the uncertainty on the distance. In further work currently in preparation, we will combine this constraint on $R_{\rm ISCO}$ from the continuum fitting with the constraints derived from the viscous timescale and from the rotational broadening of the He II $\lambda$4686 line (Tetarenko et al. 2021), in order to determine the system parameters more accurately.

## ACKNOWLEDGEMENTS

We thank the anonymous referee for useful comments. This work made use of data from the Insight-HXMT mission, a project funded by China National Space Administration (CNSA) and the Chinese Academy of Sciences (CAS). This work is supported by the National Key R&D Program of China (2021YFA0718500). We acknowledge funding support from the National Natural Science Foundation of China (NSFC) under grant Nos. 12122306 and U1838115, the CAS Pioneer Hundred Talent Program Y8291130K2 and the Scientific and technological innovation project of IHEP Y7515570U1. RS acknowledges grant number 12073029 also from the National Natural Science Foundation of China.

*Facilities: Insight-HXMT*, *Swift*

## DATA AVAILABILITY

The data for *Insight-HXMT* underlying this article is available format at *Insight-HXMT* website (http://archive.hxmt.cn/proposal; data in compressed format). *Swift* data is available in the High Energy Astrophysics Science Archive Research Center (HEASARC) at https://heasarc.gsfc. nasa.gov/db-perl/W3Browse/w3browse.pl.

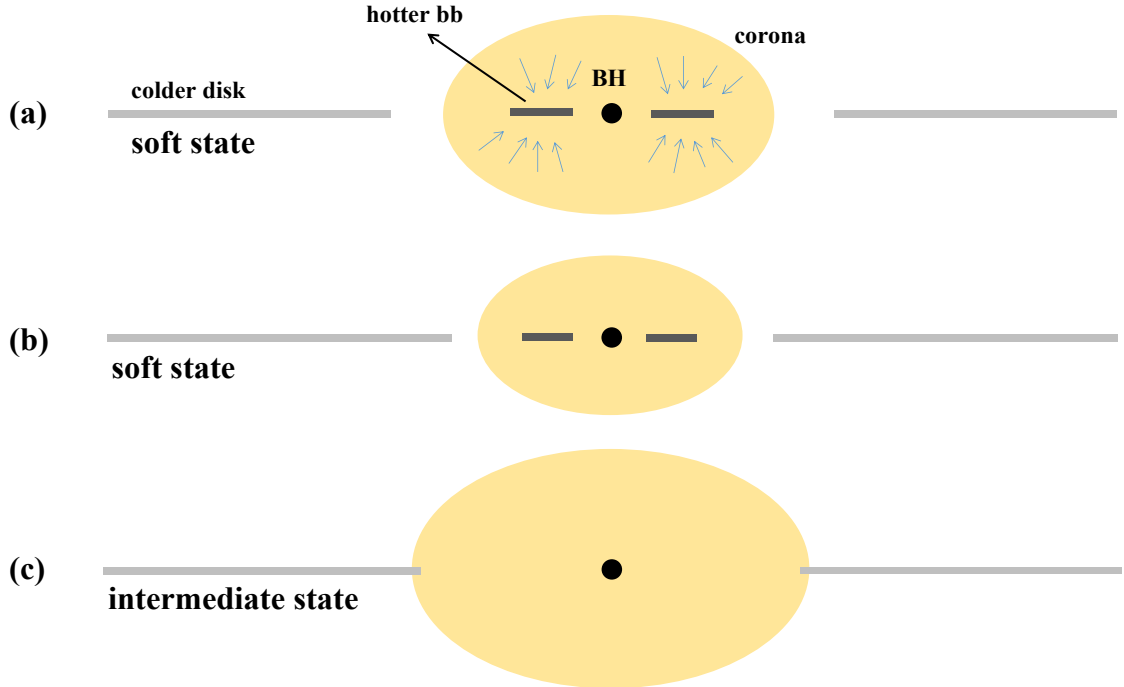

**Figure 11.** Cartoon picture of one possible physical representation of our phenomenological fitting Scenario 2 (double thermal component with a hotter blackbody near ISCO and a colder, larger disk). The hotter blackbody may be formed in situ from partial re-condensation of the hot corona (a). A gap may separate the outer disk from the condensed inner ring. As the outburst evolves, the inner radius of the colder disk gradually moves inward and decreases, and the flux of the corona decreases (b). However, the gap is never completely closed. In this sense, it is a failed (or at least incomplete) outburst. Subsequent re-evaporation of the inner ring into the hot corona spells the end of the high/soft state and the transition back to intermediate and hard states (c).

# APPENDIX A: TABLES

Table A1: *Insight-HXMT* observations log, and net count rates in the LE band.

| ObsID | ExpID | Start Time (day) | Start Time (MJD) | LE Exp. Time (s) | LE Count Rate[a] (ct s$^{-1}$) |
|---|---|---|---|---|---|
| P0214057001 | 01 | 2019-11-04 04:26:48 | 58791.19 | 1000.49 | 185.9 ± 0.4 |
|  | 02 | 2019-11-04 08:13:24 | 58791.34 | 1197.00 | 188.1 ± 0.4 |
|  | 03 | 2019-11-04 11:33:45 | 58791.48 | 1017.45 | 192.2 ± 0.4 |
|  | 04 | 2019-11-04 14:54:41 | 58791.62 | 1556.10 | 194.3 ± 0.4 |
|  | 05 | 2019-11-04 18:17:57 | 58791.76 | 1496.25 | 196.8 ± 0.4 |
|  | 06 | 2019-11-04 21:28:46 | 58791.90 | 778.05 | 196.9 ± 0.5 |
|  | 07 | 2019-11-05 00:39:35 | 58792.03 | 299.25 | 198.2 ± 0.8 |
|  | 08 | 2019-11-05 03:50:24 | 58792.16 | 478.80 | 196.9 ± 0.7 |
|  | 09 | 2019-11-05 07:01:13 | 58792.29 | 1077.30 | 198.2 ± 0.4 |
|  | 10 | 2019-11-05 10:12:02 | 58792.43 | 1316.70 | 199.7 ± 0.4 |
|  | 11 | 2019-11-05 13:22:50 | 58792.56 | 957.60 | 200.1 ± 0.5 |
|  | 12 | 2019-11-05 16:33:39 | 58792.69 | 1496.25 | 200.1 ± 0.4 |
|  | 13 | 2019-11-05 19:44:28 | 58792.82 | 1316.70 | 200.7 ± 0.4 |
| P0214057003 | 01 | 2019-11-07 00:51:17 | 58794.04 | 837.90 | 197.0 ± 0.5 |
|  | 02 | 2019-11-07 04:17:18 | 58794.18 | 527.68 | 197.1 ± 0.6 |
| P0214057004 | 01 | 2019-11-08 00:43:32 | 58795.03 | 837.90 | 191.2 ± 0.5 |
|  | 02 | 2019-11-08 04:09:43 | 58795.17 | 658.35 | 191.1 ± 0.6 |
| P0214057005 | 01 | 2019-11-09 00:36:04 | 58796.03 | 1017.45 | 186.7 ± 0.4 |
|  | 02 | 2019-11-09 04:02:05 | 58796.17 | 595.51 | 186.6 ± 0.6 |
| P0214057009 | 01 | 2019-11-11 19:29:17 | 58798.81 | 538.65 | 158.6 ± 0.6 |
|  | 02 | 2019-11-11 22:36:21 | 58798.94 | 1197.00 | 159.0 ± 0.4 |
|  | 03 | 2019-11-12 01:51:21 | 58799.08 | 1691.76 | 158.7 ± 0.3 |
| P0214057010 | 01 | 2019-11-13 06:32:34 | 58800.27 | 684.29 | 148.6 ± 0.5 |
|  | 02 | 2019-11-13 10:11:20 | 58800.43 | 1496.25 | 148.9 ± 0.3 |
|  | 03 | 2019-11-13 13:46:51 | 58800.57 | 538.65 | 148.6 ± 0.5 |
| P0214057011 | 01 | 2019-11-14 00:04:14 | 58801.00 | 778.05 | 142.1 ± 0.4 |
|  | 02 | 2019-11-14 03:20:39 | 58801.14 | 1128.17 | 142.4 ± 0.4 |
|  | 03 | 2019-11-14 06:41:53 | 58801.28 | 465.83 | 141.9 ± 0.6 |
|  | 04 | 2019-11-14 10:02:18 | 58801.42 | 1479.29 | 140.9 ± 0.3 |
|  | 05 | 2019-11-14 13:37:51 | 58801.57 | 1950.11 | 140.0 ± 0.3 |
|  | 06 | 2019-11-14 16:48:39 | 58801.70 | 1122.19 | 139.5 ± 0.4 |
|  | 07 | 2019-11-14 19:59:27 | 58801.83 | 1017.45 | 138.9 ± 0.4 |
|  | 08 | 2019-11-14 23:10:15 | 58801.97 | 1256.85 | 138.3 ± 0.3 |
|  | 09 | 2019-11-15 02:21:04 | 58802.10 | 717.20 | 137.6 ± 0.5 |
| P0214057012 | 01 | 2019-11-17 09:16:10 | 58804.39 | 1436.40 | 121.8 ± 0.3 |
|  | 02 | 2019-11-17 13:10:50 | 58804.55 | 1412.46 | 122.0 ± 0.3 |
|  | 03 | 2019-11-17 16:21:38 | 58804.68 | 1197.00 | 120.3 ± 0.3 |
|  | 04 | 2019-11-17 19:32:25 | 58804.81 | 598.50 | 120.1 ± 0.5 |
| P0214057013 | 01 | 2019-11-18 09:07:50 | 58805.38 | 1483.28 | 114.0 ± 0.3 |
|  | 02 | 2019-11-18 13:01:49 | 58805.54 | 400.00 | 111.2 ± 0.6 |
| P0214057014 | 01 | 2019-11-19 04:12:59 | 58806.18 | 1087.27 | 102.5 ± 0.3 |
| P0214057015 | 01 | 2019-11-20 00:53:24 | 58807.04 | 950.62 | 105.1 ± 0.3 |
|  | 02 | 2019-11-20 04:06:41 | 58807.17 | 858.85 | 104.4 ± 0.4 |
|  | 03 | 2019-11-20 07:27:05 | 58807.31 | 1137.15 | 105.1 ± 0.3 |
| P0214057016 | 01 | 2019-11-22 00:35:39 | 58809.03 | 610.47 | 97.8 ± 0.4 |
|  | 02 | 2019-11-22 03:48:12 | 58809.16 | 401.99 | 98.5 ± 0.5 |
|  | 03 | 2019-11-22 07:09:03 | 58809.30 | 897.75 | 97.8 ± 0.3 |
| P0214057017 | 01 | 2019-11-22 22:51:17 | 58809.95 | 1527.17 | 93.1 ± 0.3 |
|  | 02 | 2019-11-23 01:59:11 | 58810.08 | 1170.07 | 93.3 ± 0.3 |
|  | 03 | 2019-11-23 05:19:06 | 58810.22 | 837.90 | 92.0 ± 0.3 |
| P0214057018 | 01 | 2019-11-24 01:53:07 | 58811.08 | 1137.15 | 87.9 ± 0.3 |
|  | 02 | 2019-11-24 05:09:59 | 58811.22 | 1170.07 | 87.9 ± 0.3 |
|  | 03 | 2019-11-24 08:57:14 | 58811.37 | 418.95 | 87.4 ± 0.5 |
| P0214057019 | 01 | 2019-11-25 22:24:22 | 58812.93 | 1915.20 | 81.8 ± 0.2 |
|  | 02 | 2019-11-26 01:31:29 | 58813.06 | 1017.45 | 81.3 ± 0.3 |
|  | 03 | 2019-11-26 04:51:50 | 58813.20 | 1197.00 | 80.0 ± 0.3 |
| P0214057021 | 01 | 2019-11-27 07:47:50 | 58814.33 | 1915.20 | 77.4 ± 0.2 |
|  | 02 | 2019-11-27 11:42:00 | 58814.49 | 1795.50 | 77.2 ± 0.2 |
|  | 03 | 2019-11-27 14:52:52 | 58814.62 | 897.75 | 76.5 ± 0.3 |
| P0214057022 | 01 | 2019-11-28 18:46:46 | 58815.78 | 2633.40 | 72.7 ± 0.2 |
|  | 02 | 2019-11-28 22:41:51 | 58815.95 | 1470.32 | 71.9 ± 0.2 |
|  | 03 | 2019-11-29 01:52:46 | 58816.08 | 418.95 | 71.8 ± 0.4 |

[a]: Background-subtracted count rates in the 1–10 keV band.

Table A2: *Swift*/XRT observations log, and net count rates

| ObsID | Start Time (day) | Start Time (MJD) | Exp. Time (s) | Count Rate[a] (ct s$^{-1}$) |
|---|---|---|---|---|
| 00012172003 | 2019-11-09 17:47:55 | 58796.74 | 1437.38 | 257.2 ± 0.9 |
| 00012172004 | 2019-11-10 19:38:42 | 58797.82 | 1908.12 | 271.2 ± 0.7 |
| 00012172005 | 2019-11-11 14:26:32 | 58798.60 | 333.66 | 253.7 ± 1.9 |
| 00012172006 | 2019-11-12 14:33:50 | 58799.61 | 1951.92 | 240.9 ± 0.6 |
| 00012172008 | 2019-11-14 09:38:00 | 58801.40 | 2515.35 | 258.9 ± 0.6 |
| 00088999001 | 2019-11-15 01:32:20 | 58802.06 | 1857.65 | 242.7 ± 0.6 |
| 00012172009 | 2019-11-16 17:04:45 | 58803.71 | 2481.50 | 216.2 ± 0.6 |
| 00012172010 | 2019-11-17 20:15:15 | 58804.84 | 2086.44 | 179.4 ± 0.6 |
| 00012172011 | 2019-11-18 21:46:35 | 58805.91 | 1696.77 | 178.2 ± 0.6 |
| 00012172012 | 2019-11-19 07:29:26 | 58806.31 | 1951.39 | 182.9 ± 0.5 |
| 00012172013 | 2019-11-20 18:19:49 | 58807.76 | 1621.90 | 197.5 ± 0.6 |
| 00012172014 | 2019-11-21 21:28:03 | 58808.90 | 1666.83 | 176.4 ± 0.6 |
| 00012172018 | 2019-11-25 19:31:04 | 58812.81 | 1667.00 | 145.5 ± 0.4 |
| 00012172019 | 2019-11-26 05:26:48 | 58813.23 | 1185.63 | 139.9 ± 0.5 |
| 00012172020 | 2019-11-27 03:33:30 | 58814.15 | 1767.38 | 146.5 ± 0.4 |
| 00012172021 | 2019-12-01 04:42:01 | 58818.20 | 2060.56 | 124.3 ± 0.3 |
| 00012172022 | 2019-12-02 00:05:15 | 58819.00 | 1852.87 | 121.7 ± 0.3 |
| 00012172023 | 2019-12-03 20:25:01 | 58820.85 | 1671.77 | 117.9 ± 0.3 |
| 00012172024 | 2019-12-04 04:34:57 | 58821.19 | 1775.42 | 112.9 ± 0.4 |
| 00012172025 | 2019-12-05 15:26:32 | 58822.64 | 646.88 | 110.0 ± 0.5 |
| 00012172026 | 2019-12-06 23:23:16 | 58823.98 | 1715.90 | 106.6 ± 0.3 |
| 00012172027 | 2019-12-07 20:00:17 | 58824.83 | 1711.74 | 102.3 ± 0.3 |
| 00012172028 | 2019-12-08 21:35:37 | 58825.90 | 838.38 | 98.0 ± 0.4 |
| 00012172029 | 2019-12-09 15:02:07 | 58826.63 | 1661.87 | 96.6 ± 0.3 |
| 00012172030 | 2019-12-10 00:36:07 | 58827.03 | 1606.97 | 95.0 ± 0.3 |
| 00012172031 | 2019-12-11 15:01:18 | 58828.63 | 1720.06 | 88.9 ± 0.3 |
| 00012172032 | 2019-12-12 13:07:41 | 58829.55 | 2095.44 | 85.3 ± 0.2 |
| 00012172033 | 2019-12-13 14:36:56 | 58830.61 | 2111.37 | 80.3 ± 0.2 |
| 00012172034 | 2019-12-14 13:03:45 | 58831.55 | 2216.88 | 83.6 ± 0.2 |
| 00012172035 | 2019-12-15 12:49:01 | 58832.53 | 2239.59 | 79.5 ± 0.2 |
| 00012172036 | 2019-12-16 20:55:02 | 58833.87 | 1980.62 | 76.6 ± 0.2 |
| 00012172037 | 2019-12-17 14:21:22 | 58834.60 | 1985.05 | 75.2 ± 0.2 |
| 00012172038 | 2019-12-18 14:09:21 | 58835.59 | 1971.45 | 75.4 ± 0.2 |
| 00012172039 | 2019-12-19 17:11:01 | 58836.72 | 1671.83 | 70.5 ± 0.2 |
| 00012172040 | 2019-12-20 15:28:54 | 58837.65 | 2252.73 | 68.5 ± 0.2 |
| 00012172041 | 2019-12-22 15:18:03 | 58839.64 | 469.00 | 64.5 ± 0.4 |
| 00012172042 | 2019-12-23 16:45:41 | 58840.70 | 1691.67 | 62.2 ± 0.2 |
| 00012172043 | 2019-12-24 15:30:26 | 58841.65 | 2085.66 | 62.8 ± 0.2 |
| 00012172044 | 2019-12-25 16:34:55 | 58842.69 | 1842.06 | 59.9 ± 0.2 |
| 00012172045 | 2019-12-26 16:45:26 | 58843.70 | 1956.53 | 59.0 ± 0.2 |
| 00012172046 | 2019-12-27 06:55:00 | 58844.29 | 1907.00 | 55.9 ± 0.2 |
| 00012172047 | 2019-12-28 19:37:42 | 58845.82 | 2000.23 | 54.7 ± 0.2 |
| 00012172048 | 2019-12-29 21:12:46 | 58846.88 | 1253.40 | 52.3 ± 0.2 |
| 00012172049 | 2019-12-30 17:44:06 | 58847.74 | 2011.04 | 50.8 ± 0.2 |
| 00012172050 | 2019-12-31 21:10:23 | 58848.88 | 269.17 | 49.9 ± 0.5 |
| 00012172051 | 2020-01-01 17:31:50 | 58849.73 | 2061.80 | 49.1 ± 0.2 |
| 00012172052 | 2020-01-02 06:12:11 | 58850.26 | 1811.06 | 45.4 ± 0.2 |
| 00012172053 | 2020-01-03 06:04:35 | 58851.25 | 1867.54 | 46.5 ± 0.2 |
| 00012172054 | 2020-01-04 10:46:07 | 58852.45 | 1890.46 | 46.0 ± 0.2 |
| 00012172055 | 2020-01-05 09:12:55 | 58853.38 | 1736.87 | 43.5 ± 0.3 |
| 00012172056 | 2020-01-06 07:32:28 | 58854.32 | 1642.03 | 43.3 ± 0.2 |
| 00012172057 | 2020-01-07 15:24:43 | 58855.64 | 1626.93 | 41.8 ± 0.2 |
| 00012172058 | 2020-01-08 13:40:59 | 58856.57 | 1908.47 | 41.5 ± 0.2 |
| 00012172059 | 2020-01-09 02:20:07 | 58857.10 | 2035.00 | 40.4 ± 0.2 |
| 00012172060 | 2020-01-10 10:12:17 | 58858.43 | 1974.75 | 41.1 ± 0.2 |
| 00012172064 | 2020-01-16 16:32:01 | 58864.69 | 354.03 | 11.9 ± 0.2 |
| 00012172066 | 2020-01-18 14:41:40 | 58866.61 | 673.65 | 10.66 ± 0.14 |
| 00012172067 | 2020-01-21 04:46:51 | 58869.20 | 1621.34 | 4.60 ± 0.06 |
| 00012172071 | 2020-01-25 23:24:30 | 58873.98 | 1641.89 | 2.42 ± 0.04 |
| 00012172072 | 2020-01-26 12:21:15 | 58874.52 | 796.52 | 2.05 ± 0.06 |
| 00012172073 | 2020-01-27 10:31:08 | 58875.44 | 1002.96 | 2.27 ± 0.05 |
| 00012172074 | 2020-01-28 13:24:08 | 58876.56 | 1365.92 | 1.61 ± 0.04 |
| 00012172075 | 2020-01-29 00:32:43 | 58877.02 | 1565.38 | 1.74 ± 0.04 |
| 00012172076 | 2020-01-30 03:38:08 | 58878.15 | 1680.76 | 1.35 ± 0.03 |
| 00012172077 | 2020-01-31 05:20:32 | 58879.22 | 1686.06 | 1.27 ± 0.03 |

[a]:Background-subtracted 0.5–10.0 keV count rates, including corrections for bad pixels and pile-up (Evans et al. 2009).

Table A3: Best-fitting parameters to the *Insight-HXMT* data (LE plus ME, 1–35 keV band). The model is tbabs× (diskbb + powerlaw). Uncertainties are 90% confidence levels for one independent parameter.

| ObsID | ExpID | $kT_{in}$ (keV) | $\sqrt{N_{disk}}$ (km) | $\Gamma$ | $N_{pl}$ | $F_{disk}/F_{pl}$ | $\chi^2/\nu$ |
|---|---|---|---|---|---|---|---|
| P0214057001 | 01 | $0.604 \pm 0.005$ | $54.5 \pm 1.1$ | $3.6 \pm 0.2$ | $0.47 \pm 0.07$ | $10.1 \pm 1.5$ | 1016.7/1094 |
| | 02 | $0.605 \pm 0.004$ | $54.3 \pm 1.0$ | $3.49^{+0.16}_{-0.14}$ | $0.48^{+0.05}_{-0.07}$ | $9.32^{+0.19}_{-1.14}$ | 1107.4/1128 |
| | 03 | $0.604^{+0.005}_{-0.004}$ | $54.7 \pm 1.1$ | $3.26^{+0.13}_{-0.12}$ | $0.51 \pm 0.08$ | $7.5 \pm 0.8$ | 968.2/1097 |
| | 04 | $0.598 \pm 0.004$ | $56.2^{+1.0}_{-0.9}$ | $3.08 \pm 0.10$ | $0.47 \pm 0.07$ | $6.89^{+0.58}_{-0.59}$ | 1115.6/1182 |
| | 05 | $0.597^{+0.004}_{-0.003}$ | $56.6 \pm 1.0$ | $3.01^{+0.09}_{-0.10}$ | $0.48 \pm 0.07$ | $6.42^{+0.53}_{-0.54}$ | 1111.8/1205 |
| | 06 | $0.602^{+0.005}_{-0.003}$ | $55.9^{+1.3}_{-1.0}$ | $3.16^{+0.14}_{-0.12}$ | $0.45 \pm 0.09$ | $7.9 \pm 1.0$ | 1055.6/1064 |
| | 07 | $0.591^{+0.007}_{-0.006}$ | $58.3 \pm 2.1$ | $3.0 \pm 0.2$ | $0.46 \pm 0.15$ | $6.5^{+1.2}_{-1.3}$ | 743.2/908 |
| | 08 | $0.599 \pm 0.006$ | $57.3^{+1.8}_{-1.7}$ | $3.1^{+0.2}_{-0.3}$ | $0.34 \pm 0.12$ | $10.6^{+2.3}_{-2.4}$ | 985.1/963 |
| | 09 | $0.613 \pm 0.004$ | $53.8 \pm 1.0$ | $3.46^{+0.15}_{-0.13}$ | $0.54^{+0.06}_{-0.04}$ | $8.5 \pm 1.0$ | 1019.1/1117 |
| | 10 | $0.609 \pm 0.004$ | $55.0 \pm 1.0$ | $3.26 \pm 0.13$ | $0.46 \pm 0.07$ | $8.9 \pm 0.9$ | 1129.2/1162 |
| | 11 | $0.606^{+0.005}_{-0.004}$ | $55.4^{+1.2}_{-0.6}$ | $3.28^{+0.13}_{-0.12}$ | $0.52^{+0.05}_{-0.08}$ | $7.7^{+0.9}_{-0.8}$ | 932.7/1083 |
| | 12 | $0.602 \pm 0.004$ | $56.6 \pm 1.0$ | $3.13^{+0.11}_{-0.12}$ | $0.44 \pm 0.07$ | $8.1 \pm 0.8$ | 1223.9/1173 |
| | 13 | $0.605 \pm 0.004$ | $55.7^{+1.1}_{-0.8}$ | $3.20^{+0.11}_{-0.07}$ | $0.51 \pm 0.07$ | $7.3^{+0.7}_{-0.6}$ | 1061.6/1160 |
| P0214057003 | 01 | $0.600 \pm 0.005$ | $56.1 \pm 1.3$ | $3.04^{+0.10}_{-0.12}$ | $0.49 \pm 0.09$ | $6.7^{+0.7}_{-0.8}$ | 1041.0/1082 |
| | 02 | $0.608 \pm 0.006$ | $55.1^{+1.5}_{-1.4}$ | $3.5^{+0.3}_{-0.2}$ | $0.51^{+0.08}_{-0.10}$ | $9.2^{+1.6}_{-1.5}$ | 911.4/981 |
| P0214057004 | 01 | $0.597^{+0.004}_{-0.005}$ | $54.9^{+1.3}_{-1.2}$ | $2.75^{+0.08}_{-0.09}$ | $0.53 \pm 0.08$ | $4.29^{+0.36}_{-0.37}$ | 1003.6/1099 |
| | 02 | $0.599 \pm 0.005$ | $54.1 \pm 1.5$ | $2.98^{+0.09}_{-0.10}$ | $0.61 \pm 0.10$ | $4.6 \pm 0.4$ | 909.4/1042 |
| P0214057005 | 01 | $0.59^{+0.004}_{-0.005}$ | $52.3^{+1.4}_{-1.2}$ | $2.66^{+0.05}_{-0.07}$ | $0.78 \pm 0.08$ | $2.24 \pm 0.12$ | 1039.1/1172 |
| | 02 | $0.586^{+0.005}_{-0.004}$ | $55.0^{+1.3}_{-1.6}$ | $2.60^{+0.09}_{-0.08}$ | $0.58 \pm 0.09$ | $3.1 \pm 0.3$ | 949.7/1042 |
| P0214057009 | 01 | $0.566 \pm 0.005$ | $53.7 \pm 1.5$ | $2.57^{+0.07}_{-0.08}$ | $0.58^{+0.10}_{-0.09}$ | $2.4 \pm 0.2$ | 882.3/999 |
| | 02 | $0.576^{+0.004}_{-0.005}$ | $51.6 \pm 1.1$ | $2.67^{+0.06}_{-0.08}$ | $0.62 \pm 0.07$ | $2.50 \pm 0.14$ | 1043.2/1167 |
| | 03 | $0.576 \pm 0.003$ | $50.6 \pm 1.0$ | $2.62 \pm 0.04$ | $0.68 \pm 0.06$ | $2.04 \pm 0.09$ | 1246.8/1271 |
| P0214057010 | 01 | $0.572^{+0.006}_{-0.005}$ | $51.9^{+1.6}_{-1.7}$ | $2.72^{+0.10}_{-0.11}$ | $0.50^{+0.10}_{-0.09}$ | $3.2 \pm 0.3$ | 934.1/1052 |
| | 02 | $0.564 \pm 0.003$ | $54.2^{+1.0}_{-1.1}$ | $2.66 \pm 0.07$ | $0.46 \pm 0.06$ | $3.3 \pm 0.2$ | 1177.4/1205 |
| | 03 | $0.571 \pm 0.006$ | $51.9^{+1.8}_{-1.9}$ | $2.80^{+0.10}_{-0.11}$ | $0.57 \pm 0.10$ | $3.0 \pm 0.3$ | 871.2/978 |
| P0214057011 | 01 | $0.562^{+0.006}_{-0.005}$ | $54.5^{+1.7}_{-1.4}$ | $3.10^{+0.17}_{-0.12}$ | $0.47 \pm 0.09$ | $4.8 \pm 0.5$ | 953.2/1016 |
| | 02 | $0.562 \pm 0.005$ | $54.4^{+1.5}_{-1.3}$ | $3.00^{+0.06}_{-0.12}$ | $0.45 \pm 0.07$ | $4.7 \pm 0.4$ | 989.5/1095 |
| | 03 | $0.551^{+0.006}_{-0.005}$ | $58.5^{+1.8}_{-2.2}$ | $2.7^{+0.2}_{-0.3}$ | $0.26^{+0.13}_{-0.11}$ | $6.0 \pm 1.3$ | 787.5/950 |
| | 04 | $0.555 \pm 0.003$ | $56.9^{+0.4}_{-0.9}$ | $2.73^{+0.11}_{-0.12}$ | $0.31^{+0.06}_{-0.03}$ | $5.4^{+0.5}_{-0.3}$ | 1102.2/1167 |
| | 05 | $0.56 \pm 0.003$ | $55.0 \pm 1.1$ | $2.91^{+0.07}_{-0.10}$ | $0.38 \pm 0.06$ | $5.3 \pm 0.4$ | 1125.7/1219 |
| | 06 | $0.561 \pm 0.005$ | $54.5^{+1.1}_{-0.9}$ | $3.05 \pm 0.10$ | $0.43^{+0.04}_{-0.07}$ | $5.0 \pm 0.5$ | 958.3/1070 |
| | 07 | $0.555^{+0.005}_{-0.004}$ | $55.9 \pm 0.9$ | $2.85^{+0.12}_{-0.13}$ | $0.37^{+0.07}_{-0.08}$ | $4.9 \pm 0.5$ | 874.4/1051 |
| | 08 | $0.553 \pm 0.004$ | $55.5 \pm 1.3$ | $2.81 \pm 0.10$ | $0.42 \pm 0.07$ | $4.2 \pm 0.4$ | 995.6/1126 |
| | 09 | $0.565^{+0.007}_{-0.006}$ | $51.8^{+1.8}_{-1.7}$ | $3.04 \pm 0.11$ | $0.55^{+0.09}_{-0.08}$ | $3.8 \pm 0.4$ | 841.2/1011 |
| P0214057012 | 01 | $0.547^{+0.005}_{-0.002}$ | $52.7^{+1.1}_{-0.9}$ | $2.88 \pm 0.10$ | $0.44 \pm 0.06$ | $3.5 \pm 0.2$ | 1029.4/1143 |
| | 02 | $0.546^{+0.003}_{-0.004}$ | $52.0^{+0.9}_{-1.2}$ | $2.74 \pm 0.08$ | $0.47 \pm 0.06$ | $2.75^{+0.18}_{-0.12}$ | 1078.8/1147 |
| | 03 | $0.545 \pm 0.004$ | $50.9 \pm 1.3$ | $2.64^{+0.06}_{-0.07}$ | $0.50 \pm 0.06$ | $2.21 \pm 0.14$ | 1010.7/1128 |
| | 04 | $0.538 \pm 0.005$ | $52.5^{+1.5}_{-1.6}$ | $2.49^{+0.09}_{-0.10}$ | $0.43^{+0.08}_{-0.07}$ | $2.15 \pm 0.18$ | 808.1/987 |
| P0214057013 | 01 | $0.534 \pm 0.004$ | $49.5 \pm 1.3$ | $2.60^{+0.05}_{-0.07}$ | $0.59^{+0.06}_{-0.05}$ | $1.53 \pm 0.08$ | 1042.2/1204 |
| | 02 | $0.519^{+0.009}_{-0.008}$ | $50 \pm 2$ | $2.59^{+0.09}_{-0.10}$ | $0.68 \pm 0.07$ | $1.19 \pm 0.11$ | 760.6/927 |
| P0214057014 | 01 | $0.516 \pm 0.006$ | $47.1^{+1.8}_{-1.9}$ | $2.59^{+0.05}_{-0.06}$ | $0.68^{+0.07}_{-0.06}$ | $1.54 \pm 0.10$ | 1071.3/1158 |
| P0214057015 | 01 | $0.522 \pm 0.005$ | $50.8 \pm 1.5$ | $2.52 \pm 0.07$ | $0.48^{+0.07}_{-0.06}$ | $1.60^{+0.11}_{-0.10}$ | 933.7/1088 |
| | 02 | $0.533^{+0.008}_{-0.004}$ | $46 \pm 2$ | $2.73^{+0.07}_{-0.08}$ | $0.67 \pm 0.08$ | $1.32 \pm 0.09$ | 919.1/1080 |
| | 03 | $0.536^{+0.007}_{-0.006}$ | $46.3^{+1.9}_{-1.8}$ | $2.81^{+0.06}_{-0.07}$ | $0.66 \pm 0.07$ | $1.54 \pm 0.10$ | 955.2/1125 |
| P0214057016 | 01 | $0.518^{+0.008}_{-0.006}$ | $50^{+2}_{-3}$ | $2.76^{+0.09}_{-0.10}$ | $0.51^{+0.09}_{-0.08}$ | $1.93 \pm 0.18$ | 801.5/977 |
| | 02 | $0.518^{+0.007}_{-0.004}$ | $51.6^{+1.9}_{-2.6}$ | $2.70 \pm 0.12$ | $0.43^{+0.11}_{-0.09}$ | $2.2^{+0.3}_{-0.2}$ | 724.4/879 |
| | 03 | $0.526^{+0.007}_{-0.005}$ | $49.5^{+1.1}_{-1.2}$ | $2.94^{+0.09}_{-0.15}$ | $0.51^{+0.07}_{-0.06}$ | $2.31^{+0.19}_{-0.12}$ | 832.3/1034 |
| P0214057017 | 01 | $0.519^{+0.005}_{-0.006}$ | $53.3^{+1.8}_{-0.9}$ | $3.24^{+0.12}_{-0.17}$ | $0.36^{+0.03}_{-0.06}$ | $4.6 \pm 0.5$ | 969.6/1095 |
| | 02 | $0.523^{+0.007}_{-0.006}$ | $51.7 \pm 1.9$ | $3.30^{+0.14}_{-0.13}$ | $0.42^{+0.07}_{-0.06}$ | $4.1 \pm 0.4$ | 909.2/1045 |
| | 03 | $0.521^{+0.008}_{-0.007}$ | $53 \pm 2$ | $3.4 \pm 0.2$ | $0.33 \pm 0.08$ | $5.7 \pm 0.9$ | 769.3/972 |
| P0214057018 | 01 | $0.525^{+0.007}_{-0.006}$ | $50.9^{+1.4}_{-1.7}$ | $3.5 \pm 0.2$ | $0.35 \pm 0.06$ | $5.8 \pm 0.8$ | 925.9/1020 |
| | 02 | $0.526^{+0.007}_{-0.006}$ | $50.7^{+1.8}_{-1.7}$ | $3.5 \pm 0.2$ | $0.34^{+0.06}_{-0.03}$ | $5.8 \pm 0.8$ | 886.6/1030 |
| | 03 | $0.517^{+0.011}_{-0.010}$ | $53 \pm 3$ | $3.6^{+0.4}_{-0.3}$ | $0.32^{+0.10}_{-0.05}$ | $6.2 \pm 1.5$ | 727.8/832 |
| P0214057019 | 01 | $0.507 \pm 0.005$ | $53.2^{+1.6}_{-1.5}$ | $3.49^{+0.08}_{-0.14}$ | $0.34 \pm 0.05$ | $5.2 \pm 0.5$ | 1011.1/1134 |
| | 02 | $0.51 \pm 0.007$ | $53 \pm 2$ | $3.5 \pm 0.2$ | $0.28 \pm 0.07$ | $6.5 \pm 1.1$ | 790.9/983 |
| | 03 | $0.518^{+0.006}_{-0.007}$ | $50.6^{+1.8}_{-1.6}$ | $3.9^{+0.4}_{-0.3}$ | $0.33 \pm 0.06$ | $7.1 \pm 1.3$ | 914.8/1024 |
| P0214057021 | 01 | $0.499^{+0.006}_{-0.005}$ | $53.0^{+1.8}_{-1.7}$ | $3.32^{+0.13}_{-0.12}$ | $0.35 \pm 0.05$ | $4.2 \pm 0.4$ | 981.1/1129 |
| | 02 | $0.498 \pm 0.005$ | $54.2^{+1.7}_{-1.6}$ | $3.16^{+0.14}_{-0.15}$ | $0.25 \pm 0.05$ | $5.3 \pm 0.6$ | 957.3/1106 |
| | 03 | $0.502^{+0.008}_{-0.007}$ | $51.8^{+2.6}_{-1.2}$ | $3.33^{+0.17}_{-0.10}$ | $0.33 \pm 0.08$ | $4.2 \pm 0.6$ | 778.0/955 |
| P0214057022 | 01 | $0.496^{+0.005}_{-0.004}$ | $52.7 \pm 1.4$ | $3.35 \pm 0.12$ | $0.29 \pm 0.04$ | $5.0 \pm 0.5$ | 1079.1/1213 |
| | 02 | $0.496 \pm 0.006$ | $52.8 \pm 1.9$ | $3.4 \pm 0.2$ | $0.26 \pm 0.05$ | $5.5 \pm 0.8$ | 837.5/1055 |
| | 03 | $0.494^{+0.011}_{-0.010}$ | $53 \pm 3$ | $3.2 \pm 0.3$ | $0.25^{+0.10}_{-0.09}$ | $5.2 \pm 1.3$ | 612.1/777 |

Notes: the absorption column was fixed at $N_{\rm H} = 2 \times 10^{21}$ cm$^{-2}$; $kT_{\rm in}$ is the peak color temperature of the disk; $\sqrt{N_{\rm disk}} = R_{\rm in}$ (km) $* d_{10}^{-1} * \sqrt{\cos(i)}$, is the square root of the normalization of the `diskbb` model; $\Gamma$ is the photon index; $N_{\rm pl}$ is the normalization of the `powerlaw` model in XSPEC, defined as photons keV$^{-1}$ cm$^{-2}$ s$^{-1}$ at 1 keV; $F_{\rm diskbb}/F_{\rm pl}$ is the ratio of unabsorbed disk to power-law fluxes in the 1–35 keV band.

Table A4: Best-fitting parameters to the *Swift*/XRT data (0.5–10 keV band). The model is
tbabs× (bbodyrad$_{\rm colder}$ + diskbb$_{\rm hotter}$ + powerlaw) (Scenario 1) for the high/soft state,
tbabs× (diskbb + powerlaw) for the intermediate state, and tbabs× powerlaw for the
hard state. Uncertainties are 90% confidence levels for one independent parameter.

| ObsID | $R$pileup[a] (pixels) | $N_{\rm H}^{b}$ ($10^{22}$ cm$^{-2}$) | $kT_{\rm BB}^{c}$ (keV) | $\sqrt{N_{\rm BB}}^{d}$ ($10^3$ km) | $kT_{\rm in}^{e}$ (keV) | $\sqrt{N_{\rm disk}}^{f}$ (km) | $\Gamma^{g}$ | $N_{\rm pl}^{h}$ ($10^{-2}$) | $\chi^2/\nu$ |
|---|---|---|---|---|---|---|---|---|---|
| 00012172003 | 2 | $0.26 \pm 0.04$ | $0.106^{+0.005}_{-0.004}$ | $1.9^{+0.6}_{-0.4}$ | $0.650^{+0.015}_{-0.014}$ | $32.8^{+1.5}_{-1.4}$ | $1.7^{+0.4}_{-0.5}$ | $10^{+10}_{-6}$ | 519.9/451 |
| 00012172004 | 3 | $0.26 \pm 0.03$ | $0.107 \pm 0.003$ | $1.9^{+1.4}_{-0.3}$ | $0.659^{+0.011}_{-0.013}$ | $34.0^{+1.4}_{-0.6}$ | $1.1^{+0.4}_{**}$ | $3.4^{+4.1}_{-0.7}$ | 504.7/479 |
| 00012172005 | 3 | $0.32^{+0.08}_{-0.05}$ | $0.105^{+0.003}_{-0.008}$ | $2.6^{+1.9}_{-0.4}$ | $0.64^{+0.02}_{-0.03}$ | $40^{+5}_{-3}$ | $1.1^{+1.2}_{**}$ | $3.5^{+30.0}_{-0.9}$ | 294.6/294 |
| 00012172006 | 3 | $0.24 \pm 0.04$ | $0.109 \pm 0.004$ | $1.8^{+0.5}_{-0.3}$ | $0.621^{+0.010}_{-0.013}$ | $39.6^{+1.9}_{-1.4}$ | $1.2^{+0.5}_{**}$ | $3^{+6}_{-1}$ | 461.0/432 |
| 00012172008 | 3 | $0.30 \pm 0.02$ | $0.101 \pm 0.002$ | $2.9^{+0.4}_{-0.3}$ | $0.618^{+0.006}_{-0.007}$ | $43.0^{+1.2}_{-1.1}$ | [1.0] | $1.26^{+1.84}_{-0.15}$ | 583.8/451 |
| 00088999001 | 3 | $0.31 \pm 0.03$ | $0.097 \pm 0.002$ | $3.4^{+0.5}_{-0.4}$ | $0.635 \pm 0.007$ | $41.8 \pm 1.2$ | [1.0] | $1.05^{+0.72}_{-0.17}$ | 609.1/435 |
| 00012172009 | 2 | $0.24 \pm 0.03$ | $0.106 \pm 0.003$ | $2.0^{+0.4}_{-0.3}$ | $0.615^{+0.010}_{-0.011}$ | $41.3^{+1.7}_{-1.4}$ | $1.2^{+0.4}_{**}$ | $3.6^{+4.6}_{-1.5}$ | 630.1/459 |
| 00012172010 | 2 | $0.24 \pm 0.03$ | $0.109 \pm 0.004$ | $1.7^{+0.4}_{-0.3}$ | $0.605^{+0.010}_{-0.013}$ | $36.2^{+1.8}_{-1.4}$ | $1.2^{+0.4}_{**}$ | $3.1^{+4.0}_{-1.2}$ | 502.8/443 |
| 00012172011 | 0 | $0.17 \pm 0.05$ | $0.116^{+0.009}_{-0.007}$ | $1.3^{+0.7}_{-0.4}$ | $0.57 \pm 0.02$ | $43 \pm 3$ | $1.8 \pm 0.3$ | $21^{+15}_{-10}$ | 499.1/429 |
| 00012172012 | 3 | $0.20 \pm 0.05$ | $0.112^{+0.007}_{-0.006}$ | $1.5^{+0.6}_{-0.4}$ | $0.60 \pm 0.02$ | $40 \pm 2$ | $1.7 \pm 0.4$ | $14^{+13}_{-7}$ | 380.1/435 |
| 00012172013 | 3 | $0.18^{+0.05}_{-0.04}$ | $0.119^{+0.009}_{-0.007}$ | $1.0^{+0.5}_{-0.3}$ | $0.59 \pm 0.02$ | $35.2^{+2.2}_{-1.9}$ | $1.7^{+0.4}_{-0.5}$ | $8^{+10}_{-5}$ | 430.6/414 |
| 00012172014 | 2 | $0.19 \pm 0.04$ | $0.113^{+0.006}_{-0.005}$ | $1.3^{+0.5}_{-0.3}$ | $0.592 \pm 0.013$ | $37.6^{+1.8}_{-1.7}$ | $1.8 \pm 0.5$ | $6^{+9}_{-4}$ | 471.8/421 |
| 00012172018 | 0 | $0.22 \pm 0.03$ | $0.110^{+0.004}_{-0.003}$ | $1.4^{+0.3}_{-0.2}$ | $0.550^{+0.007}_{-0.008}$ | $39.2^{+1.5}_{-1.4}$ | [1.0] | $0.28^{+1.18}_{-0.07}$ | 432.9/363 |
| 00012172019 | 2 | $0.25 \pm 0.04$ | $0.106 \pm 0.005$ | $1.7^{+0.6}_{-0.4}$ | $0.559^{+0.010}_{-0.014}$ | $39.2^{+2.4}_{-1.8}$ | [1.0] | $0.41^{+6.40}_{-0.11}$ | 349.3/338 |
| 00012172020 | 1 | $0.23 \pm 0.03$ | $0.106 \pm 0.003$ | $1.7^{+0.4}_{-0.3}$ | $0.567^{+0.007}_{-0.008}$ | $40.1^{+1.5}_{-1.3}$ | [1.0] | $0.63^{+1.28}_{-0.09}$ | 493.7/392 |
| 00012172021 | 1 | $0.27 \pm 0.03$ | $0.104 \pm 0.003$ | $2.0^{+0.4}_{-0.3}$ | $0.540 \pm 0.006$ | $44.3 \pm 1.5$ | [1.0] | $0.32^{+0.50}_{-0.07}$ | 481.8/366 |
| 00012172022 | 1 | $0.26 \pm 0.03$ | $0.105 \pm 0.003$ | $2.1^{+0.4}_{-0.3}$ | $0.540^{+0.006}_{-0.008}$ | $45.8^{+1.8}_{-1.5}$ | [1.0] | $0.33^{+1.63}_{-0.08}$ | 426.5/365 |
| 00012172023 | 1 | $0.21^{+0.04}_{-0.02}$ | $0.109 \pm 0.004$ | $1.4^{+0.4}_{-0.3}$ | $0.532^{+0.008}_{-0.010}$ | $39.8^{+2.0}_{-0.7}$ | $1.16^{+0.93}_{**}$ | $0.5^{+2.2}_{-0.2}$ | 368.8/359 |
| 00012172024 | 0 | $0.20 \pm 0.04$ | $0.112 \pm 0.006$ | $1.3^{+0.5}_{-0.3}$ | $0.526 \pm 0.011$ | $45^{+3}_{-2}$ | $1.8^{+1.0}_{**}$ | $1.5^{+7.4}_{-1.2}$ | 358.3/330 |
| 00012172025 | 0 | $0.240 \pm 0.051$ | $0.107^{+0.006}_{-0.005}$ | $1.7^{+0.7}_{-0.4}$ | $0.529^{+0.011}_{-0.015}$ | $43^{+3}_{-2}$ | [1.0] | $0.36^{+6.77}_{-0.11}$ | 298.9/298 |
| 00012172026 | 0 | $0.24 \pm 0.03$ | $0.107^{+0.004}_{-0.003}$ | $1.7^{+0.4}_{-0.3}$ | $0.521 \pm 0.007$ | $42.3^{+1.8}_{-1.6}$ | [1.0] | $0.38^{+0.91}_{-0.07}$ | 434.1/348 |
| 00012172027 | 0 | $0.17 \pm 0.04$ | $0.112^{+0.006}_{-0.005}$ | $1.2^{+0.4}_{-0.3}$ | $0.510 \pm 0.010$ | $45 \pm 2$ | $1.9^{+0.6}_{-0.8}$ | $2.6^{+5.5}_{-1.9}$ | 431.0/362 |
| 00012172028 | 0 | $0.20 \pm 0.05$ | $0.109^{+0.006}_{-0.005}$ | $1.3^{+0.5}_{-0.3}$ | $0.521^{+0.010}_{-0.011}$ | $43.1^{+1.7}_{-2.2}$ | [1.0] | $0.33^{+2.31}_{-0.10}$ | 323.4/310 |
| 00012172029 | 0 | $0.23 \pm 0.04$ | $0.107 \pm 0.004$ | $1.6^{+0.5}_{-0.3}$ | $0.508 \pm 0.009$ | $44.9^{+2.1}_{-1.9}$ | $1.9^{+0.8}$ | $1.7^{+5.5}_{-1.4}$ | 402.8/353 |
| 00012172030 | 0 | $0.19^{+0.03}_{-0.04}$ | $0.111^{+0.005}_{-0.006}$ | $1.2^{+0.3}_{-0.2}$ | $0.518 \pm 0.007$ | $42.4^{+1.8}_{-1.7}$ | [1.0] | $0.38^{+0.46}_{-0.07}$ | 403.7/348 |
| 00012172031 | 0 | $0.25^{+0.03}_{-0.04}$ | $0.105^{+0.004}_{-0.004}$ | $1.7^{+0.5}_{-0.3}$ | $0.498^{+0.008}_{-0.009}$ | $45.3^{+2.2}_{-1.9}$ | $1.3^{+1.0}_{-0.3}$ | $0.5^{+2.3}_{-0.3}$ | 456.0/336 |
| 00012172032 | 0 | $0.15 \pm 0.04$ | $0.119^{+0.007}_{-0.006}$ | $0.9^{+0.3}_{-0.2}$ | $0.500^{+0.008}_{-0.010}$ | $43.9^{+2.4}_{-1.3}$ | $1.3^{+0.9}_{-0.3}$ | $0.6^{+2.4}_{-0.3}$ | 382.9/338 |
| 00012172033 | 0 | $0.20 \pm 0.03$ | $0.114^{+0.005}_{-0.004}$ | $1.2^{+0.3}_{-0.2}$ | $0.490 \pm 0.007$ | $44.1 \pm 1.8$ | [1.0] | $0.23^{+0.55}_{-0.05}$ | 355.7/325 |
| 00012172034 | 0 | $0.16 \pm 0.03$ | $0.116^{+0.004}_{-0.005}$ | $1.0^{+0.3}_{-0.2}$ | $0.495^{+0.007}_{-0.009}$ | $44.2^{+2.2}_{-1.8}$ | $1.2^{+0.9}_{**}$ | $0.4^{+1.6}_{-0.2}$ | 406.0/343 |
| 00012172035 | 0 | $0.16 \pm 0.03$ | $0.118^{+0.006}_{-0.005}$ | $0.9^{+0.3}_{-0.2}$ | $0.491^{+0.006}_{-0.008}$ | $42.4^{+2.0}_{-1.6}$ | [1.0] | $0.27^{+0.82}_{-0.04}$ | 395.7/343 |
| 00012172036 | 0 | $0.17 \pm 0.03$ | $0.117^{+0.006}_{-0.005}$ | $1.0^{+0.3}_{-0.2}$ | $0.491^{+0.006}_{-0.007}$ | $41.9^{+1.9}_{-1.6}$ | [1.0] | $0.15^{+0.96}_{-0.04}$ | 342.9/328 |
| 00012172037 | 0 | $0.17^{+0.03}_{-0.04}$ | $0.113 \pm 0.005$ | $1.0^{+0.3}_{-0.2}$ | $0.486^{+0.007}_{-0.009}$ | $43.1^{+1.2}_{-1.8}$ | [1.0] | $0.18^{+1.37}_{-0.04}$ | 371.1/326 |
| 00012172038 | 0 | $0.20 \pm 0.04$ | $0.107 \pm 0.005$ | $1.4^{+0.5}_{-0.3}$ | $0.472 \pm 0.010$ | $47.1^{+2.7}_{-2.5}$ | [1.0] | $1.2^{+3.6}_{-0.9}$ | 346.0/314 |
| 00012172039 | 0 | $0.20 \pm 0.04$ | $0.114^{+0.006}_{-0.005}$ | $1.1^{+0.4}_{-0.2}$ | $0.469^{+0.008}_{-0.010}$ | $44^{+3}_{-2}$ | [1.0] | $0.24^{+1.31}_{-0.05}$ | 294.2/306 |
| 00012172040 | 0 | $0.15 \pm 0.04$ | $0.119^{+0.008}_{-0.006}$ | $0.9^{+0.3}_{-0.2}$ | $0.468 \pm 0.010$ | $44^{+3}_{-2}$ | $1.6^{+0.7}_{**}$ | $0.8^{+1.8}_{-0.5}$ | 372.6/317 |
| 00012172041 | 0 | $0.10^{+0.08}_{-0.07}$ | $0.129^{+0.024}_{-0.013}$ | $0.6^{+0.7}_{-0.2}$ | $0.46 \pm 0.02$ | $41^{+6}_{-4}$ | [1.0] | $0.29^{+4.13}_{-0.11}$ | 226.8/218 |
| 00012172042 | 0 | $0.16 \pm 0.04$ | $0.121^{+0.008}_{-0.006}$ | $0.8^{+0.3}_{-0.2}$ | $0.467^{+0.008}_{-0.009}$ | $41^{+3}_{-2}$ | [1.0] | $0.31^{+0.57}_{-0.05}$ | 342.0/299 |
| 00012172043 | 0 | $0.15 \pm 0.04$ | $0.118 \pm 0.007$ | $0.8^{+0.4}_{-0.2}$ | $0.455^{+0.010}_{-0.011}$ | $43.2^{+3.1}_{-1.6}$ | $1.3^{+1.0}_{**}$ | $0.3^{+1.5}_{-0.2}$ | 385.0/292 |
| 00012172044 | 0 | $0.17 \pm 0.05$ | $0.113^{+0.008}_{-0.007}$ | $1.0^{+0.5}_{-0.3}$ | $0.440 \pm 0.012$ | $46^{+4}_{-3}$ | $1.8^{+0.7}_{**}$ | $1.2^{+2.9}_{-0.9}$ | 291.9/295 |
| 00012172045 | 0 | $0.18^{+0.05}_{-0.04}$ | $0.117^{+0.008}_{-0.007}$ | $1.1^{+0.5}_{-0.3}$ | $0.444 \pm 0.014$ | $48 \pm 4$ | $2.0^{+0.8}_{-0.9}$ | $1.8^{+5.2}_{-1.4}$ | 300.2/284 |
| 00012172046 | 0 | $0.16 \pm 0.05$ | $0.115^{+0.009}_{-0.007}$ | $0.9^{+0.4}_{-0.2}$ | $0.453 \pm 0.012$ | $42 \pm 3$ | $1.5^{+0.7}_{**}$ | $0.7^{+1.8}_{-0.4}$ | 333.4/299 |
| 00012172047 | 0 | $0.11^{+0.04}_{-0.03}$ | $0.128^{+0.008}_{-0.009}$ | $0.58^{+0.29}_{-0.13}$ | $0.452^{+0.011}_{-0.014}$ | $40^{+4}_{-3}$ | $1.3^{+1.0}_{**}$ | $0.39^{+1.63}_{-0.17}$ | 271.3/293 |
| 00012172048 | 0 | $0.16 \pm 0.07$ | $0.116^{+0.015}_{-0.011}$ | $0.8^{+0.8}_{-0.3}$ | $0.43 \pm 0.02$ | $45^{+6}_{-5}$ | $2.3^{+0.9}_{-1.1}$ | $1.8^{+6.1}_{-1.5}$ | 258.6/243 |
| 00012172049 | 0 | $0.14 \pm 0.04$ | $0.123^{+0.010}_{-0.008}$ | $0.7^{+0.3}_{-0.2}$ | $0.441^{+0.009}_{-0.008}$ | $42 \pm 3$ | [1.0] | $0.23^{+0.49}_{-0.04}$ | 262.2/272 |
| 00012172050 | 0 | [0.2] | $0.109^{+0.008}_{-0.007}$ | $1.3^{+0.31}_{-0.19}$ | $0.42 \pm -0.02$ | $53^{+9}_{-6}$ | [1.0] | $0.4^{+19.7}_{-0.2}$ | 129.3/162 |
| 00012172051 | 0 | $0.16 \pm 0.05$ | $0.119^{+0.011}_{-0.009}$ | $0.8^{+0.5}_{-0.2}$ | $0.419 \pm 0.015$ | $45^{+5}_{-4}$ | $2.2^{+0.7}_{-0.8}$ | $1.8^{+4.0}_{-1.3}$ | 286.0/278 |
| 00012172052 | 0 | $0.17 \pm 0.05$ | $0.113 \pm 0.008$ | $0.9^{+0.5}_{-0.2}$ | $0.424^{+0.011}_{-0.013}$ | $43^{+4}_{-2}$ | $1.2^{+0.9}_{**}$ | $0.31^{+1.12}_{-0.14}$ | 282.2/258 |
| 00012172053 | 0 | $0.14 \pm 0.05$ | $0.119^{+0.012}_{-0.009}$ | $0.7^{+0.4}_{-0.2}$ | $0.423^{+0.013}_{-0.014}$ | $43 \pm 4$ | [1.0] | $0.7^{+2.2}_{-0.6}$ | 316.0/268 |
| 00012172054 | 0 | $0.09^{+0.05}_{-0.04}$ | $0.128^{+0.014}_{-0.011}$ | $0.48^{+0.30}_{-0.14}$ | $0.425^{+0.009}_{-0.012}$ | $42^{+4}_{-3}$ | [1.0] | $0.15^{+0.64}_{-0.03}$ | 273.4/262 |
| 00012172055 | 0 | $0.12^{+0.10}_{-0.09}$ | $0.12^{+0.04}_{-0.02}$ | $0.5^{+1.1}_{-0.2}$ | $0.39 \pm 0.02$ | $44^{+9}_{-6}$ | $1.6^{+1.2}_{**}$ | $0.5^{+2.9}_{-0.3}$ | 204.8/199 |
| 00012172056 | 0 | $0.14^{+0.05}_{-0.04}$ | $0.123^{+0.011}_{-0.009}$ | $0.6^{+0.3}_{-0.2}$ | $0.417^{+0.010}_{-0.011}$ | $43^{+4}_{-3}$ | [1.0] | $0.21^{+0.48}_{-0.03}$ | 261.1/249 |
| 00012172057 | 0 | $0.13^{+0.07}_{-0.06}$ | $0.122^{+0.016}_{-0.012}$ | $0.59^{+0.53}_{-0.19}$ | $0.428^{+0.011}_{-0.020}$ | $37^{+5}_{-4}$ | $1.6^{+1.3}_{**}$ | $0.5^{+3.5}_{-0.4}$ | 221.4/232 |
| 00012172058 | 0 | $0.12^{+0.04}_{-0.02}$ | $0.130^{+0.006}_{-0.010}$ | $0.53^{+0.29}_{-0.13}$ | $0.418^{+0.010}_{-0.015}$ | $41^{+5}_{-3}$ | [1.0] | $0.16^{+1.06}_{-0.03}$ | 255.6/253 |
| 00012172059 | 0 | $0.12^{+0.04}_{-0.03}$ | $0.131^{+0.008}_{-0.010}$ | $0.53^{+0.27}_{-0.11}$ | $0.423^{+0.013}_{-0.015}$ | $38^{+5}_{-2}$ | $1.2^{+0.8}_{**}$ | $0.26^{+0.77}_{-0.09}$ | 291.8/256 |
| 00012172060 | 0 | $0.08^{+0.05}_{-0.04}$ | $0.14 \pm 0.02$ | $0.37^{+0.26}_{-0.10}$ | $0.43 \pm 0.03$ | $35^{+6}_{-5}$ | $1.9^{+0.5}_{-0.6}$ | $2.1^{+3.1}_{-1.3}$ | 326.9/280 |
| 00012172064 | 0 | $0.06^{+0.12}_{-0.06}$ | ... | ... | $0.16 \pm 0.03$ | $210^{+610}_{-80}$ | $2.1 \pm 0.2$ | $4.7^{+1.3}_{-1.2}$ | 118.9/107 |
| 00012172066 | 0 | $0.06^{+0.09}_{-0.06}$ | ... | ... | $0.15 \pm 0.02$ | $220^{+300}_{-70}$ | $2.02 \pm 0.15$ | $4.1 \pm 0.7$ | 145.8/144 |
| 00012172067 | 0 | $0.11^{+0.12}_{-0.11}$ | ... | ... | $0.12 \pm 0.02$ | $330^{+640}_{-140}$ | $1.78 \pm 0.11$ | $2.1 \pm 0.3$ | 175.7/178 |
| 00012172071 | 0 | [0.09] | ... | ... | $0.13^{+0.03}_{-0.02}$ | $120^{+110}_{-40}$ | $1.70^{+0.07}_{-0.10}$ | $1.33^{+0.08}_{-0.12}$ | 114.8/139 |
| 00012172072 | 0 | [0.09] | ... | ... | $0.11^{+0.03}_{-0.02}$ | $260^{+530}_{-90}$ | $1.65^{+0.11}_{-0.13}$ | $1.18^{+0.11}_{-0.09}$ | 48.7/54 |
| 00012172073 | 0 | $0.2^{+0.3}_{-0.2}$ | ... | ... | $0.10^{+0.04}_{-0.03}$ | $340^{+330}_{-170}$ | $1.8 \pm 0.2$ | $1.4^{+0.4}_{-0.3}$ | 74.4/76 |
| 00012172074 | 0 | $0.06^{+0.28}_{-0.06}$ | ... | ... | $0.10^{+0.04}_{-0.03}$ | $220^{+330}_{-100}$ | $1.66^{+0.24}_{-0.14}$ | $0.87^{+0.27}_{-0.08}$ | 47.8/59 |
| 00012172075 | 0 | [0.09] | ... | ... | ... | ... | $1.95 \pm 0.07$ | $1.21 \pm 0.05$ | 130.9/95 |
| 00012172076 | 0 | [0.09] | ... | ... | ... | ... | $1.79^{+0.08}_{-0.07}$ | $0.88 \pm 0.05$ | 106.2/76 |
| 00012172077 | 0 | [0.09] | ... | ... | ... | ... | $1.82^{+0.09}_{-0.08}$ | $0.79 \pm 0.04$ | 105.7/71 |

$^a$: radius (in pixels) of the central circle around the source in the XRT image that was excised from our spectral data analysis because of pileup.
$^b$: for epochs when $N_{\rm H}$ is not well constrained from spectral modelling, we assumed a fixed $N_{\rm H} = 2 \times 10^{21}$ cm$^{-2}$ in the soft state and $N_{\rm H} = 9 \times 10^{20}$ cm$^{-2}$ in the intermediate and hard states.
$^c$: blackbody temperature.
$^d$: normalization of the `bbodyrad` model: $\sqrt{N_{\rm BB}} = R_{\rm in}$ (km) $* d_{10}^{-1}$.
$^e$: peak colour temperature of the disk-blackbody component.
$^f$: normalization of the `diskbb` model: $\sqrt{N_{\rm disk}} = R_{\rm in}$ (km) $* d_{10}^{-1} * \sqrt{\cos(i)}$.
$^g$: photon index. We set $\Gamma = 1$ as the lowest permitted value of the photon index for all XSPEC fits. For epochs when the best-fitting value of $\Gamma$ would go below that limit, we assumed a fixed $\Gamma = 1$. At other epochs, we find a best-fitting value of $\Gamma > 1$ but the lower limit of the 90% confidence intervals is undetermined (below 1.0): in that case, the lower limit is marked with $**$.
$^h$: normalization of the `powerlaw` model, defined as photons keV$^{-1}$ cm$^{-2}$ s$^{-1}$ at 1 keV.

Table A5: Best-fitting parameters to the *Swift*/XRT data (0.5–10 keV band), for the high/soft state epochs only. The model is tbabs× (bbodyrad$_{hotter}$ + diskbb$_{colder}$ + powerlaw) (Scenario 2). Uncertainties are 90% confidence levels for one independent parameter.

| ObsID | $N_H$ ($10^{22}$ cm$^{-2}$) | $kT_{BB}$ (keV) | $\sqrt{N_{BB}}$ (km) | $kT_{in}$ (keV) | $\sqrt{N_{disk}}$ ($10^2$ km) | $\Gamma$ | $N_{pl}$ | $\chi^2/\nu$ |
|---|---|---|---|---|---|---|---|---|
| 00012172003 | $0.24 \pm 0.05$ | $0.467 \pm 0.008$ | $64^{+3}_{-4}$ | $0.15 \pm 0.02$ | $7.0^{+5}_{-2.0}$ | $2.5 \pm 0.2$ | $0.5 \pm 0.2$ | 495.4/450 |
| 00012172004 | $0.21 \pm 0.04$ | $0.466^{+0.007}_{-0.006}$ | $71^{+2}_{-3}$ | $0.164^{+0.014}_{-0.013}$ | $6.1^{+2.0}_{-1.5}$ | $2.2^{+0.2}_{-0.3}$ | $0.35^{+0.16}_{-0.13}$ | 451.0/478 |
| 00012172005 | $0.29^{+0.11}_{-0.09}$ | $0.46 \pm 0.02$ | $80 \pm 9$ | $0.15 \pm 0.03$ | $10^{+15}_{-4}$ | $2.5^{+0.4}_{-0.7}$ | $0.6^{+0.6}_{-0.4}$ | 289.8/293 |
| 00012172006 | $0.19 \pm 0.04$ | $0.452^{+0.009}_{-0.007}$ | $77 \pm 3$ | $0.18 \pm 0.02$ | $5.1^{+2.5}_{-1.2}$ | $2.3^{+0.3}_{-0.4}$ | $0.3 \pm 0.2$ | 434.0/431 |
| 00012172008 | $0.22^{+0.04}_{-0.03}$ | $0.454 \pm 0.005$ | $83.3^{+2.4}_{-2.6}$ | $0.161^{+0.012}_{-0.013}$ | $7.1^{+2.8}_{-1.4}$ | $2.6 \pm 0.3$ | $0.3 \pm 0.2$ | 512.9/450 |
| 00088999001 | $0.17 \pm 0.03$ | $0.469^{+0.006}_{-0.005}$ | $83 \pm 2$ | $0.174 \pm 0.012$ | $5.3^{+1.7}_{-1.1}$ | $2.1^{+0.4}_{-0.5}$ | $0.13^{+0.14}_{-0.08}$ | 494.9/434 |
| 00012172009 | $0.18 \pm 0.04$ | $0.447^{+0.007}_{-0.006}$ | $81 \pm 3$ | $0.170^{+0.015}_{-0.014}$ | $5.5^{+2.1}_{-1.2}$ | $2.3^{+0.2}_{-0.3}$ | $0.34^{+0.17}_{-0.14}$ | 568.2/458 |
| 00012172010 | $0.19 \pm 0.04$ | $0.440^{+0.008}_{-0.007}$ | $71.4 \pm 2.7$ | $0.170^{+0.014}_{-0.013}$ | $5.3^{+1.9}_{-1.1}$ | $2.2 \pm 0.3$ | $0.21^{+0.13}_{-0.09}$ | 460.2/442 |
| 00012172011 | $0.17 \pm 0.04$ | $0.419^{+0.012}_{-0.010}$ | $82 \pm 5$ | $0.17 \pm 0.02$ | $5.3^{+3.1}_{-1.5}$ | $2.3^{+0.2}_{-0.3}$ | $0.5 \pm 0.2$ | 474.9/428 |
| 00012172012 | $0.18^{+0.05}_{-0.04}$ | $0.433^{+0.008}_{-0.009}$ | $77.3 \pm 4$ | $0.17 \pm 0.02$ | $5.8^{+3.5}_{-1.6}$ | $2.3^{+0.2}_{-0.3}$ | $0.5 \pm 0.2$ | 358.3/434 |
| 00012172013 | $0.17^{+0.05}_{-0.04}$ | $0.433^{+0.013}_{-0.010}$ | $67 \pm 4$ | $0.18 \pm 0.02$ | $4.1^{+2.4}_{-1.1}$ | $2.3^{+0.3}_{-0.4}$ | $0.30^{+0.20}_{-0.14}$ | 427.4/413 |
| 00012172014 | $0.19 \pm 0.05$ | $0.436^{+0.008}_{-0.007}$ | $70 \pm 4$ | $0.17 \pm 0.02$ | $4.5^{+2.9}_{-1.2}$ | $2.7^{+0.2}_{-0.3}$ | $0.4 \pm 0.2$ | 446.7/420 |
| 00012172018 | $0.16^{+0.04}_{-0.03}$ | $0.420 \pm 0.007$ | $70 \pm 3$ | $0.185 \pm 0.015$ | $3.7^{+1.2}_{-0.7}$ | $2.6^{+0.5}_{-0.7}$ | $0.08^{+0.13}_{-0.05}$ | 399.7/362 |
| 00012172019 | $0.19^{+0.10}_{-0.06}$ | $0.421^{+0.010}_{-0.008}$ | $71 \pm 4$ | $0.17^{+0.02}_{-0.03}$ | $4.6^{+6.3}_{-1.4}$ | $2.9^{+0.6}_{-0.8}$ | $0.19^{+0.36}_{-0.14}$ | 346.9/337 |
| 00012172020 | $0.17^{+0.04}_{-0.03}$ | $0.426^{+0.007}_{-0.006}$ | $75^{+3}_{-2}$ | $0.176 \pm 0.013$ | $4.5^{+1.5}_{-0.9}$ | $2.4^{+0.4}_{-0.5}$ | $0.11^{+0.13}_{-0.07}$ | 446.9/391 |
| 00012172021 | $0.19 \pm 0.03$ | $0.415 \pm 0.006$ | $80 \pm 3$ | $0.172 \pm 0.011$ | $5.3^{+1.5}_{-0.9}$ | $2.2^{+0.7}_{-0.8}$ | $0.05^{+0.11}_{-0.04}$ | 421.5/365 |
| 00012172022 | $0.21^{+0.04}_{-0.03}$ | $0.411^{+0.006}_{-0.005}$ | $83 \pm 3$ | $0.166^{+0.011}_{-0.012}$ | $6.2^{+2.0}_{-1.2}$ | $2.7^{+0.5}_{-0.6}$ | $0.11^{+0.15}_{-0.08}$ | 392.7/364 |
| 00012172023 | $0.17^{+0.05}_{-0.04}$ | $0.405^{+0.008}_{-0.007}$ | $71 \pm 3$ | $0.18 \pm 0.02$ | $4.1^{+1.7}_{-0.9}$ | $2.7^{+0.5}_{-0.6}$ | $0.10^{+0.14}_{-0.07}$ | 352.3/358 |
| 00012172024 | $0.17^{+0.06}_{-0.04}$ | $0.404^{+0.009}_{-0.008}$ | $78 \pm 4$ | $0.18 \pm 0.02$ | $4.4^{+2.2}_{-1.0}$ | $2.8^{+0.5}_{-0.7}$ | $0.14^{+0.20}_{-0.10}$ | 348.2/329 |
| 00012172025 | $0.20^{+0.08}_{-0.06}$ | $0.402^{+0.011}_{-0.009}$ | $77^{+5}_{-4}$ | $0.17 \pm 0.02$ | $5^{+5}_{-2}$ | $2.8^{+0.7}_{-1.0}$ | $0.14^{+0.30}_{-0.11}$ | 290.5/297 |
| 00012172026 | $0.19^{+0.04}_{-0.03}$ | $0.398 \pm 0.007$ | $76 \pm 3$ | $0.167 \pm 0.012$ | $5.3^{+1.8}_{-1.1}$ | $2.4^{+0.6}_{-0.7}$ | $0.06^{+0.10}_{-0.04}$ | 406.0/347 |
| 00012172027 | $0.15^{+0.05}_{-0.04}$ | $0.393^{+0.008}_{-0.007}$ | $77 \pm 4$ | $0.18 \pm 0.02$ | $3.7^{+1.6}_{-0.8}$ | $2.8^{+0.4}_{-0.5}$ | $0.15^{+0.15}_{-0.09}$ | 417.2/361 |
| 00012172028 | $0.13^{+0.06}_{-0.05}$ | $0.402^{+0.011}_{-0.010}$ | $75 \pm 5$ | $0.19 \pm 0.02$ | $3.4^{+2.0}_{-0.9}$ | $2.4^{+0.8}_{-1.2}$ | $0.060^{+0.18}_{-0.05}$ | 304.7/309 |
| 00012172029 | $0.21^{+0.06}_{-0.04}$ | $0.391^{+0.007}_{-0.006}$ | $77 \pm 3$ | $0.164^{+0.015}_{-0.017}$ | $5.5^{+2.8}_{-1.2}$ | $3.1^{+0.4}_{-0.5}$ | $0.18^{+0.19}_{-0.11}$ | 395.1/352 |
| 00012172030 | $0.13 \pm 0.03$ | $0.402^{+0.008}_{-0.007}$ | $74^{+3}_{-4}$ | $0.188^{+0.015}_{-0.013}$ | $3.4^{+0.7}_{-0.7}$ | $1.9^{+0.7}_{-0.8}$ | $0.025^{+0.057}_{-0.019}$ | 366.4/347 |
| 00012172031 | $0.22 \pm 0.04$ | $0.384 \pm 0.006$ | $79 \pm 3$ | $0.160^{+0.011}_{-0.012}$ | $6.1^{+2.2}_{-1.2}$ | $2.8^{+0.6}_{-0.7}$ | $0.09^{+0.14}_{-0.06}$ | 413.8/335 |
| 00012172032 | $0.14^{+0.04}_{-0.03}$ | $0.388 \pm 0.008$ | $75 \pm 4$ | $0.188^{+0.016}_{-0.014}$ | $3.5^{+1.2}_{-0.7}$ | $2.5^{+0.5}_{-0.7}$ | $0.06^{+0.09}_{-0.04}$ | 369.7/337 |
| 00012172033 | $0.16 \pm 0.003$ | $0.386 \pm 0.008$ | $74 \pm 4$ | $0.180^{+0.012}_{-0.011}$ | $4.2^{+1.2}_{-0.8}$ | $2.0^{+0.9}_{**}$ | $0.017^{+0.063}_{-0.014}$ | 334.9/324 |
| 00012172034 | $0.15 \pm 0.03$ | $0.386 \pm 0.007$ | $75 \pm 3$ | $0.183^{+0.013}_{-0.012}$ | $3.9^{+1.1}_{-0.7}$ | $2.6^{+0.5}_{**}$ | $0.06^{+0.08}_{-0.04}$ | 392.0/342 |
| 00012172035 | $0.15 \pm 0.03$ | $0.384^{+0.008}_{-0.007}$ | $71 \pm 4$ | $0.187^{+0.013}_{-0.012}$ | $3.5^{+1.0}_{-0.6}$ | $2.2^{+0.6}_{-0.7}$ | $0.03^{+0.05}_{-0.02}$ | 379.0/342 |
| 00012172036 | $0.14^{+0.04}_{-0.03}$ | $0.385^{+0.008}_{-0.007}$ | $70^{+4}_{-3}$ | $0.186^{+0.014}_{-0.013}$ | $3.5^{+1.1}_{-0.7}$ | $2.6^{+0.6}_{-1.0}$ | $0.04^{+0.08}_{-0.03}$ | 323.6/327 |
| 00012172037 | $0.14^{+0.05}_{-0.04}$ | $0.381^{+0.009}_{-0.008}$ | $71 \pm 4$ | $0.18 \pm 0.02$ | $3.4^{+1.4}_{-0.7}$ | $2.7^{+0.6}_{-0.9}$ | $0.05^{+0.09}_{-0.04}$ | 367.0/325 |
| 00012172038 | $0.17^{+0.05}_{-0.04}$ | $0.366 \pm 0.007$ | $80 \pm 4$ | $0.164^{+0.015}_{-0.014}$ | $4.9^{+2.2}_{-1.1}$ | $2.9^{+0.5}_{-0.6}$ | $0.10^{+0.12}_{-0.06}$ | 318.4/313 |
| 00012172039 | $0.19^{+0.04}_{-0.03}$ | $0.367^{+0.010}_{-0.009}$ | $74 \pm 5$ | $0.170 \pm 0.014$ | $4.7^{+2.0}_{-1.0}$ | $2.5^{+0.8}_{-1.1}$ | $0.04^{+0.10}_{-0.03}$ | 298.9/305 |
| 00012172040 | $0.15 \pm 0.03$ | $0.368 \pm 0.008$ | $73 \pm 4$ | $0.180^{+0.014}_{-0.013}$ | $3.8^{+1.3}_{-0.7}$ | $2.4 \pm 0.5$ | $0.04^{+0.05}_{-0.02}$ | 362.4/316 |
| 00012172041 | $0.09 \pm 0.05$ | $0.38 \pm 0.02$ | $62^{+11}_{-10}$ | $0.21^{+0.04}_{-0.03}$ | $2.3^{+1.4}_{-0.7}$ | [1.0] | $0.0035^{+0.0722}_{-0.0012}$ | 225.2/217 |
| 00012172042 | $0.15 \pm 0.03$ | $0.37 \pm 0.01$ | $66 \pm 5$ | $0.182^{+0.014}_{-0.013}$ | $3.6^{+1.2}_{-0.7}$ | $1.6^{+0.8}_{**}$ | $0.010^{+0.027}_{-0.007}$ | 331.5/298 |
| 00012172043 | $0.14^{+0.04}_{-0.03}$ | $0.360 \pm 0.009$ | $70 \pm 5$ | $0.180^{+0.016}_{-0.015}$ | $3.5^{+1.4}_{-0.8}$ | $2.3^{+0.6}_{-0.8}$ | $0.023^{+0.044}_{-0.014}$ | 376.2/291 |
| 00012172044 | $0.16 \pm 0.04$ | $0.349 \pm 0.010$ | $74^{+6}_{-5}$ | $0.17 \pm 0.02$ | $4.1^{+2.0}_{-1.0}$ | $2.5 \pm 0.6$ | $0.04^{+0.07}_{-0.03}$ | 294.0/294 |
| 00012172045 | $0.19 \pm 0.04$ | $0.350^{+0.011}_{-0.010}$ | $78 \pm 6$ | $0.164^{+0.015}_{-0.014}$ | $5.6^{+2.5}_{-1.3}$ | $2.7^{+0.6}_{-0.7}$ | $0.07^{+0.11}_{-0.05}$ | 296.7/283 |
| 00012172046 | $0.14 \pm 0.04$ | $0.357 \pm 0.010$ | $69^{+6}_{-5}$ | $0.18 \pm 0.02$ | $3.4^{+1.8}_{-0.8}$ | $2.3 \pm 0.6$ | $0.03^{+0.06}_{-0.02}$ | 324.8/298 |
| 00012172047 | $0.13^{+0.04}_{-0.03}$ | $0.360^{+0.013}_{-0.012}$ | $63 \pm 6$ | $0.19 \pm 0.02$ | $3.0^{+1.3}_{-0.7}$ | $2.1^{+0.7}_{-0.9}$ | $0.018^{+0.044}_{-0.014}$ | 275.8/292 |
| 00012172048 | $0.16^{+0.07}_{-0.05}$ | $0.343^{+0.016}_{-0.014}$ | $69^{+9}_{-8}$ | $0.17^{+0.03}_{-0.02}$ | $3.6^{+3.1}_{-1.1}$ | $2.7^{+0.7}_{-0.8}$ | $0.04^{+0.09}_{-0.03}$ | 260.3/242 |
| 00012172049 | $0.14^{+0.04}_{-0.03}$ | $0.359^{+0.012}_{-0.013}$ | $63^{+7}_{-4}$ | $0.186^{+0.015}_{-0.017}$ | $3.2^{+1.4}_{-0.7}$ | $1.3^{+1.0}_{**}$ | $0.005^{+0.020}_{-0.002}$ | 265.6/271 |
| 00012172050 | [0.15]$^b$ | $0.35 \pm 0.02$ | $79^{+15}_{-10}$ | $0.172^{+0.014}_{-0.016}$ | $4.4^{+1.2}_{-0.7}$ | [1.0] | $0.005^{+0.274}_{-0.002}$ | 129.5/161 |
| 00012172051 | $0.17^{+0.05}_{-0.04}$ | $0.338^{+0.014}_{-0.013}$ | $68 \pm 7$ | $0.17 \pm 0.02$ | $3.8^{+2.1}_{-1.0}$ | $2.7^{+0.5}_{-0.7}$ | $0.04^{+0.06}_{-0.03}$ | 296.5/277 |
| 00012172052 | $0.17^{+0.05}_{-0.04}$ | $0.339 \pm 0.011$ | $68^{+7}_{-6}$ | $0.17 \pm 0.02$ | $4.0^{+2.2}_{-1.1}$ | $2.0 \pm 0.8$ | $0.012^{+0.030}_{-0.009}$ | 277.6/257 |
| 00012172053 | $0.13 \pm 0.04$ | $0.343^{+0.013}_{-0.012}$ | $65^{+7}_{-6}$ | $0.18 \pm 0.02$ | $3.0^{+1.6}_{-0.8}$ | $2.3^{+0.7}_{-0.6}$ | $0.020^{+0.039}_{-0.014}$ | 317.7/267 |
| 00012172054 | $0.10^{+0.04}_{-0.03}$ | $0.347 \pm 0.013$ | $62^{+8}_{-7}$ | $0.20 \pm 0.02$ | $2.4^{+1.3}_{-0.6}$ | $1.7^{+1.0}_{**}$ | $0.005^{+0.023}_{-0.004}$ | 273.6/261 |
| 00012172055 | $0.11^{+0.08}_{-0.07}$ | $0.33^{+0.03}_{-0.02}$ | $61 \pm 15$ | $0.19^{+0.06}_{-0.04}$ | $2.2^{+3.6}_{-0.9}$ | $1.9^{+1.2}_{**}$ | $0.008^{+0.047}_{-0.007}$ | 203.6/198 |
| 00012172056 | $0.16 \pm 0.04$ | $0.339^{+0.012}_{-0.013}$ | $64^{+8}_{-7}$ | $0.176^{+0.015}_{-0.016}$ | $3.5^{+1.7}_{-0.8}$ | $1.4^{+1.0}_{**}$ | $0.004^{+0.016}_{-0.002}$ | 263.3/248 |
| 00012172057 | $0.13^{+0.06}_{-0.05}$ | $0.34 \pm 0.02$ | $56^{+9}_{-7}$ | $0.18^{+0.03}_{-0.02}$ | $2.9^{+2.5}_{-0.8}$ | $2.3^{+0.9}_{-1.3}$ | $0.017^{+0.060}_{-0.015}$ | 223.4/231 |
| 00012172058 | $0.15^{+0.04}_{-0.03}$ | $0.339 \pm 0.014$ | $61^{+8}_{-7}$ | $0.18 \pm 0.02$ | $3.2^{+1.4}_{-0.7}$ | $1.8^{+1.0}_{**}$ | $0.007^{+0.029}_{-0.005}$ | 255.8/252 |
| 00012172059 | $0.14^{+0.04}_{-0.03}$ | $0.346^{+0.015}_{-0.014}$ | $55^{+8}_{-6}$ | $0.19 \pm 0.02$ | $2.9^{+1.3}_{-0.6}$ | $1.6^{+0.8}_{**}$ | $0.006^{+0.015}_{-0.004}$ | 296.4/255 |
| 00012172060 | $0.13^{+0.05}_{-0.03}$ | $0.35 \pm 0.02$ | $51^{+11}_{-8}$ | $0.20 \pm 0.02$ | $2.5^{+1.6}_{-0.6}$ | $2.1 \pm 0.5$ | $0.03^{+0.04}_{-0.02}$ | 326.3/279 |

Notes: columns and fitting limits are defined as in Table A4. Parameter values frozen during the fitting process are in square brackets.

Table A6: Best-fitting parameters to the *Swift*/XRT data (0.5–10 keV band), for the high/soft state epochs only. The model is tbabs×absori × (apec + diskbb + powerlaw) (Scenario 3). Uncertainties are 90% confidence levels for one independent parameter.

| ObsID | $N_H^a$ ($10^{22}$ cm$^{-2}$) | $N_{H_{ab}}^b$ ($10^{22}$ cm$^{-2}$) | $\xi^c$ | $kT^d$ (keV) | $N_{apec}^e$ | $kT_{in}$ (keV) | $\sqrt{N_{disk}}$ (km) | $\Gamma$ | $N_{pl}$ ($10^{-2}$) | $\chi^2/\nu$ |
|---|---|---|---|---|---|---|---|---|---|---|
| 00012172003 | $0.05^{+0.04}_{-0.03}$ | $0.58^{+0.16}_{-0.15}$ | $28^{+40}_{-17}$ | $0.69 \pm 0.06$ | $0.13^{+0.05}_{-0.04}$ | $0.63 \pm 0.02$ | $35 \pm 2$ | $2.0^{+0.4}_{-0.5}$ | $18^{+19}_{-11}$ | 498.1/448 |
| 00012172004 | $0.047^{+0.017}_{-0.014}$ | $1.0^{+0.3}_{-0.2}$ | $100^{+110}_{-42}$ | $0.80 \pm 0.07$ | $0.09 \pm 0.02$ | $0.63 \pm 0.02$ | $39^{+3}_{-2}$ | $1.4 \pm 0.4$ | $7^{+8}_{-4}$ | 472.0/476 |
| 00012172005 | $0.09^{+0.11}_{-0.06}$ | $0.9^{+0.5}_{-0.4}$ | $40^{+520}_{-30}$ | $0.74^{+0.13}_{-0.11}$ | $0.21^{+0.27}_{-0.10}$ | $0.60 \pm 0.04$ | $46^{+8}_{-6}$ | $2.0^{+1.0}_{-0.8}$ | $20^{+92}_{-19}$ | 285.0/291 |
| 00012172006 | $0.038^{+0.011}_{-0.014}$ | $1.3^{+0.4}_{-0.5}$ | $600^{+800}_{-400}$ | $0.86^{+0.05}_{-0.11}$ | $0.09^{+0.04}_{-0.03}$ | $0.609^{+0.014}_{-0.012}$ | $43 \pm 3$ | $0.7^{+0.5}_{-0.4}$ | $1.4^{+2.5}_{-0.7}$ | 452.4/429 |
| 00012172008 | $0.034^{+0.012}_{-0.019}$ | $0.9^{+0.3}_{-0.2}$ | $250^{+530}_{-180}$ | $0.68^{+0.06}_{-0.07}$ | $0.12 \pm 0.03$ | $0.607 \pm 0.009$ | $46.2^{+2.1}_{-1.9}$ | $0.7^{+0.6}_{-0.5}$ | $0.8^{+1.8}_{-0.5}$ | 568.3/448 |
| 00088999001 | [0.02] | $1.2 \pm 0.2$ | $600 \pm 200$ | $0.76^{+0.08}_{-0.07}$ | $0.20 \pm 0.04$ | $0.609 \pm 0.008$ | $48.9 \pm 2.0$ | $0.25^{+0.14}_{-0.10}$ | $0.31^{+0.09}_{-0.06}$ | 512.6/434 |
| 00012172009 | $0.042^{+0.019}_{-0.015}$ | $0.8 \pm 0.2$ | $60^{+80}_{-30}$ | $0.75^{+0.08}_{-0.07}$ | $0.10 \pm 0.03$ | $0.591 \pm 0.014$ | $46 \pm 3$ | $1.6^{+0.4}_{-0.5}$ | $7^{+9}_{-4}$ | 567.3/456 |
| 00012172010 | $0.036^{+0.018}_{-0.015}$ | $0.9 \pm 0.3$ | $90^{+110}_{-40}$ | $0.78 \pm 0.08$ | $0.08 \pm 0.02$ | $0.58 \pm 0.02$ | $41 \pm 3$ | $1.5 \pm 0.4$ | $5^{+7}_{-3}$ | 480.1/440 |
| 00012172011 | $0.03^{+0.04}_{-0.03}$ | $0.4^{+0.4}_{-0.3}$ | $60^{+100}_{-30}$ | $0.76^{+0.12}_{-0.13}$ | $0.09^{+0.06}_{-0.04}$ | $0.56 \pm 0.02$ | $45^{+5}_{-4}$ | $1.9^{+0.3}_{-0.4}$ | $25^{+21}_{-13}$ | 484.1/426 |
| 00012172012 | $0.02^{+0.04}_{-0.02}$ | $0.5 \pm 0.2$ | $30 \pm 30$ | $0.70^{+0.09}_{-0.12}$ | $0.11^{+0.06}_{-0.04}$ | $0.58 \pm 0.02$ | $42 \pm 4$ | $1.9 \pm 0.4$ | $20^{+20}_{-11}$ | 368.3/432 |
| 00012172013 | $0.03^{+0.03}_{-0.02}$ | $1.0 \pm 0.3$ | $170^{+100}_{-60}$ | [0.8] | $0.044^{+0.015}_{-0.013}$ | $0.57 \pm 0.02$ | $40 \pm 3$ | $1.9^{+0.3}_{-0.4}$ | $12^{+10}_{-6}$ | 431.7/412 |
| 00012172014 | $0.02^{+0.05}_{-0.02}$ | $0.31^{+0.15}_{-0.09}$ | $10^{+49}_{-10}$ | $0.65^{+0.07}_{-0.05}$ | $0.10^{+0.07}_{-0.04}$ | $0.59 \pm 0.02$ | $38^{+3}_{-2}$ | $2.0 \pm 0.6$ | $10^{+21}_{-7}$ | 453.0/418 |
| 00012172018 | [0.02] | $1.67^{+0.07}_{-0.08}$ | $1500^{+200}_{-700}$ | $0.853^{+0.035}_{-0.017}$ | $0.117 \pm 0.009$ | $0.53 \pm 0.02$ | $46.8 \pm 0.2$ | $0.21 \pm 0.02$ | $0.077 \pm 0.014$ | 401.9/362 |
| 00012172019 | [0.02] | $0.8 \pm 0.4$ | $700^{+4300}_{-300}$ | $0.66^{+0.141}_{-0.17}$ | $0.06 \pm 0.02$ | $0.555^{+0.010}_{-0.012}$ | $41 \pm 3$ | $0.4^{+1.1}_{-0.2}$ | $0.15^{+0.86}_{-0.06}$ | 349.6/336 |
| 00012172020 | $0.013 \pm 0.003$ | $1.69^{+0.07}_{-0.06}$ | $1700^{+700}_{-1100}$ | $0.86 \pm 0.02$ | $0.132 \pm 0.009$ | $0.549 \pm 0.002$ | $46.3 \pm 0.2$ | $0.130 \pm 0.013$ | $0.13 \pm 0.02$ | 466.6/389 |
| 00012172021 | $0.037^{+0.015}_{-0.007}$ | $1.86^{+0.10}_{-0.47}$ | $1000^{+4000}_{-6}$ | $0.86^{+0.03}_{-0.06}$ | $0.15^{+0.02}_{-0.04}$ | $0.513^{+0.010}_{-0.009}$ | $55^{+4}_{-3}$ | $0.36^{+0.37}_{-0.02}$ | $0.12^{+0.12}_{-0.02}$ | 438.7/363 |
| 00012172022 | $0.03 \pm 0.02$ | $1.4^{+0.5}_{-0.4}$ | $700^{+1200}_{-500}$ | $0.79 \pm 0.07$ | $0.12 \pm 0.05$ | $0.5201^{+0.0060}_{-0.0007}$ | $53 \pm 4$ | $0.5^{+0.8}_{-0.3}$ | $0.16^{+0.56}_{-0.08}$ | 407.4/362 |
| 00012172023 | $0.02 \pm 0.02$ | $0.6^{+0.3}_{-0.2}$ | $60^{+230}_{-40}$ | $0.70^{+0.09}_{-0.11}$ | $0.05^{+0.03}_{-0.02}$ | $0.519^{+0.019}_{-0.012}$ | $43 \pm 3$ | $1.6^{+0.8}_{-1.0}$ | $1.3^{+4.3}_{-1.1}$ | 359.0/356 |
| 00012172024 | [0.02] | $1.64 \pm 0.11$ | $1500^{+400}_{-800}$ | $0.87^{+0.02}_{-0.04}$ | $0.116 \pm 0.011$ | $0.511 \pm 0.003$ | $51.0^{+0.3}_{-0.2}$ | $0.30^{+0.03}_{-0.02}$ | $0.10 \pm 0.02$ | 351.0/329 |
| 00012172025 | $0.03^{+0.04}_{-0.03}$ | $0.8^{+0.5}_{-0.3}$ | $60^{+540}_{-30}$ | $0.71^{+0.11}_{-0.13}$ | $0.07^{+0.06}_{-0.03}$ | $0.51 \pm 0.02$ | $48 \pm 5$ | $1.8^{+1.3}_{-1.6}$ | $1.7^{+14.6}_{-1.6}$ | 287.0/295 |
| 00012172026 | $0.013 \pm 0.003$ | $1.57^{+0.07}_{-0.30}$ | $1200^{+3800}_{-600}$ | $0.83 \pm 0.02$ | $0.120 \pm 0.009$ | $0.499 \pm 0.002$ | $50.3^{+1.1}_{-0.2}$ | $0.2^{+0.4}_{-0.2}$ | $0.11 \pm 0.02$ | 415.8/345 |
| 00012172027 | [0.02] | $0.4^{+0.4}_{-0.2}$ | $80^{+70}_{-40}$ | $0.64^{+0.18}_{-0.09}$ | $0.05^{+0.03}_{-0.02}$ | $0.503^{+0.007}_{-0.006}$ | $46^{+3}_{-2}$ | $2.2^{+0.3}_{-0.6}$ | $5 \pm 3$ | 429.9/360 |
| 00012172028 | $0.020^{+0.016}_{-0.009}$ | $1.36^{+0.10}_{-0.37}$ | $1000^{+1400}_{-600}$ | $0.86^{+0.05}_{-0.14}$ | $0.08^{+0.02}_{-0.03}$ | $0.502^{+0.003}_{-0.013}$ | $50.0^{+5.1}_{-0.3}$ | $0.4^{+0.5}_{-0.3}$ | $0.12^{+0.18}_{-0.03}$ | 305.7/308 |
| 00012172029 | [0.02] | $0.43^{+0.22}_{-0.14}$ | $50^{+60}_{-20}$ | $0.60 \pm 0.09$ | $0.06^{+0.04}_{-0.02}$ | $0.506^{+0.007}_{-0.007}$ | $45.3^{+1.9}_{-2.1}$ | $2.1^{+0.4}_{-0.7}$ | $2.1^{+2.3}_{-1.5}$ | 411.2/351 |
| 00012172030 | [0.02] | $1.6^{+0.2}_{-0.3}$ | $1000 \pm 400$ | $0.87^{+0.05}_{-0.03}$ | $0.10^{+0.02}_{-0.03}$ | $0.493^{+0.008}_{-0.007}$ | $52 \pm 3$ | $0.40^{+0.26}_{-0.05}$ | $0.15^{+0.09}_{-0.02}$ | 366.0/347 |
| 00012172031 | $0.04 \pm 0.02$ | $1.2^{+0.7}_{-0.5}$ | $400^{+4500}_{-40}$ | $0.77^{+0.08}_{-0.09}$ | $0.08^{+0.04}_{-0.03}$ | $0.484^{+0.010}_{-0.011}$ | $51^{+5}_{-4}$ | $0.7^{+0.9}_{-0.5}$ | $0.15^{+0.69}_{-0.09}$ | 431.0/333 |
| 00012172032 | $0.018 \pm 0.013$ | $1.1 \pm 0.5$ | $230^{+580}_{-120}$ | $0.84^{+0.08}_{-0.10}$ | $0.05^{+0.03}_{-0.02}$ | $0.485 \pm 0.010$ | $50^{+4}_{-3}$ | $1.4^{+0.8}_{-0.7}$ | $0.7^{+1.9}_{-0.5}$ | 368.9/335 |
| 00012172033 | [0.02] | $2.44 \pm 0.11$ | $1500^{+300}_{-800}$ | $0.91 \pm 0.02$ | $0.145 \pm 0.009$ | $0.460 \pm 0.002$ | $56.5 \pm 0.3$ | $0.351^{+0.018}_{-0.017}$ | $0.084 \pm 0.014$ | 333.7/324 |
| 00012172034 | $0.014^{+0.009}_{-0.014}$ | $1.5^{+0.4}_{-0.5}$ | $500^{+4400}_{-300}$ | $0.86^{+0.05}_{-0.08}$ | $0.07^{+0.04}_{-0.02}$ | $0.477^{+0.010}_{-0.007}$ | $52^{+3}_{-4}$ | $0.9^{+0.6}_{-0.4}$ | $0.22^{+0.49}_{-0.13}$ | 390.9/340 |
| 00012172035 | $0.020^{+0.009}_{-0.011}$ | $1.5^{+0.4}_{-0.3}$ | $410^{+570}_{-180}$ | $0.86 \pm 0.05$ | $0.050^{+0.038}_{-0.012}$ | $0.473 \pm 0.007$ | $49 \pm 3$ | $1.1^{+0.4}_{-0.5}$ | $0.3^{+0.4}_{-0.2}$ | 389.0/340 |
| 00012172036 | [0.02] | $1.4^{+0.5}_{-0.3}$ | $400^{+600}_{-300}$ | $0.86^{+0.05}_{-0.04}$ | $0.062^{+0.038}_{-0.013}$ | $0.470^{+0.005}_{-0.008}$ | $49.8^{+3.5}_{-1.8}$ | $0.9^{+0.6}_{-0.4}$ | $0.15^{+0.33}_{-0.08}$ | 327.1/326 |
| 00012172037 | [0.02] | $1.8^{+0.3}_{-0.5}$ | $900^{+800}_{-600}$ | $0.88^{+0.06}_{-0.08}$ | $0.07^{+0.02}_{-0.03}$ | $0.466 \pm 0.006$ | $51 \pm 2$ | $0.61^{+0.54}_{-0.12}$ | $0.10^{+0.16}_{-0.02}$ | 355.7/324 |
| 00012172038 | [0.02] | $0.5 \pm 0.2$ | $64^{+80}_{-30}$ | $0.68^{+0.09}_{-0.08}$ | $0.06^{+0.03}_{-0.02}$ | $0.460^{+0.008}_{-0.006}$ | $51 \pm 3$ | $2.3^{+0.4}_{-0.6}$ | $3^{+3}_{-2}$ | 333.4/312 |
| 00012172039 | [0.02] | $0.36^{+0.15}_{-0.10}$ | $110^{+190}_{-60}$ | [0.6] | $0.07 \pm 0.03$ | $0.452^{+0.008}_{-0.007}$ | $48 \pm 3$ | $2.6^{+0.2}_{-0.4}$ | $4.0^{+1.7}_{-1.9}$ | 314.1/305 |
| 00012172040 | $0.018^{+0.016}_{-0.015}$ | $0.9^{+0.5}_{-0.4}$ | $130^{+190}_{-70}$ | $0.83^{+0.10}_{-0.11}$ | $0.039^{+0.013}_{-0.012}$ | $0.449 \pm 0.011$ | $51 \pm 4$ | $1.8 \pm 0.6$ | $1.4^{+2.4}_{-1.0}$ | 380.9/314 |
| 00012172041 | [0.02] | $3.1 \pm 0.3$ | $1700^{+800}_{-1300}$ | [1.0] | $0.17 \pm 0.02$ | $0.412 \pm 0.004$ | $64.8 \pm 0.7$ | $0.33 \pm 0.03$ | $0.12 \pm 0.03$ | 223.9/217 |
| 00012172042 | [0.02] | $2.2 \pm 0.7$ | $600^{+600}_{-300}$ | $0.93^{+0.05}_{-0.08}$ | $0.08^{+0.05}_{-0.04}$ | $0.436^{+0.005}_{-0.007}$ | $53^{+3}_{-2}$ | $0.9^{+0.6}_{-0.4}$ | $0.30^{+0.61}_{-0.14}$ | 342.3/297 |
| 00012172043 | [0.02] | $1.5 \pm 0.5$ | $400^{+500}_{-300}$ | [0.9] | $0.05^{+0.04}_{-0.03}$ | $0.431^{+0.005}_{-0.009}$ | $54^{+5}_{-3}$ | $0.9^{+1.1}_{-0.5}$ | $0.20^{+1.07}_{-0.12}$ | 375.5/291 |
| 00012172044 | $0.014^{+0.016}_{-0.014}$ | $0.9^{+0.4}_{-0.3}$ | $190^{+200}_{-80}$ | [0.8] | $0.024^{+0.010}_{-0.008}$ | $0.432^{+0.011}_{-0.012}$ | $50^{+5}_{-4}$ | $1.7 \pm 0.6$ | $1.0^{+1.9}_{-0.7}$ | 305.5/293 |
| 00012172045 | [0.02] | $1.0^{+0.6}_{-0.4}$ | $500^{+1000}_{-300}$ | $0.76^{+0.08}_{-0.12}$ | $0.047^{+0.014}_{-0.012}$ | $0.422^{+0.006}_{-0.006}$ | $57^{+3}_{-4}$ | $2.4^{+0.3}_{-0.6}$ | $4^{+2}_{-3}$ | 304.2/282 |
| 00012172046 | [0.02] | $0.8^{+0.7}_{-0.4}$ | $130^{+260}_{-60}$ | $0.78^{+0.12}_{-0.16}$ | $0.026^{+0.014}_{-0.010}$ | $0.435^{+0.006}_{-0.006}$ | $48.3^{+1.9}_{-3.0}$ | $2.0^{+0.6}_{-0.9}$ | $1.5^{+2.4}_{-1.2}$ | 330.9/297 |
| 00012172047 | [0.02] | $1.2^{+0.4}_{-0.3}$ | $400^{+900}_{-200}$ | [0.8] | $0.030^{+0.018}_{-0.008}$ | $0.417^{+0.006}_{-0.005}$ | $52^{+2}_{-3}$ | $2.3^{+0.3}_{-0.6}$ | $2.4^{+1.7}_{-1.5}$ | 291.9/292 |
| 00012172048 | [0.02] | $1.9^{+1.1}_{-0.8}$ | $370^{+460}_{-160}$ | [0.9] | $0.027^{+0.018}_{-0.007}$ | $0.420^{+0.006}_{-0.007}$ | $50.6^{+2.1}_{-1.8}$ | $1.6^{+0.8}_{-0.6}$ | $0.5^{+1.7}_{-0.4}$ | 266.3/242 |
| 00012172049 | [0.02] | $0.38^{+0.74}_{-0.14}$ | $230^{+3030}_{-140}$ | [0.6] | $0.05^{+0.02}_{-0.03}$ | $0.410^{+0.009}_{-0.008}$ | $50 \pm 4$ | $2.7^{+0.2}_{-0.7}$ | $4.2^{+1.6}_{-2.8}$ | 281.1/271 |
| 00012172050 | [0.02] | $1.9^{+3.3}_{-1.9}$ | $600^{+4400}_{-600}$ | $0.7^{+0.3}_{-0.4}$ | $0.04^{+0.06}_{-0.02}$ | $0.41 \pm 0.02$ | $59^{+10}_{-7}$ | $1.080^{+2.0}_{-0.7}$ | $0.496^{+2.5}_{-0.3}$ | 126.4/159 |
| 00012172051 | [0.02] | $1.4^{+0.6}_{-0.5}$ | $600^{+1300}_{-300}$ | [0.7] | $0.020^{+0.006}_{-0.005}$ | $0.406 \pm 0.005$ | $51.0^{+1.8}_{-1.6}$ | $2.2^{+0.2}_{-0.4}$ | $2.0 \pm 0.9$ | 302.5/277 |
| 00012172052 | [0.02] | $0.9^{+1.7}_{-0.5}$ | $100^{+760}_{-60}$ | $0.8 \pm 0.2$ | $0.031^{+0.014}_{-0.011}$ | $0.403 \pm 0.007$ | $51^{+2}_{-3}$ | $1.7 \pm 1.0$ | $0.7^{+2.2}_{-0.6}$ | 284.2/256 |
| 00012172053 | $0.018 \pm 0.014$ | $1.9^{+0.6}_{-0.5}$ | $320^{+220}_{-90}$ | [0.9] | $0.021^{+0.012}_{-0.009}$ | $0.409 \pm 0.009$ | $51^{+4}_{-3}$ | $1.7^{+0.6}_{-0.5}$ | $0.6^{+1.0}_{-0.4}$ | 311.5/266 |
| 00012172054 | $0.012^{+0.021}_{-0.012}$ | $1.0^{+0.6}_{-0.5}$ | $150^{+4850}_{-80}$ | $0.89^{+0.10}_{-0.16}$ | $0.025^{+0.014}_{-0.010}$ | $0.397^{+0.010}_{-0.013}$ | $54^{+6}_{-4}$ | $1.9^{+0.8}_{-1.0}$ | $0.7^{+2.1}_{-0.6}$ | 275.8/259 |
| 00012172055 | $0.02^{+0.02}_{**}$ | $1.6^{+1.2}_{-1.1}$ | $360^{+640}_{-180}$ | $0.9^{+0.2}_{-0.3}$ | $0.015^{+0.024}_{-0.011}$ | $0.382 \pm 0.014$ | $50^{+8}_{-6}$ | $1.5^{+0.9}_{-0.8}$ | $0.4^{+1.4}_{-0.3}$ | 198.8/196 |
| 00012172056 | $0.03 \pm 0.02$ | $2.6^{+0.6}_{-0.7}$ | $700^{+1500}_{-500}$ | [0.9] | $0.09 \pm 0.04$ | $0.378^{+0.013}_{-0.015}$ | $64^{+12}_{-7}$ | $0.7^{+0.8}_{-0.2}$ | $0.14^{+0.38}_{-0.05}$ | 256.7/247 |
| 00012172057 | [0.02] | $1.07^{+0.16}_{-0.59}$ | $1000^{+4000}_{-700}$ | [0.6] | $0.012^{+0.010}_{-0.004}$ | $0.399^{+0.004}_{-0.006}$ | $46 \pm 3$ | $2.5^{+0.2}_{-0.4}$ | $2.4^{+1.9}_{-1.1}$ | 232.3/231 |
| 00012172058 | [0.02] | $2.1^{+0.9}_{-0.8}$ | $290^{+450}_{-90}$ | $0.90^{+0.07}_{-0.06}$ | $0.04^{+0.03}_{-0.02}$ | $0.383 \pm 0.005$ | $57 \pm 3$ | $1.6^{+0.9}_{-0.7}$ | $0.5^{+1.6}_{-0.3}$ | 264.3/251 |
| 00012172059 | [0.02] | $2.5^{+1.0}_{-0.7}$ | $300^{+300}_{-100}$ | $0.87^{+0.09}_{-0.05}$ | $0.038^{+0.027}_{-0.009}$ | $0.387 \pm 0.005$ | $53.5^{+2.2}_{-0.9}$ | $1.6^{+0.5}_{-0.6}$ | $0.6^{+0.8}_{-0.4}$ | 313.1/254 |
| 00012172060 | $0.013^{+0.021}_{-0.013}$ | $2.7^{+1.6}_{-1.2}$ | $470^{+4400}_{-140}$ | $0.91^{+0.07}_{-0.15}$ | $0.036^{+0.021}_{-0.014}$ | $0.399^{+0.013}_{-0.016}$ | $48^{+6}_{-4}$ | $1.9^{+0.5}_{-0.4}$ | $2.3^{+2.7}_{-1.2}$ | 320.3/277 |

$^a$: column density of the neutral absorber.
$^b$: column density of the ionized absorber (`absori` model).
$^c$: ionization parameter of the ionized absorber ($\xi = L/nR^2$).
$^d$: temperature of the thermal-plasma emitter (`apec` model).
$^e$: normalization of the `apec` component, defined as $N_{\rm apec} = \left[10^{-14}/(4\pi d^2)\right] \int n_e\, n_{\rm H}\, dV$.

This paper has been typeset from a TEX/LATEX file prepared by the author.